\documentclass[twocolumn, amssymb, aps, prb, superscriptaddress, nofootinbib]{revtex4-2} 

\usepackage{amsmath}
\usepackage{tabularx}
\usepackage{hyperref}
\hypersetup{colorlinks=true, citecolor=blue, urlcolor=blue, linkcolor=blue} \usepackage{siunitx}
\usepackage{graphicx}
\usepackage{bm}
\usepackage{xcolor}

\usepackage{ifthen}
\usepackage{braket}
\usepackage{multirow}
\usepackage{soul}
\usepackage{svg}
\usepackage[normalem]{ulem} 
\usepackage{booktabs}
\usepackage{colortbl}
\usepackage{mathtools}

\usepackage[most]{tcolorbox}

\usepackage[symbol]{footmisc}

\definecolor{purple}{rgb}{0.75,0.0,0.75}
\definecolor{darkgreen}{RGB}{0, 136, 58}
\definecolor{darkblue}{RGB}{0, 0, 139}
\definecolor{darkgreen}{RGB}{25, 125, 25}

\def\bk{{\bf k}}
\def\bq{{\bf q}}

\def\bG{{\bf G}}
\def\bkp{{{\bf k}^\prime}}
\def\bqp{{{\bf q}^\prime}}
\def\ba{{\bf a}}
\def\br{{\bf r}}
\def\bR{{\bf R}}
\def\bT{{\bf T}}

\def\Rp{{{\bf R}_p}}
\def\Rpp{{{\bf R}_{p^\prime}}}
\def\Rppp{{{\bf R}_{p^{\prime\prime}}}}

\def\a{\alpha}
\def\b{\beta}
\def\o{\omega}
\def\O{\Omega}
\def\k{\kappa}
\def\ap{\alpha^\prime}

\def\kp{\kappa^\prime}

\def\sc{\text{sc}}
\def\uc{\text{uc}}
\def\dag{\dagger}

\def\sU{\mathcal{U}}

\def\bv{{\bf v}}

\def\bRhp{{\bf R}_{p}^{H}}
\def\bRhpE{{\bf R}_{p}^{H \text{(E)}}}
\def\bRhpV{{\bf R}_{p}^{H \text{(V)}}}

\def\bRdp{{\bf R}_{p}^{D}}
\def\bRdpE{{\bf R}_{p}^{D \text{(E)}}}
\def\bRdpV{{\bf R}_{p}^{D \text{(V)}}}
\def\bRdpp{{\bf R}_{p^\prime}^{D}}

\def\bRgo{{\bf R}_{0}^{g}}
\def\bRgp{{\bf R}_{p}^{g}}
\def\bRgpp{{\bf R}_{p^\prime}^{g}}
\def\bRgppp{{\bf R}_{p^{\prime\prime}}^{g}}

\def\bRep{{\bf R}_{p}^{\text{e}}}
\def\bRepE{{\bf R}_{p}^{\text{e(E)}}}
\def\bRepV{{\bf R}_{p}^{\text{e(V)}}}

\def\bReppp{{\bf R}_{p''}^{\text{e}}}
\def\bReo{{\bf R}_{0}^{\text{e}}}
\def\bRphp{{\bf R}_{p}^{\text{ph}}}
\def\bRphpE{{\bf R}_{p}^{\text{ph(E)}}}
\def\bRphpV{{\bf R}_{p}^{\text{ph(V)}}}
\def\bRphpp{{\bf R}_{p'}^{\text{ph}}}

\def\bRpho{{\bf R}_{0}^{\text{ph}}}

\def\ppsi{\tilde{\psi}}
\def\bRgoE{{\bRgo}^{\text{(E)}}}
\def\bRgpE{{\bRgp}^{\text{(E)}}}
\def\bRgppE{{\bf R}_{p^{\prime}}^{g \text{(E)}}}
\def\bRgoV{{\bRgo}^{\text{(V)}}}
\def\bRgpV{{\bRgp}^{\text{(V)}}}
\def\bRgpppV{{\bf R}_{p^{\prime\prime}}^{g \text{(V)}}}

\newcolumntype{L}{>{\raggedright\arraybackslash}X} 
\newcolumntype{R}{>{\raggedleft\arraybackslash}X} 
\newcolumntype{C}{>{\centering\arraybackslash}X} 
\newcolumntype{?}{!{\vrule width 1pt}} 
\newcolumntype{:}{!{\color{gray}\vrule}} 
\begin{document}

\title{EPW-VASP interface for first-principles calculations of electron-phonon interactions}

\author{Danylo Radevych\footnotemark[2]{$^\dagger$}}
\thanks{These authors contributed equally.}
\affiliation{
Department of Physics, Applied Physics and Astronomy, Binghamton University-SUNY,
PO Box 6000, Binghamton, New York 13902, USA
}

\author{Aidan Thorn}
\thanks{These authors contributed equally.}
\affiliation{
Department of Physics, Applied Physics and Astronomy, Binghamton University-SUNY,
PO Box 6000, Binghamton, New York 13902, USA
}

\author{Manuel Engel}
\affiliation{University of Vienna, Faculty of Physics and Center for Computational Materials Physics, A-1090 Vienna, Austria}
\affiliation{VASP Software GmbH, A-1090, Vienna, Austria}

\author{Aleksey N. Kolmogorov}
\affiliation{
Department of Physics, Applied Physics and Astronomy, Binghamton University-SUNY,
PO Box 6000, Binghamton, New York 13902, USA
}

\author{Sabyasachi Tiwari}
\affiliation{
Oden Institute for Computational Engineering and Sciences, The University of Texas at Austin, Austin, TX 78712, USA
}
\affiliation{
Department of Physics, The University of Texas at Austin, Austin, TX 78712, USA
}

\author{Georg Kresse}
\affiliation{University of Vienna, Faculty of Physics and Center for Computational Materials Physics, A-1090 Vienna, Austria}
\affiliation{VASP Software GmbH, A-1090, Vienna, Austria}

\author{Feliciano Giustino}
\affiliation{
Oden Institute for Computational Engineering and Sciences, The University of Texas at Austin, Austin, TX 78712, USA
}
\affiliation{
Department of Physics, The University of Texas at Austin, Austin, TX 78712, USA
}

\author{Elena R. Margine\footnotemark[2]{$^\ddagger$}}
\affiliation{
Department of Physics, Applied Physics and Astronomy, Binghamton University-SUNY,
PO Box 6000, Binghamton, New York 13902, USA
}

\date{\today}

\begin{abstract}
We present an interface between the Vienna \textit{Ab initio} Simulation Package (VASP) and the EPW software for calculating materials properties governed by electron-phonon (e-ph) interactions. Computation of the e-ph matrix elements with the finite-difference supercell approach in VASP and their fine-grid interpolation in EPW enable accurate modeling of temperature-dependent materials properties and phonon-assisted quantum processes with VASP's extensive library of exchange-correlation functionals and pseudopotentials. We demonstrate the functionality of the EPW-VASP interface by examining the superconducting gap and critical temperature in MgB$_2$ using the anisotropic Migdal-Eliashberg equations, and the carrier mobility in cubic BN using the \textit{ab initio} Boltzmann transport equation.
\end{abstract}

\maketitle

\footnotetext[2]{Corresponding author: \href{mailto:dradevych@binghamton.edu}{dradevych@binghamton.edu}}
\footnotetext[3]{Corresponding author: \href{mailto:rmargine@binghamton.edu}{rmargine@binghamton.edu}}

\section{Introduction}
\label{sec:intro}

Electron-phonon interactions (EPIs) are fundamental to a wide range of physical phenomena that underpin advanced materials properties, including electrical and thermal transport, optical absorption, superconductivity, and excitonic effects~\cite{Giustino2017, Ziman1960}.
First-principles calculations of EPIs have thus become central to accelerating the discovery and optimization of materials for next-generation technologies \cite{Samsonidze2018,  Kolmogorov2010, Duan2014, Sohier2018, Ponce2021, Ha2024}.
In parallel, the growing adoption of data-driven methodologies has created a pressing need for high-quality, reproducible EPI datasets to enable design of functional materials with machine learning models \cite{Samsonidze2018, Sohier2018,  Ha2024, Luo2024, Li2024, Zhong2024, Haldar2024, Cerqueira2024, Gibson2025, Madika2025}. Meeting this demand requires interoperable software ecosystems that support seamless integration across electronic structure codes, data formats, and workflows.

In this work, we have extended the interoperability of EPW~\cite{Noffsinger2010,Ponce2016,Lee2023}, a widely used code for computing EPIs~\cite{Giustino2017,Stefanucci2023}. EPW is a part of the Quantum ESPRESSO (QE) materials simulation suite~\cite{Giannozzi2017}, and utilizes electron-phonon matrix elements calculated via density functional perturbation theory (DFPT)~\cite{Baroni1987, Savrasov1992, Gonze1997, Baroni2001} to model a wide range of materials properties. These include charge carrier mobility under electric and magnetic fields using the Boltzmann transport equation~\cite{Ponce2018,Ponce2020,Ponce2023a,Ponce2023b,Leveillee2023,Cuco2024,Ha2024,Ha2025,Lihm2025,Liu2025}; conventional superconductivity within the anisotropic multi-band Eliashberg theory~\cite{Margine2013,Lucrezi2024,Mishra2024,Tomassetti2024,Mori2024,Gochitashvili2025,Mishra2025}; optical absorption processes treated via second-order time-dependent perturbation
theory~\cite{Noffsinger2012,ZhangX2022,Bushick2023} and many-body quasi-degenerate perturbation theory~\cite{Tiwari2024,Yang2025}; and the formation of polarons and excitonic polarons without requiring supercells~\cite{Sio2019a, Sio2019b, Lafuente2022a, Lafuente2022b,Dai2024a,Dai2024b}. EPW also incorporates advanced treatments of electron-phonon interactions, including long-range contributions arising from dipole and quadrupole contributions in polar materials~\cite{Verdi2015, Sjakste2015, Brunin2020a, Brunin2020b, Park2020, Jhalani2020, Ponce2021, Royo2021, Ponce2023a, Ponce2023b, Sohier2016, Deng2021, Sio2022, Zhang2022}, self-consistent on-site Hubbard contributions for strongly correlated systems~\cite{Yang2025}, vertex corrections for nonadiabatic superconductors~\cite{Mishra2025}, and spin-polarized effects for magnetic materials~\cite{Alvarez2025}.

The presently introduced interface with another major plane-wave density functional theory (DFT) package, the Vienna \textit{ab initio} simulation package (VASP)~\cite{Kresse1993, Kresse1994, Kresse1996a, Kresse1996b}, provides access to a wider range of features available through the electron-phonon matrix elements computed using the finite-difference supercell approach~\cite{Engel2020, Engel2022}. In particular, this combination enables the use of the projector augmented-wave (PAW) formalism~\cite{Blochl1994, Kresse1999}, hybrid functionals~\cite{Heyd2003}, and meta-generalized gradient approximation (meta-GGA) functionals such as the strongly-constrained and appropriately-normed (SCAN)~\cite{Sun2015} density functional and its variant r$^2$SCAN~\cite{Furness2020}.
Given that a large fraction of publicly available materials databases~\cite{Jain2013, Curtarolo2013, Saal2013, Kirklin2015, Zhou2019, Choudhary2020, Schmidt2022, Merchant2023}
are based on VASP, the present interface substantially enhances the interoperability and applicability of EPW.

To validate the functionality of the new interface, we examined the superconducting properties of MgB$_2$ using the anisotropic Eliashberg formalism and the electronic transport properties of cubic-BN using the Boltzmann transport equation implemented in EPW. The results obtained with the PAW Perdew-Burke-Ernzerhof (PBE) \cite{Perdew1996} functional via the EPW-VASP workflow show excellent agreement with reference calculations performed using norm-conserving (NC) PBE pseudopotentials in the EPW-QE workflow under comparable computational settings and grids. Furthermore, we assessed the effect of exchange-correlation functional choice on the superconducting properties of MgB$_2$ by employing the r$^2$SCAN meta-GGA functional.

The manuscript is organized as follows: Section \ref{sec:methodology} outlines the theoretical formalism underlying the interface. Section \ref{sec:usage} describes the workflow for performing calculations. Section \ref{sec:results} demonstrates the capabilities of the interface through example calculations for MgB$_2$ and c-BN, including comparisons with previous computational and experimental findings. Section \ref{sec:conclusion} concludes the manuscript with a summary of the key features and an overview of planned future developments. The Appendices provide additional technical details: Appendix~\ref{sec:WS} discusses the Wigner-Seitz construction; Appendix~\ref{sec:paw} details the computation of electron–phonon matrix elements within the PAW framework; and Appendix~\ref{sec:conventions} summarizes the differences in conventions between VASP and EPW.

\section{Methodology}
\label{sec:methodology}

Evaluating materials properties related to the electron–phonon interaction typically requires Brillouin zone (BZ) integrations over electron and phonon grids with very fine sampling.  However, this becomes impractical for direct DFPT calculations for a large number of phonon wave vectors or the finite-difference approach for very large supercells. The challenge has been overcome by computing the electron-phonon matrix elements in real space~\cite{Engel2020,Engel2022} using the interpolation method based on maximally-localized Wannier functions (MLWFs) introduced by Giustino \textit{et al.} in the EPW code within the DFPT framework~\cite{Giustino2007}. The EPW-VASP interface builds upon (i) the recent advancements in the VASP code for calculating EPIs via the PAW method and finite-displacement approach in real space~\cite{Engel2020, Engel2022}, and (ii) the expanded functionalities of the EPW code for computing materials properties related to electron–phonon interactions~\cite{Giustino2017, Lee2023}.

In this section, we first outline the procedure to obtain the electronic Hamiltonian and the electron-phonon matrix elements in the Wannier representation along with the interatomic force constants using the finite-difference supercell scheme in VASP. Next, we describe how these real-space quantities are employed to obtain their counterparts on dense electron and phonon momenta in Bloch space via Wannier-Fourier interpolation, using EPW. As the theoretical background and methodology have been thoroughly detailed in Refs.~\cite{Giustino2007, Giustino2017, Lee2023, Engel2020, Engel2022}, only the main equations relevant to the present discussion are presented here. The conventions and notations in this section follow those in Refs.~\cite{Giustino2017, Lee2023}.

\subsection{Wannier interpolation in VASP}

Electronic structure and lattice dynamics in crystalline solids are  modeled as infinitely extended systems using a Born–von Kármán (BvK) supercell. In this approach, periodic boundary conditions are applied to a supercell containing $N_p = N_1 \times N_2 \times N_3$ primitive unit cells, arranged according to the underlying Bravais lattice. The $p$th unit cell ($p = 1, \dots, N_p$) is identified by a direct lattice vector $\Rp = \sum_i p_i \ba_i$, with $\ba_i$ ($i=1,2,3$) the primitive lattice vectors and $p_i\in[0,N_i)$.  Exploiting this periodicity, Bloch's theorem enables the description of electronic states and lattice vibrations in terms of wavevectors defined over the crystal's Brillouin zone. The reciprocal lattice associated with the BvK supercell defines a uniform mesh of $N_p$ $\bk$- and $\bq$-points, corresponding to the electron and phonon wavevectors within the BZ. Note that the number of unit cells $N_p$ can be different for electrons and phonons.

Using Bloch's theorem, the eigenfunctions $\psi_{n\bk}(\br)$, with  band index $n$ and crystal momentum $\bk$, of the Kohn-Sham (KS) Hamiltonian can be expressed as
\begin{align}
    \psi_{n\bk}(\br) = \frac{1}{\sqrt{N_p}} u_{n\bk}(\br)\,e^{i{\bk \cdot \br}},
\label{Bwf}
\end{align}
where $u_{n\bk}(\br)$ is the Bloch-periodic component of the KS wavefunction. The wave function $\psi_{n\bk}(\br)$ is taken to be normalized to one in the BvK supercell,
while the periodic part $u_{n\bk}(\br)$ is normalized to one in the primitive unit cell~\cite{Giustino2017,Lee2023}.
Given an isolated Bloch band $m$, the Wannier function $w_{mp} (\br)$ associated with the unit cell identified by $\Rp$ is defined in terms of Bloch states $\psi_{n\bk} (\br)$ as~\cite{Marzari2012,Pizzi2020,Marrazzo2024}:
\begin{equation}
\label{eq:WanOrb}
w_{mp} (\br) = w_{m0}(\br - \Rp) = \frac{1}{\sqrt{N_p}} \sum_{n\bk} e^{-i\bk\cdot\Rp} U_{nm\bk} \psi_{n\bk} (\br).
\end{equation}
The inverse relation of Eq.~\eqref{eq:WanOrb} is obtained by a standard inverse Fourier transform:
\begin{equation}
\label{eq:WanOrbinv}
\psi_{n\bk} (\br) = \frac{1}{\sqrt{N_p}} \sum_{mp} e^{i\bk\cdot\Rp} U^\dag_{mn\bk} w_{mp} (\br).
\end{equation}
Here $U_{nm\bk}$ is a unitary matrix determined by minimizing the spatial spread of the functions $w_{mp}(\br)$, which transforms the Bloch wavefunctions to a Wannier gauge
\begin{equation}
\label{eq:PsiW}
\psi^{\rm W}_{m\bk} (\br) = \sum_{n} U_{nm\bk} \psi_{n\bk} (\br).
\end{equation}

The degree of localization  directly impacts the accuracy and efficiency of the interpolation procedure as more localized WFs decay faster in real space, allowing smaller BvK supercells to include all nonvanishing matrix elements of the Hamiltonian and other operators.

The Hamiltonian matrix elements in the Wannier basis are given by~\cite{Pizzi2020}
\begin{equation}
\label{eq:EPWHamWanOrb}
\quad H_{mn}(\Rp) = \braket{w_{m0}|\hat{H}|w_{np}} = \int_{\sc} d\br w^{*}_{m0}(\br) \hat{H} w_{n0}(\br - \Rp),
\end{equation}
where the subscript “sc” indicates that the integral is over the BvK supercell. Using Eq.~\eqref{eq:WanOrb}, these elements can be represented in terms of the DFT  KS Hamiltonian in the Bloch representation as
\begin{equation}
\label{eq:EPWHamWan}
H_{mn}(\Rp) = \frac{1}{N_p} \sum_{m^\prime n^\prime \bk} e^{-i\bk\cdot\Rp} U^\dag_{mm^\prime \bk} H_{m^\prime n^\prime} (\bk) U_{n^\prime n \bk},
\end{equation}
where $H_{m n}(\bk)=\delta_{mn} \epsilon_{n\bk}$. The eigenstates $\ket{\psi_{n{\bk}}}$ and eigenvalues $\epsilon_{n\bk}$ of the KS Hamiltonian are used to determine the unitary matrix for the transformation to MLWFs.

The electron-phonon matrix elements are the central quantity in the evaluation of properties governed by EPIs, describing the scattering of an electron from a Bloch eigenstate $\ket{\psi_{n{\bk}}}$ to $\ket{\psi_{m{\bk + \bq}}}$, where $\bk$ is the electron momentum, $\bq$ is the phonon momentum transfer, and $n$, $m$ are band indices. Within DFT, these electron-phonon matrix elements can be computed as~\cite{Giustino2007, Giustino2017}
\begin{equation}
\label{eq:EPWelph}
g_{mn\nu}(\bk,\bq) = \bra{\psi_{m\bk + \bq}} \Delta_{\bq\nu}\hat{H} \ket{\psi_{n\bk}}_{\sc}.
\end{equation}
The term $\Delta_{\bq\nu}\hat{H}$ is the first-order change of the KS Hamiltonian with respect to a collective ionic displacement corresponding to a phonon
with momentum $\bq$ and branch index $\nu$, and can be expressed as
\begin{equation}
\label{eq:EPWdisplacement}
\begin{aligned}
\Delta_{\bq\nu}\hat{H} = \sum_{\k\a p} \sqrt{\frac{\hbar}{2 M_{\k} \o_{\bq\nu}}} e^{i\bq\cdot\Rp} e_{\k\a,\nu}(\bq) \frac{\partial \hat{H}}{\partial \tau_{\k \a p}}.
\end{aligned}
\end{equation}
Here, $\tau_{\k \a p}$ is the position of atom $\k$ in unit cell $p$ along the Cartesian direction $\a$ (${\bm\tau}_{\k p} = {\bm \tau}_{\k} + \Rp$), and $M_\k$ is the mass of atom $\k$. The derivative $\frac{\partial \hat{H}}{\partial \tau_{\k\a p}}$ corresponds to individual atomic displacements and can be evaluated using the finite-displacement approach with supercells \cite{Frank1995, Kresse1995, Parlinski1997, Togo2015}. Finally, $\o_{\bq \nu}$ is the frequency and $e_{\k\a,\nu}(\bq)$ is the normal mode of vibration or polarization vector of the phonon mode characterized by $\bq$ and $\nu$. These quantities are eigenvalues and eigenvectors of the dynamical matrix $D_{\k\a,\kp \ap}(\bq)$, which is obtained from the Fourier transform of the matrix of interatomic force constants (IFCs) $C_{\k\a p, \kp \ap p'}$:
\begin{equation}
\label{eq:IFCs}
C_{\k\a p, \kp \ap p'} = \frac{\partial^2 V}{\partial{\tau_{\k\a p}} \partial{\tau_{\kp \ap p'}}}.
\end{equation}
\begin{equation}
\label{eq:EPWDynmat}
D_{\k\a,\kp \ap}(\bq) = \frac{1}{\sqrt{M_\k M_{\kp}}} \sum_{p} C_{\k\a 0, \kp \ap p} e^{i\bq\cdot\Rp},
\end{equation}
\begin{equation}
\label{eq:EPWIFCs}
C_{\k\a 0,\kp \ap p} \equiv D_{\k\a,\kp \ap}(\Rp) = \frac{\partial^2 V}{\partial{\tau_{\k\a 0}} \partial{\tau_{\kp \ap p}}}. 
\end{equation}
The matrix of IFCs is defined as the second derivative of the total potential energy, $V$, with respect to atomic displacements evaluated at the equilibrium configuration \cite{Giustino2017}. In VASP, the interatomic force constants are computed in the supercell using a finite-difference scheme, and one makes the approximation that all force constants outside of the Wigner-Seitz cell associated with the BvK cell are zero.

To express the electron-phonon matrix element in the Bloch representation in terms of the corresponding element in the Wannier representation, we insert Eqs.~\eqref{eq:WanOrbinv} and \eqref{eq:EPWdisplacement} inside Eq.~\eqref{eq:EPWelph}. After rearranging the terms, we find
\begin{equation}
\label{eq:EPWeph}
\begin{aligned}
  g_{mn\nu}(\bk, \bq)
  &=
  \sum_{p p'} e^{i(\bk \cdot\Rp + \bq\cdot\Rpp)} \sum_{m' n'}
  \sum_{\k\a}
  \sqrt{\frac{\hbar}{2 M_\k \o_{\bq\nu}}} \\
  &
  \times
  U_{mm'\bk+\bq}  g_{m'n'\k\a}(\Rp, \Rpp) U^\dag_{n'n\bk}
  e_{\k\a,\nu}(\bq),
\end{aligned}
\end{equation}
having introduced the electron-phonon matrix element in the Wannier representation
\begin{equation}
\label{eq:EPWElphWan}
  g_{mn\k\a}(\Rp, \Rpp) = \braket{w_{m0}|\frac{\partial \hat{H}}{\partial \tau_{\k \a p'}}|w_{np}}.
\end{equation}

In VASP, Eq.~\eqref{eq:EPWElphWan} is computed similarly to the interatomic force constants, using finite atomic displacements in a large supercell. After each displacement step, a self-consistent electronic minimization is performed and the action of the resulting perturbed Hamiltonian on the Wannier orbitals is computed.
This procedure is implemented entirely in the PAW method. It is crucial to note that there are two non-equivalent definitions of the electron-phonon matrix element in the PAW method.
In this paper, we only consider the so-called ``all-electron'' formulation~\cite{Chaput2019}, which is formally equivalent to Eq.~\eqref{eq:EPWElphWan} in the limit of a complete PAW basis.
However, matrix elements in the other, so-called ``pseudo'' formulation~\cite{Engel2020} can also be passed to the EPW-VASP interface. In general, using these non-Hermitian pseudo matrix elements is not advised, but there are use cases where they offer computational advantages compared to the all-electron formulation. More details on the PAW formalism are provided in Appendix~\ref{sec:paw}. A detailed comparison between the all-electron approach and pseudo approach is given in Ref.~\cite{Engel2022}.

\subsection{Wannier to Bloch transform in EPW}

\begin{figure*}[t]
\centering
\includegraphics[width=1\textwidth]{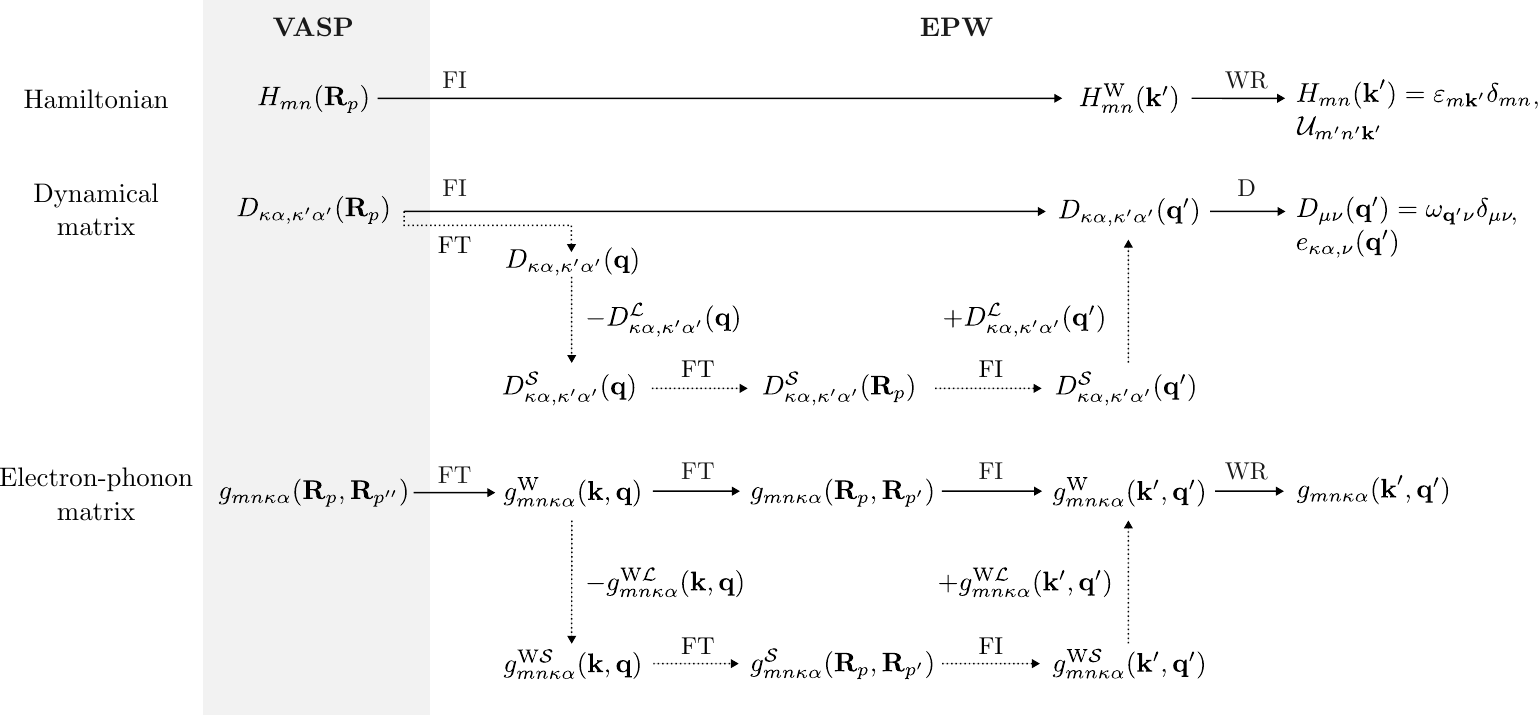}
\caption{General workflow for transformations from Wannier to Bloch space in the EPW-VASP interface. Solid arrows indicate the default route; the alternative route along the dotted arrows is followed when long-range effects are included. Quantities in the Wannier gauge are marked with a superscript ``W''. The symbols $\mathcal{L}$ and $\mathcal{S}$ denote long- and short-range contributions, respectively. Abbreviations: ``FT'' - standard Fourier transformation between commensurate $(\Rp, \Rpp)$ and coarse $(\bk, \bq)$  grids; ``FI'' - Fourier interpolation from $(\Rp, \Rpp)$ to fine $(\bkp, \bqp)$ grids; ``WR'' - Wannier rotation from the Wannier to the Bloch gauge via the unitary matrices $\sU_{\bkp}$, $\sU_{\bkp + \bqp}$ defined in Eqs.~\eqref{eq:EPWHamBloch} and \eqref{eq:ElphBloch};
``D'' - diagonalization of the dynamical matrix. For the electron-phonon matrix, two sequential Fourier transformations are first carried out to obtain matrix elements on a differently ordered $(\Rp, \Rpp)$ grid [as opposed to $(\Rp, \Rppp)$], required by the difference in conventions between VASP and EPW discussed in Section~\ref{sec:conventions}.}
\label{fig:general_scheme}
\end{figure*}

The real-space Hamiltonian, IFCs, and electron-phonon matrix elements --- computed via Eqs.~\eqref{eq:EPWHamWan}, \eqref{eq:EPWIFCs}, and \eqref{eq:EPWElphWan} --- are the central quantities calculated in VASP and passed to EPW. Here, we outline the procedure carried out in EPW to transform these quantities onto fine electron and phonon momentum grids in the Bloch representation, following the general workflow shown in Fig.~\ref{fig:general_scheme}.

To start, we first do an inverse Fourier transform of the Hamiltonian $H_{mn}(\Rp)$ from the Wannier to the Bloch representation, and then diagonalize the resulting matrix through a unitary rotation matrix $\sU_{nm\bkp}$
\begin{equation}
\label{eq:EPWHamBloch}
H_{mn}(\bkp)  = \sum_{m' n'} \sU^{\dag}_{m m' \bkp} \biggl[
\sum_{p} e^{i \bkp \cdot \Rp} H_{m' n'}(\Rp) \biggr] \sU_{n' n\bkp}.
\end{equation}
The summation runs over the lattice vectors $\Rp$ belonging to the Wigner-Seitz supercell centered at the origin, as discussed in Appendix~\ref{sec:WS}. The term in the square brackets
\begin{equation}
\label{eq:EPWHamBlochW}
H^{\text{W}}_{m'n'}(\bkp)  =
\sum_{p} e^{i \bkp \cdot \Rp} H_{m' n'}(\Rp)
\end{equation}
represents the Hamiltonian in the Bloch basis in the Wannier gauge, and is the only known quantity at this stage. Note that the Hamiltonian $H^{\text{W}}_{mn}(\bkp)$ in the Wannier gauge is not diagonal, as opposed to the Hamiltonian $H_{mn}(\bkp)$ in the Bloch gauge which is diagonal by construction, $H_{m n}(\bkp)=\delta_{mn} \epsilon_{n\bkp}$. Diagonalizing the aforementioned matrix $H^{\text{W}}_{mn}(\bkp)$ yields the eigenvalues $\epsilon_{n\bkp}$ and the unitary matrix $\sU_{n m\bkp}$ at any arbitrary wave vector $\bkp$ in the BZ. The columns of $\sU_{n m\bkp}$ are the  eigenvectors of the new Bloch states.

For phonons, the calculation proceeds along the path of solid arrows shown in Fig.~\ref{fig:general_scheme}. Using the IFCs $D_{\k\a,\kp \ap}(\Rp) \equiv C_{\k\a 0,\kp \ap p}$ generated from the supercell approach with VASP according to Eq.~\eqref{eq:EPWIFCs}, the eigenfrequencies and eigenmodes at arbitrary momenta $\bqp$ are found using the same two steps as for the Hamiltonian. First, we inverse Fourier transform the real-space dynamical matrix to Bloch space and, then, diagonalize the resulting matrix
\begin{align}
\label{eq:DynmatBloch}
D_{\mu\nu}(\bqp) & = \frac{1}{\sqrt{M_{\k}M_{\kp}}} \sum_{\k\a\kp\ap} e^\dag_{\k\a,\mu}(\bqp)  \nonumber \\
& \times \left[
\sum_{p} e^{i \bqp \cdot \Rp} D_{\k\a,\kp\ap}(\Rp)\right] e_{\kp \ap, \nu}(\bqp).
\end{align}
As for the electrons, the known quantity in Eq.~\eqref{eq:DynmatBloch} is the term within the square brackets

\begin{align}
\label{eq:DynmatBlochW}
D_{\k\a,\kp\ap}(\bqp) =
\sum_{p} e^{i \bqp \cdot \Rp} D_{\k\a,\kp\ap}(\Rp),
\end{align}
and the $D_{\mu\nu}(\bqp)$ matrix on the left-hand side of Eq.~\eqref{eq:DynmatBloch} is diagonal by construction, $D_{\mu\nu}(\bqp) = \o^{2}_{\bqp \nu} \delta_{\mu\nu}$. Diagonalizing $D_{\k\a,\kp\ap}(\bqp)$ yields the eigenmodes $e_{\k \a, \nu}(\bqp)$ and squared eigenfrequencies $\o^{2}_{\bqp \nu}$.

Finally, the electron-phonon matrix elements for an arbitrary pair of ($\bkp$, $\bqp$) points on fine grids can be obtained using the unitary matrix $\sU_{nm \bkp}$ that interpolates the Bloch Hamiltonian in Eq.~\eqref{eq:EPWHamBloch}
\begin{align}
\label{eq:ElphBloch}
  g_{mn\k \a}(\bkp, \bqp)
  &= \sum_{m' n'}
  \sU^\dag_{mm'\bkp+\bqp}
  g^{\text{W}}_{m'n'\k\a}(\bkp, \bqp)
  \sU_{n'n\bkp},
\end{align}
where
\begin{equation}
  \label{eq:tg_ft}
  g^{\text{W}}_{mn\k\a}(\bkp, \bqp) =
  \sum_{p p'}
  e^{i \left(\bkp \cdot \Rp + \bqp \cdot \Rpp\right)}
  g_{mn\k\a}(\Rp, \Rpp).
\end{equation}
is the electron-phonon matrix element computed in the Wannier gauge. The conversion from the atomic indices $(\k, \a)$ to the phonon mode index $\nu$ is trivial, as it involves summations over $(\k, \a)$ using the phonon eigenvectors $e_{\k\a, \nu}(\bqp)$ and eigenvalues $\o_{\bqp \nu}$ of the dynamical matrix in Eq.~\eqref{eq:DynmatBloch}:
\begin{align}\label{eq:gka_to_gknu}
    g_{mn\nu}(\bkp, \bqp)
    &=
    \sum_{\k\a} \sqrt{\frac{\hbar}{2 M_\k \o_{\bqp\nu}}} g_{mn\k\a}(\bkp, \bqp) e_{\k\a, \nu}(\bqp).
\end{align}
The Wannier-gauge matrix elements, $g^{\text{W}}_{mn\k\a}(\bkp, \bqp)$ and $g^{\text{W}}_{mn\nu}(\bkp, \bqp)$, transform in exactly the same way.

At this stage, it is assumed that the electron-phonon  interactions are short-ranged in real space, meaning that $g_{mn\k\a}(\Rp, \Rpp)$ decays rapidly with both $\Rp$ and $\Rpp$. Under this assumption, it is sufficient to compute the matrix elements on only a small set of ($\Rp$, $\Rpp$) lattice vectors to fully capture the coupling between electrons and phonons. For materials with nonvanishing dynamical dipoles, however, this interpolation procedure must be modified to account for long-range interactions, as discussed in Sec.~\ref{sec:long-range}.

Following the solid-arrow path in Fig.~\ref{fig:general_scheme}, the electron-phonon matrix elements $g(\Rp, \Rppp)$, generated in VASP, are Fourier transformed to obtain the electron-phonon matrix elements $g^{\text{W}}_{mn\k\a}(\bk, \bq)$ on a commensurate $(\bk, \bq)$ grid in the Wannier gauge, according to Eq.~\eqref{eq:EPWElphWan}. An inverse Fourier transform is then applied to produce $g(\Rp, \Rpp)$ on a differently ordered $(\Rp, \Rpp)$ grid [as opposed to $(\Rp, \Rppp)$], required by the difference in conventions between VASP and EPW discussed in Section~\ref{sec:conventions}. Subsequently, the matrix elements $g_{mn\k\a}(\Rp, \Rpp)$ are Fourier interpolated to a fine $(\bkp, \bqp)$ grid in the Wannier gauge, yielding $g^{\text{W}}_{mn\k\a}(\bkp, \bqp)$ via Eq.~\eqref{eq:tg_ft}. Finally, a rotation to the Bloch gauge is carried out using Eq.~\eqref{eq:ElphBloch}, resulting in $g_{mn\k\a}(\bkp, \bqp)$.

\subsection{Long-range electron-phonon interactions in polar materials}
\label{sec:long-range}

Long-range electron-phonon interactions, arising from dipole fields (decaying as $\sim1/r^2$) in polar materials~\cite{Vogl1976} give rise to nonanalytic behavior in the perturbation Hamiltonian in the long-wavelength limit $\bq \rightarrow 0$~\cite{Frohlich1954,Verdi2015,Sjakste2015,Brunin2020a,Brunin2020b,Park2020,Jhalani2020,Ponce2021,Ponce2023b}.
Mathematically, the presence of these electron-phonon
interactions is typically indicated by atoms having a nonzero Born effective charge tensor ${\bf Z}_\k$ \cite{Born1998}. The dipole field diverges as $1/q$ in the limit of $\bq \rightarrow 0$, leading to the well-known longitudinal optical–transverse optical (LO-TO) mode splitting. Dipolar contributions violate the locality assumption required for Wannier interpolation, resulting in divergent or discontinuous electron-phonon matrix elements near $\bq=0$, which cannot be interpolated directly. These long-range effects, however, can be treated analytically in both bulk and 2D materials~\cite{Verdi2015, Sjakste2015, Brunin2020a, Brunin2020b, Park2020, Jhalani2020, Ponce2021, Royo2021, Ponce2023a, Ponce2023b, Sohier2016, Deng2021, Sio2022, Zhang2022}.

In this section, we outline the procedure used within the EPW-VASP interface to leverage the existing EPW capabilities for treating long-range polar contributions for both the dynamical matrix and the electron-phonon matrix elements for 3D materials~\cite{Lee2023}.
Since VASP provides IFCs and real-space electron-phonon matrix elements computed via Eqs.~\eqref{eq:EPWIFCs} and \eqref{eq:EPWElphWan}, the first step in EPW is to convert these quantities from the real-space supercell ($\Rp$, $\Rpp$) to the corresponding coarse BZ grid ($\bk$, $\bq$).
Then, the strategy is to separate the dynamical matrix and electron-phonon matrix elements into short-range ($\mathcal{S}$) and long-range ($\mathcal{L}$) parts,
\begin{equation}
\label{eq:D_LR}
    D_{\k\a, \kp\ap}(\bq) =
    D^{\mathcal{S}}_{\k\a, \kp\ap}(\bq) +
    D^{\mathcal{L}}_{\k\a, \kp\ap}(\bq),
\end{equation}
\begin{equation}
\label{eq:g_LR0}
    g_{mn\k\a}(\bk,\bq) =
    g^{\mathcal{S}}_{mn\k\a}(\bk,\bq) +
    g^{\mathcal{L}}_{mn\k\a}(\bk,\bq),
\end{equation}
and remove the long-range term before performing the Fourier transform from the coarse BZ grid ($\bk$, $\bq$) to the Wannier representation ($\Rp$, $\Rpp$) in the corresponding real-space supercell. Next, the short-range components are inverse Fourier transformed from the Wannier representation to the Bloch representation on a fine ($\bkp$, $\bqp$) grid, after which the long-range contributions are calculated via analytical expressions and added back on the same fine grid. The entire procedure described in this section is summarized in Fig.~\ref{fig:general_scheme}.

For phonons, when dipole corrections are included, the long-range part of the dynamical matrix, consisting of dipole-dipole ($DD$) contributions, is computed using the corresponding terms of the analytic expressions described in Refs.~\cite{Gonze1997, Royo2020, Ponce2021}:

\begin{equation}
  \label{eq:D_long-range_1}
  \begin{aligned}
    D&^{\mathcal{L}}_{\k\a,\kp\ap}(\bq)
    =
    \frac{4\pi}{\O_{\uc} \sqrt{M_\k M_{\kp}}} \frac{e^2}{4 \pi \varepsilon_0} 
    \\
    &\times
    \Bigg[
    \sum_{{\bf G}\ne -\bq}
    D^{\mathcal{L}, DD}_{\k\a, \kp\ap}(\bq + \bG)
    -
    \sum_{\k^{\prime\prime}}
    \sum_{{\bf G}\ne \mathbf{0}}
    D^{\mathcal{L}, DD}_{\kp\a, \k^{\prime\prime}\ap} (\bG)
    \Bigg],
  \end{aligned}
\end{equation}
with the tensor $D^{\mathcal{L}, DD}_{\k\a\kp\ap}(\bq)$ being defined as
\begin{equation}
  \label{eq:D_long-range_2}
  \begin{aligned}
    D^{\mathcal{L}, DD}_{\k\a,\kp\ap}(\bq)
    &=
    e^{-\frac{\bq \cdot\bm\epsilon^\infty \cdot \bq}{4\Lambda^2}}
    \frac{e^{i \bq \cdot \left(\bm{\tau}_\k - \bm{\tau}_{\kp}\right)}}
    {\bq \cdot \boldsymbol{\epsilon}^{\infty} \cdot \bq}
    \\
    &\times
    \Bigg[
    \left( \bq \cdot \mathbf{Z}^\ast_{\k}\right)_{\a}
    \cdot
    \left( \bq \cdot \mathbf{Z}^\ast_{\kp}\right)_{\ap}
    \Bigg].
  \end{aligned}
\end{equation}
In Eqs.~\eqref{eq:D_long-range_1} and \eqref{eq:D_long-range_2}, $\Omega_{\uc}$ is the volume of the unit cell, $\boldsymbol{\epsilon}^\infty$ is the dielectric tensor of the material, $\mathbf{Z_{\k}^*}$ is the Born effective charge tensor \cite{Born1998} of atom $\k$ with elements $Z^\ast_{\k, \a\b}$,
and $\bm{\tau}_\kappa$ is the position of atom $\k$.
Products of vectors ${\bf a}, {\bf c}$ and tensor ${\bf B}$ are defined as
${\bf a} \cdot {\bf B} \cdot {\bf c} = \sum_{\a} \sum_{\b} a_{\a} B_{\a \b} c_{\b}$
and $\left({\bf B} \cdot {\bf c} \right)_\a = \sum_\b B_{\a\b} c_\b$.
In numerical calculations, a Gaussian filter, $\exp{[-(\bq+{\bG})\cdot\bm\epsilon^\infty\!\cdot(\bq+{\bf G})/(4\Lambda^2)]}$
\cite{Gonze1997, Ewald1921}, with optimized parameter
$\Lambda = 2 \pi/a_{\text{lat}}$ \cite{Lee2023}, where
$a_{\text{lat}}$ is the lattice constant, is applied to truncate the summation over reciprocal lattice vectors $\bG$ when the filter value becomes sufficiently small (e.g., below $e^{-14} \approx 10^{-6}$).

Following the dotted-arrow path in Fig.~\ref{fig:general_scheme}, the real-space matrix $D_{\k\a, \kp\ap}(\Rp)$, computed with VASP, is Fourier transformed to obtain the dynamical matrix $D_{\k\a, \kp\ap}(\bq)$ on a commensurate coarse $\bq$ grid, using Eq.~\eqref{eq:EPWDynmat}.
Then the long-range contribution $D^{\mathcal{L}}_{\k\a, \kp\ap}(\bq)$ is calculated on the coarse $\bq$ grid according to Eqs.~(\ref{eq:D_long-range_1}) and (\ref{eq:D_long-range_2}), and subtracted from $D_{\k\a, \kp\ap}(\bq)$ to yield the short-range dynamical matrix $D^{\mathcal{S}}_{\k\a, \kp\ap}(\bq)$. This short-range matrix is Fourier transformed to real space to obtain $D^{\mathcal{S}}_{\k\a, \kp\ap}(\Rp)$, which is then Fourier interpolated onto a fine $\bqp$ grid to give $D^{\mathcal{S}}_{\k\a, \kp\ap}(\bqp)$. Finally, the long-range contribution $D^{\mathcal{L}}_{\k\a, \kp\ap}(\bqp)$  is recalculated on the fine $\bqp$ grid using the same analytical expressions and added back to reconstruct the full dynamical matrix $D_{\k\a, \kp\ap}(\bqp)$. We note that this procedure involves some approximation: the IFCs from VASP already contain long-range contributions, therefore these should be subtracted directly in real space before moving to reciprocal space. The effect of this approximation will be investigated in future work.

In EPW, the subtraction and addition of the long-range contributions to the electron-phonon matrix elements are carried out in the Bloch basis in the Wannier gauge, $g^{\text{W}}_{mn\k\a}(\bk, \bq)$, given in Eq.~\eqref{eq:tg_ft}. We note that based on Eq.~\eqref{eq:gka_to_gknu}, it is important to apply the long-range corrections consistently to both the dynamical matrix and the electron-phonon matrix elements. If $e_{\k\a, \nu}(\bqp)$ and $\o_{\bqp \nu}$ remain uncorrected, the resulting $g_{mn\nu}(\bkp, \bqp)$ and $g^{\text{W}}_{mn\nu}(\bkp, \bqp)$ will be inaccurate, even if $g_{mn\k\a}(\bk, \bq)$ and $g^{\text{W}}_{mn\k\a}(\bkp, \bqp)$ are corrected.

According to Refs.~\cite{Verdi2015, Sjakste2015, Ponce2021, Ponce2023b, Lee2023}, the long-range polar contribution to the electron-phonon matrix elements is calculated in the Wannier gauge as
\begin{equation}
  \label{eq:elph-dipole}
  \begin{aligned}
    g&_{mn\k\a}^{\text{W}\mathcal{L}}(\bk,\bq)
    =
    i\frac{4\pi}{
    \O_{\uc}}
    \frac{e^2}{4 \pi \varepsilon_0}
    \sum_{{\bf G}\ne -\bq}
        e^{-\frac{(\bq+{\bf G})\cdot\bm\epsilon^\infty \cdot(\bq+{\bf G})}{4\Lambda^2}}
     \\
    &\times
       e^{-i \left(\bq + \bG \right) \cdot \bm{\tau}_\kappa}
    \frac{
    \left[(\bq+{\bf G})\cdot{\bf Z}^*_\k \right]_\a
    }
    {(\bq+{\bf G})\cdot\bm\epsilon^\infty\!\cdot(\bq+{\bf G})}
    \delta_{mn},
  \end{aligned}
\end{equation}
where $\epsilon_0$ is the vacuum permittivity. In the Wannier gauge, the $g_{mn\k\a}^{\text{W}\mathcal{L}}(\bk,\bq)$ elements are diagonal in the Wannier indices ($m, n$), differing only in the atomic indices $(\k, \a)$, as indicated by $\delta_{mn}$. This Kronecker-delta function arises from the orthonormality of the Wannier functions, which is used in the derivation of Eq.~\eqref{eq:elph-dipole} \cite{Verdi2015, Giustino2017, Ponce2021, Lee2023}.

Following the dotted-arrow path in Fig.~\ref{fig:general_scheme}, the electron-phonon matrix elements $g_{mn\k\a}(\Rp, \Rppp)$, computed with VASP, are first Fourier transformed from real space to the Wannier gauge to obtain $g^{\text{W}}_{mn\k\a}(\bk, \bq)$ on the coarse $(\bk, \bq)$ grid, using Eq.~\eqref{eq:tg_ft}. The long-range component $g^{\text{W}\mathcal{L}}_{mn\k\a}(\bk, \bq)$ is then computed on the same coarse $(\bk, \bq)$ grid according to
Eq.~\eqref{eq:elph-dipole},
and subtracted from $g^{\text{W}}_{mn\k\a}(\bk, \bq)$ to yield the short-range contribution $g^{\text{W}\mathcal{S}}_{mn\k\a}(\bk, \bq)$. This short-range matrix undergoes a Fourier transformation back to real space to obtain $g^{\mathcal{S}}_{mn\k\a}(\Rp, \Rpp)$, and then Fourier interpolated to a fine $(\bkp, \bqp)$ grid in the Wannier gauge, yielding $g^{\text{W}\mathcal{S}}_{mn\k\a}(\bkp, \bqp)$. Next, the long-range component $g^{\text{W}\mathcal{L}}_{mn\k\a}(\bkp, \bqp)$ is recalculated on the fine $(\bkp, \bqp)$ grid using the same analytical expressions and added back to reconstruct the full matrix elements $g^{\text{W}}_{mn\k\a}(\bkp, \bqp)$ in the Wannier gauge. Finally, these matrices are rotated from the Wannier gauge to the Bloch gauge to obtain $g_{mn\k\a}(\bkp, \bqp)$, using Eq.~\eqref{eq:ElphBloch}. As discussed for the dynamical matrices, also in the case of the matrix elements one should remove long-range components directly from the real-space matrix elements from VASP. In the present implementation, for convenience, we carry out this subtraction in the Bloch representation within EPW. We will investigate the impact of this approximation in future work.

The EPW code also implements long-range corrections for 2D materials, which has been the focus of several studies. For instance, a unified framework for polar electron-phonon interactions was proposed to enable a smooth transition from three to two dimensions~\cite{Sio2022}. This formalism recovers established approaches in both limits, reducing to the 3D~\cite{Verdi2015,Sjakste2015} and 2D~\cite{Sohier2016,Deng2021} treatment. An alternative methodology has also been developed in a series of works~\cite{Zhang2022,Ponce2023a,Ponce2023b}, building upon a general formalism for treating long-range electrostatic interactions in 2D crystals~\cite{Royo2021}.

\subsection{\label{sec:vme} Band velocities}

\begin{figure*}[t]
\centering
\includegraphics[width=0.8\textwidth]{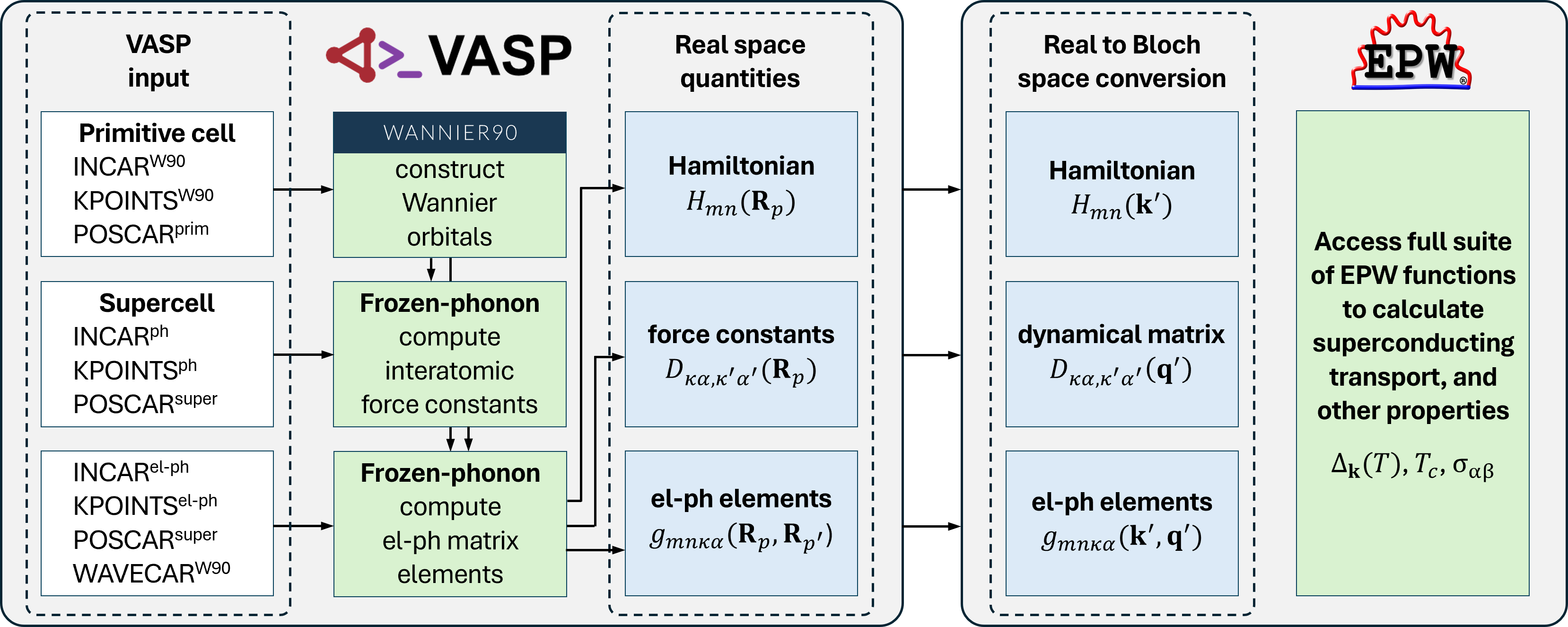}
\caption{Flowchart illustrating the required input files and data transfer between VASP and EPW using the EPW-VASP interface. On the left side, the electron-phonon matrix elements, Hamiltonian, and IFCs computed in VASP with the finite displacement approach in real space are rewritten to an HDF5 file in the Wannier representation. On the right side, EPW reads the HDF5 file and interpolates the quantities onto a fine Bloch grid for subsequent superconductivity and transport calculations.}
\label{fig:flowchart}
\end{figure*}

The calculation of drift and Hall mobility through the Boltzmann transport equation requires only diagonal matrix elements of a general velocity operator that is defined as the commutator of the Hamiltonian with the position operator $\hat{\br}$ \cite{Wang2006}:
\begin{equation}
  \hat{\bv} = \frac{i}{\hbar} \left[ \hat{H}, \hat{\br} \right].
  \label{eq:v_operator}
\end{equation}

Following Refs.~\cite{Blount1962, Wang2006, Yates2007}, these band velocities can be expressed
as diagonal elements of a gradient of the Bloch Hamiltonian,
\begin{equation}
  \begin{aligned}
    \bv_{n \bk}
    &\equiv
    \bra{\psi_{n \bk}}
    \hat{\bv}
    \ket{\psi_{n \bk}}
    =
    \frac{1}{\hbar}
    \bm{\nabla}_{\bk}
    H_{mn}(\bk) \delta_{mn}.
    \\
  \end{aligned}
  \label{eq:vme_diag}
\end{equation}
where the nabla-operator is given by
\begin{equation}
  \bm{\nabla}_{\bk} = \sum_{i \in \{x, y, z\}} \hat{{\bf i}} \frac{\partial}{\partial k_i}.
  \label{eq:nablakp}
\end{equation}

First, the Hamiltonian in Wannier space is interpolated to obtain the
gradient of the Hamiltonian in the Wannier gauge on a fine $\bkp$ grid, as

\begin{equation}
  \begin{aligned}
    &
    \bm{\nabla}_{\bkp}
    H^{\text{W}}_{mn}(\bkp)
    =
    i \sum_{p}
    \Rp
    e^{i \bkp \cdot \Rp}
    H_{mn}(\Rp).
  \end{aligned}
  \label{eq:gradhk}
\end{equation}
Next, this gradient is transformed to the Bloch gauge via rotations using the matrices $\sU_{mn\bkp} = U^\dag_{mn\bkp}$, which are obtained by diagonalizing the Hamiltonian $H^{\text{W}}_{mn}(\bkp)$ in the Wannier gauge to yield the Bloch-gauge Hamiltonian $H_{mn}(\bkp)$ [Eqs.~\eqref{eq:EPWHamBloch} and \eqref{eq:EPWHamBlochW}],
\begin{equation}
  \begin{aligned}
    &
    \bm{\nabla}_{\bkp}
    H_{mn}(\bkp)
    =
    \sum_{m^\prime n^\prime}
    \mathcal{U}^\dag_{m m^\prime\bkp}
    \left[
    \bm{\nabla}_{\bkp}
    H^{\text{W}}_{m^\prime n^\prime}(\bkp)
    \right]
    \mathcal{U}_{n^\prime n\bkp}.
  \end{aligned}
  \label{eq:hmnbkp}
\end{equation}
Then band velocities on the fine $\bkp$ grid are obtained from this gradient
by following Eq.~\eqref{eq:vme_diag}.

\section{Computational workflow}
\label{sec:usage}

This section outlines the overall workflow for using the EPW-VASP interface, from initial setup to final output. The workflow is illustrated in Fig.~\ref{fig:flowchart}. Communication between EPW and VASP is handled through the exchange of real-space quantities stored in Hierarchical Data Format version 5 (HDF5)~\cite{HDF5}. The \texttt{elphon\_wannier.h5} file generated by VASP contains the following key information that characterizes the system of interest:
\begin{itemize}
  \item Crystal structure data, including lattice vectors and atomic positions for both the primitive cell and the supercell.
  \item Wannier function centers within the primitive unit cell.
  \item Real-space Hamiltonian in the Wannier representation.
  \item Interatomic force constants (IFCs).
  \item Real-space electron–phonon matrix elements in the Wannier representation.
  \item Dielectric tensor and Born effective charges, if long-range dipole corrections are enabled in the VASP calculation.
\end{itemize}

The first step of the workflow is to compute the Hamiltonian, IFCs, and electron-phonon matrix elements in the Wannier representation by performing DFT calculations with VASP. To construct a set of Wannier functions, a DFT calculation must first be carried out for the primitive cell using a regular $\Gamma$-centered $\bk$-point grid, with appropriate flags set in the VASP input files to call the Wannier90 library. This step produces a \texttt{CONTCAR\_SUPER} file containing the supercell expansion of the primitive cell, and the \texttt{WAVECAR.wan} file storing the Wannier orbitals in the supercell. The \texttt{CONTCAR\_SUPER} file must be renamed to \texttt{POSCAR}  for use in the subsequent phonon and electron-phonon calculations. The same \texttt{WAVECAR.wan} must also be used in the electron-phonon calculation to ensure consistency with the wannierization step.

In the current version of VASP, electron-phonon calculations are restricted to using only the $\Gamma$-point for the electronic states. However, when finer Brillouin zone sampling is needed to accurately capture phonon dispersions, a preliminary standalone supercell-based phonon calculation can be performed using finite displacements and a $\bk$-point grid. The resulting \texttt{phonon\_data.h5} file, which contains the IFCs, can then be imported into the subsequent electron-phonon calculation. This approach enables VASP to reuse the high-quality IFCs while keeping the actual electron-phonon calculations limited to the $\Gamma$-point. The main output from this procedure is the \texttt{elphon\_wannier.h5} file, which contains all quantities required for subsequent EPW calculations.

To activate the interface, the \texttt{elphon\_wannier.h5} file generated by VASP is placed in the EPW working directory and the flag \texttt{lvasp = .true.} is set in the EPW input file. All other input parameters relevant to superconductivity and transport calculations can be configured as in standard EPW workflows with Quantum ESPRESSO.

\begin{figure*}[ht!]
\centering
\includegraphics[width=1\textwidth]{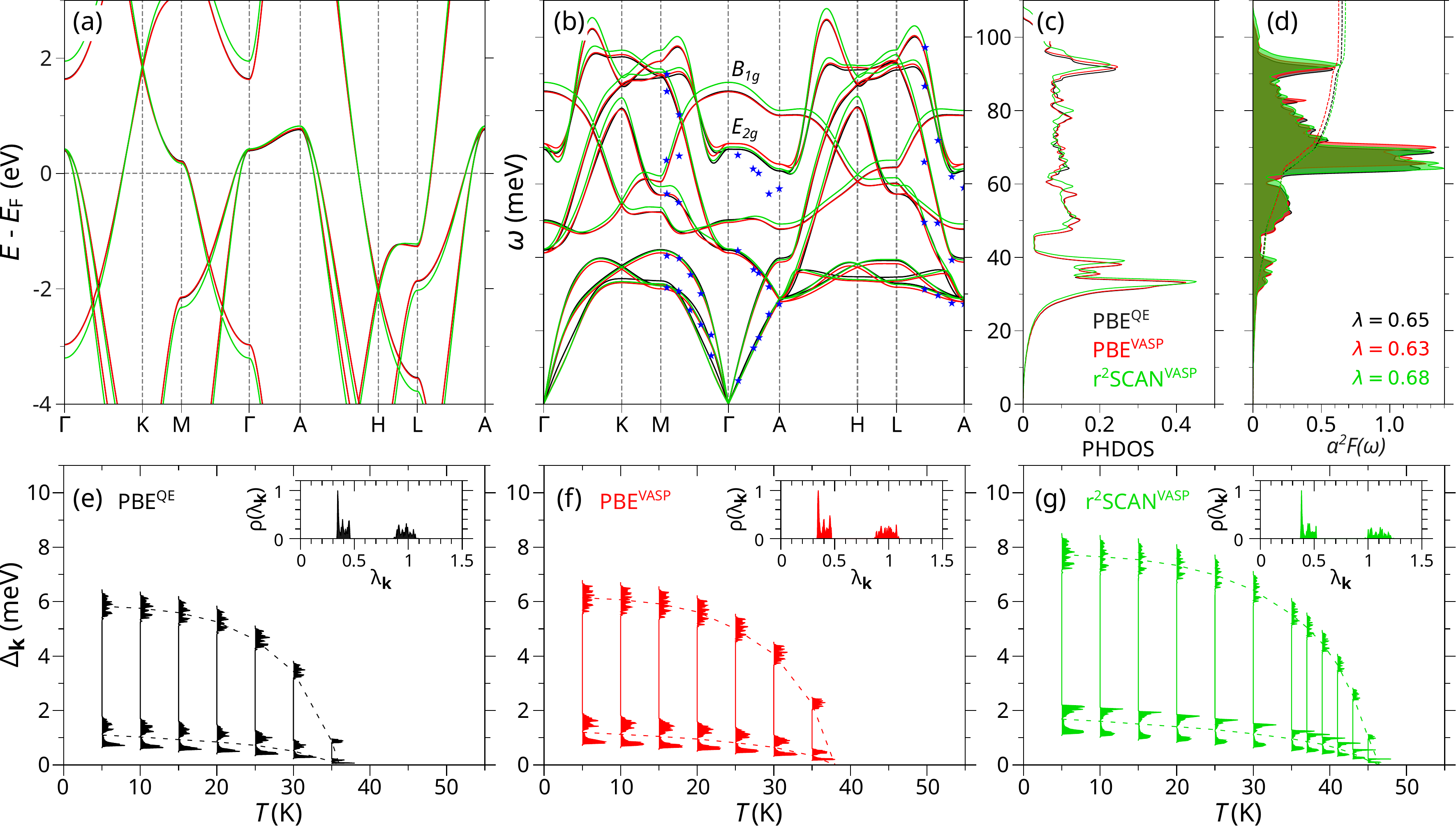}
\caption{MgB$_2$ properties calculated using QE-PBE (black), VASP-PBE (red), and VASP-r$^2$SCAN (green). The top panels show (a) band structure, (b) phonon dispersion, (c) phonon density of states, and (d) Eliashberg spectral function $\a^2F$ with integrated electron-phonon coupling strength $\lambda$. The bottom panels display energy distributions of the superconducting gap $\Delta_{\bk}$ as a function of temperature, with insets illustrating histograms of the electron-phonon coupling strength $\lambda_{\bk}$. In (b), the blue stars are inelastic X-ray scattering experimental data at 300~K from Ref.~\cite{Shukla2003}.}
\label{fig:MgB2}
\end{figure*}

\section{Examples}
\label{sec:results}

In this section, we showcase the capabilities of the EPW-VASP interface by computing the superconducting gap and critical temperature of MgB$_2$ using the anisotropic Migdal-Eliashberg (aME) equations, and the carrier mobility of cubic BN (c-BN) using the \textit{ab initio} Boltzmann transport equation (aiBTE). Details of the aME and BTE formalisms, as well as their implementation in EPW, are provided in Ref.~\cite{Lee2023}.

\subsection{Two-gap superconductivity in MgB$_2$}
\label{sec:super}

Magnesium diboride is a simple binary compound that attracted widespread interest following the discovery of its superconducting properties, with a critical temperature ($T_{\rm c}$) of 39~K~\cite{Nagamatsu2001}, which remains the highest among conventional, phonon-mediated superconductors at ambient pressure.
It exhibits a two-gap structure, arising from distinct $\sigma$ and $\pi$ bands at the Fermi level, making it an ideal system for studying electron-phonon interactions and multi-band superconductivity with different exchange-correlations functionals.

We performed DFT calculations with VASP~\cite{Kresse1993,Kresse1996a} using PAW~\cite{Kresse1999} potentials with the standard PBE functional \cite{Perdew1996} as well as the r$^2$SCAN meta-GGA. The lattice parameters and atomic positions were relaxed until the total energy was converged within 10$^{-8}$~eV and the maximum force on each atom was less than 10$^{-3}$~eV/\AA. The electronic structure was described using a plane-wave cutoff energy of 415~eV and an MP smearing of 0.27~eV. For structural relaxation, we adopted the primitive unit cell with a $\Gamma$-centered 30$\times$30$\times$24 $\bk$-mesh. The phonon calculations were performed with the finite displacement method. We used a 6$\times$6$\times$4 supercell with a $\Gamma$-centered 5$\times$5$\times$6 $\bk$-point mesh. The Wannier interpolation was performed on a uniform $\Gamma$-centered 6$\times$6$\times$4 $\bk$-grid with the Wannier90 code~\cite{Pizzi2020} in library mode. As projections for the maximally localized Wannier functions, we used one $p_z$ orbital for each B atom and three $s$ orbitals placed in the middle of the B-B bonds (see Fig.~S4 of the Supplementary Information (SI)~\cite{SI}). The real-space Hamiltonian, IFCs, and electron-phonon matrix elements obtained with VASP were read by EPW and Fourier transformed to Bloch representation on dense uniform $\bk$- and $\bq$-point grids to be used in the subsequent superconductivity calculations~\cite{Giustino2007, Ponce2016, Margine2013, Lee2023}. The anisotropic Migdal-Eliashberg equations~\cite{Margine2013} were solved on 60$\times$60$\times$60 $\bk$- and $\bq$-point grids using an energy window of $\pm 0.2$~eV around the Fermi level, a Matsubara frequency cutoff of 1.0~eV, Gaussian smearing values of 50~meV for electrons and 0.5~meV for phonons, and a Coulomb $\mu^*$ parameter of 0.20.

For comparison, we also performed DFT calculations with QE using the PBE functional and optimized norm-conserving Vanderbilt pseudopotentials (ONCVPSP) \cite{Hamann2013} from the Pseudo Dojo library~\cite{Setten2018}. The lattice parameters and atomic positions were relaxed until the total energy was converged within 10$^{-4}$~Ry and the maximum force on each atom was less than 10$^{-3}$~Ry/\AA. The electronic structure was described using a plane-wave cutoff energy of 100 Ry, a Methfessel-Paxton smearing of 0.02~Ry, and a $\Gamma$-centered 30$\times$30$\times$24 $\bk$-point mesh. The dynamical matrices and the linear variation of the self-consistent potential were calculated within DFPT~\cite{Baroni2001} on a 6$\times$6$\times$4 $\bq$-mesh. The Hamiltonian, dynamical matrices, and deformation potential on the coarse $\bk$ and $\bq$ grids generated with QE were used in EPW for the subsequent electron-phonon calculations. To carry out Wannier-Fourier interpolation on fine BZ grids and to solve the anisotropic Migdal-Eliashberg equations, we used the same computational settings as those described above for the EPW-VASP interface.

The three sets of DFT calculations produced the following lattice parameters: $a =$~3.075~\AA\,and $c =$~3.526~\AA\, (QE-PBE), $a =$~3.073~\AA\, and $c =$~3.525~\AA\, (VASP-PBE), and $a =$~3.064~\AA\, and $c =$~3.514~\AA\, (VASP-r$^2$SCAN). These values are in good agreement with the corresponding experimental lattice constants $a = \text{3.086}$~\AA\,\, and $c = \text{3.524}$~\AA~\cite{Nagamatsu2001}. Figure~\ref{fig:MgB2} shows a comparison of electronic, vibrational, and superconducting results from the three calculations. Consistent with recent findings~\cite{Wang2024}, the r$^2$SCAN functional predicts a slightly larger electronic bandwidth compared to PBE. Notably, the band splitting between the upper $\pi$ and lower $\sigma$ bands at the M-point is 2.5~eV for r$^2$SCAN and 2.4~eV for PBE.

The phonon dispersions agree well in the acoustic region, but more pronounced differences appear in the optical branches. In particular, the $B_{1g}$ mode at the $\Gamma$ point has a higher frequency for r$^2$SCAN, whereas the two PBE calculations are nearly identical. At the same time, the two degenerate $E_{2g}$ modes show large variations between the two PBEs, with QE-PBE being the softest, while r$^2$SCAN results lie in between. The frequency of the $E_{2g}$ phonon branches is very sensitive to the electronic $\bk$-mesh as well as the phonon $\bq$-grid in DFPT or the supercell size in the finite-displacement approach. To properly capture the $E_{2g}$ mode splitting along the $\Gamma-$M and A$-$H directions, at least a 6$\times$6 in-plane $\bq$-grid or corresponding supercell expansion, together with a sufficiently dense $\bk$-point mesh, are required. In our tests (see Secs.~I.A-B of SI~\cite{SI} and Figs.~S1 and S2 therein), the 24$\times$24$\times$16 unit-cell $\bk$-mesh, which corresponds to the 4$\times$4$\times$4 $\bk$-mesh for the 6$\times$6$\times$4 supercell, was sufficient to reproduce the splitting with the considered functionals. A finer 30$\times$30$\times$24 unit-cell $\bk$-point mesh, corresponding to a 5$\times$5$\times$6 $\bk$-mesh for the supercell, lowered the $E_{2g}$ modes by about 1~meV, bringing them closer to the experimental reference data, shown with blue stars in Fig.~\ref{fig:MgB2}(b).

The differences in the dispersive distribution of the $E_{2g}$ branch are further reflected in the isotropic Eliashberg spectral function $\a^2F(\omega)$, shown in Fig.~\ref{fig:MgB2}(d), where the position of the dominant peak shifts noticeably among the three calculations. The greater softening of the $E_{2g}$ mode in r$^2$SCAN leads to generally higher absolute values of the corresponding electron-phonon matrix elements (Fig.~S6 of SI \cite{SI}) and stronger electron-phonon coupling strength $\lambda=0.68$, in close agreement with Ref.~\cite{Wang2024}. In contrast, this mode is hardest in the PAW PBE calculation with VASP, resulting in a reduced coupling constant of 0.63. The NC PBE with QE yields an intermediate phonon softening and a corresponding $\lambda$ of approximately 0.65, consistent with previous work using the PBE functional~\cite{Kafle2022}. Finally, we solve the anisotropic Migdal-Eliashberg equations to capture the two-gap nature of MgB$_2$. Consistent with the larger electron-phonon coupling strength for the $\sigma$ states found using r$^2$SCAN (see the inset of Fig.~\ref{fig:MgB2}(e)-(g)), the $\Delta_{\sigma}$ gap is shifted upward by approximately 1.5~meV in the low-temperature limit. For a Coulomb parameter $\mu^\ast = \text{0.2}$, the gaps close at a critical temperature of approximately 36~K with QE-PBE, 38~K with VASP-PBE, and 46~K with VASP-r$^2$SCAN, in good agreement with the experimental value of 39~K~\cite{Nagamatsu2001}.

\begin{figure*}[ht!]
\centering
\includegraphics[width=1\textwidth]{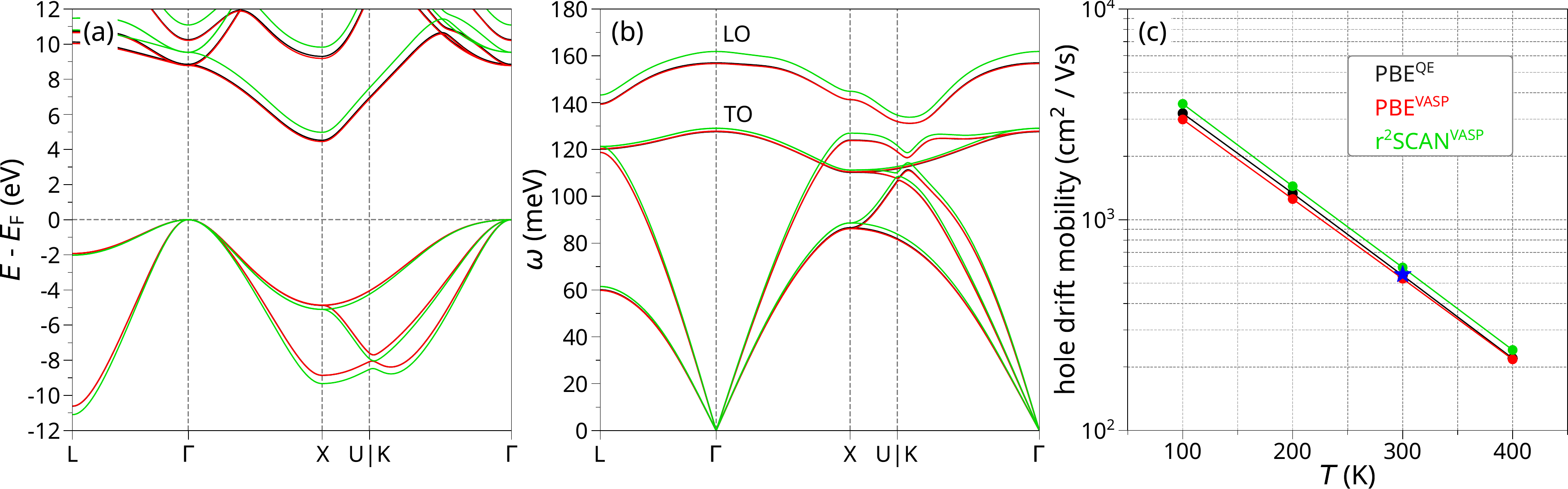}
\caption{Cubic BN properties calculated using QE-PBE (black), VASP-PBE (red), and VASP-r$^2$SCAN (green). The panels show (a) band structure, (b) phonon dispersion, and (c) hole drift mobility calculated with aiBTE including long-range dipole corrections.
In (c), the blue star marks the 300~K result from Ref.~\cite{Ponce2021}.
}
\label{fig:cBN}
\end{figure*}

\subsection{Charge transport in cubic BN}
\label{sec:transp}

Cubic boron nitride is a wide-band-gap, polar semiconductor that exhibits strong ionicity and robust thermal and mechanical stability, making it a material of considerable interest for high-power and high-frequency electronic applications. Its polar nature gives rise to long-range Fr\"{o}hlich electron-phonon interactions, which play a critical role in determining carrier mobility. These characteristics make c-BN an ideal testbed for showcasing the capabilities of the EPW-VASP interface in computing transport properties and validating its ability to handle long-range effects in polar materials.

We performed DFT calculations with VASP~\cite{Kresse1993,Kresse1996a} using PAW~\cite{Kresse1999} potentials with the standard PBE functional \cite{Perdew1996}. The lattice parameters and atomic positions were relaxed until the total energy was converged within 10$^{-8}$~eV and the maximum force on each atom was less than 10$^{-3}$~eV/\AA. The electronic structure was described using a plane-wave cutoff energy of 520~eV and a Fermi smearing of 0.001~eV were employed. For structural relaxation, we adopted the primitive unit cell with a 18$\times$18$\times$18 $\bk$-mesh. The phonon calculations were performed with the finite displacement method. We used a 6$\times$6$\times$6 supercell with a $\Gamma$-centered 3$\times$3$\times$3 $\bk$-point mesh. The Wannier interpolation was performed on a uniform $\Gamma$-centered 6$\times$6$\times$6 $\bk$-mesh with the Wannier90 code~\cite{Pizzi2020} through the VASP-Wannier90 interface. As projections for the maximally localized Wannier functions, we used single $p_x$, $p_y$, and $p_z$ orbitals on a N atom (Fig.~S5 of SI~\cite{SI}). The real-space Hamiltonian, IFCs, and electron-phonon matrix elements obtained with VASP were read by EPW and Fourier transformed to Bloch representation on dense uniform $\bk$- and $\bq$-point grids (Fig.~S7 of SI \cite{SI}) to be used in the subsequent transport calculations~\cite{Giustino2007, Ponce2016, Margine2013, Lee2023}. We solved the aiBTE on 120$\times$120$\times$120 $\bk$- and $\bq$-point grids using an energy window of $\pm 0.3$~eV around the reference energy, set to the band edge, and employed the adaptive Gaussian smearing implemented in EPW to approximate the Dirac delta functions appearing in the electron-phonon scattering terms. Long-range dipole corrections to the dynamical and electron-phonon matrices
were included.

For comparison, we also performed DFT calculations with QE using the PBE functional and ONCVPSP~\cite{Hamann2013} from the Pseudo Dojo library~\cite{Setten2018}. The lattice parameters and atomic positions were relaxed until the total energy was converged within 10$^{-4}$~Ry and the maximum force on each atom was less than 10$^{-3}$~Ry/\AA. Fixed occupations were employed, along with a plane-wave cutoff of 100~Ry and a $\Gamma$-centered 18$\times$18$\times$18 $\bk$-mesh. The dynamical matrices and the linear variation of the self-consistent potential were calculated within DFPT~\cite{Baroni2001} on a 6$\times$6$\times$6 $\bq$-mesh, which was found to be sufficient to produce the convergent phonon dispersion (see Fig.~S3 of SI \cite{SI}). The Hamiltonian, dynamical matrices, and deformation potential on the coarse $\bk$ and $\bq$ grids generated with QE were used in EPW for the subsequent electron-phonon calculations. To carry out Wannier-Fourier interpolation on fine BZ grids and to solve the Boltzmann transport equations, we used the same computational settings as those described above for the EPW-VASP interface.

The DFT calculations yielded lattice parameters of 3.623~\AA{} (QE-PBE), 3.625~\AA{} (VASP-PBE), and 3.613~\AA{} (VASP-r$^2$SCAN), all in good agreement with the experimental value of 3.614~\AA\ \cite{Madelung1991}. Dipole contributions to phonons were estimated using the isotropic Born effective charges $\text{Z}_\text{N}^* = -\text{Z}_\text{B}^*$ and a high-frequency isotropic relative dielectric constant $\epsilon^\infty$, as calculated in QE and VASP. Corresponding values were $\text{Z}_\text{N}^* = 1.91e$ and $\epsilon^\infty = 4.55$ (QE-PBE), $\text{Z}_\text{N}^* = 1.91e$ and $\epsilon^\infty = 4.60$ (VASP-PBE), and $\text{Z}_\text{N}^* = 2.03e$ and $\epsilon^\infty = 4.53$ (VASP-r$^2$SCAN), in close agreement with $\text{Z}_\text{N}^* = 1.91e$ and $\epsilon^\infty = 4.54$ values from a previous work~\cite{Ponce2021}.

\begin{table}[h]
\begin{tabular}{ccc}
\hline\hline
Source & Hole Hall mobility & Hall Factor \\
       &    (cm$^2$~/~Vs)   &             \\
\hline
QE-PBE & 398 & 0.73 \\
VASP-PBE & 418 &  0.80 \\
VASP-r$^2$SCAN & 469 &  0.79 \\
\hline
Ref.~\cite{Ponce2021} &  410 &  0.75 \\
\hline\hline
\end{tabular}
\caption{Hole Hall mobilities and Hall factors for cubic BN at 300~K obtained from transport calculations that include long-range dipole corrections, using QE-PBE, VASP-PBE, and VASP-r$^2$SCAN. For comparison, results from Ref.~\cite{Ponce2021} are also included.}
\label{tab:cbn-mob}
\end{table}

Figure~\ref{fig:cBN}(a)-(b) compares the electronic band structure and the phonon dispersion computed with VASP and QE. The band gaps are 4.5~eV for both PBE functionals and 5.0~eV for r$^2$SCAN, and the LO-TO splitting is approximately 30~meV, in excellent agreement with previous PBE-level calculations~\cite{Ponce2021}. Figure~\ref{fig:cBN}(c) shows the hole drift mobility as a function of temperature, evaluated with the aiBTE for a hole carrier density of 10$^{-13}$ cm$^{-3}$. There is good agreement between the two codes, with the 300~K point from Ref.~\cite{Ponce2021} also matching our data. Table~\ref{tab:cbn-mob} summarizes the Hall factor and hole Hall mobility at 300~K from this work alongside Ref.~\cite{Ponce2021}. The overall agreement is good; the residual differences stem primarily from different $\bk$- and $\bq$-mesh densities employed in Ref.~\cite{Ponce2021}, namely coarse meshes of 14$\times$14$\times$14 ($\bk$) and 7$\times$7$\times$7 ($\bq$) and fine meshes of 180$\times$180$\times$180 ($ \bk'$ and $\bq'$). Detailed convergence analyses in Refs.~\cite{Ponce2021,Lee2023} indicate that coarse and fine $\bk$ and $\bq$ grids, typically exceeding 14$\times$14$\times$14 and 250$\times$250$\times$250, respectively, are required to reach full convergence.

\section{Conclusion}
\label{sec:conclusion}

In this article, we introduce the EPW-VASP interface for first-principles calculations of advanced materials properties defined by electron-phonon interactions. The interface enables the use of real-space Hamiltonians, interatomic force constants, and electron-phonon matrix elements computed within the supercell approach in VASP, employing the projector augmented-wave method and Wannier functions. The EPW code performs the necessary Fourier interpolation to fine grids and evaluates relevant materials properties. We demonstrate the capabilities of the EPW-VASP interface through two representative examples: the calculation of anisotropic superconducting properties using Migdal-Eliashberg theory and charge transport properties in polar materials using the Boltzmann transport equation.

Several features are currently under testing or planned for future development, including support for two-dimensional systems, introduction of long-range quadrupole corrections to the dynamical and electron-phonon matrices in the PAW formalism, as well as calculations of phonon-assisted optical processes, and polaronic effects. Additionally, extending the interface to support electron-phonon calculations with hybrid functionals, made possible by the universality of the finite-difference approach used in VASP, would be highly desirable. Finally, the EPW-VASP interface can be used in conjunction with the recently developed EPWpy program, which provides a high-level Python interface to streamline automated workflows for high-throughput calculations of materials properties.

During the preparation of this manuscript, related work by Poliukhin \textit{et al.}~\cite{Poliukhin2025} introduced a finite-displacement supercell approach that constructs electron–phonon matrix elements via a projectability scheme using the eigenvalues and wavefunction overlaps of pristine and displaced structures. This general method is implemented in the ElePhAny code, which interfaces Quantum ESPRESSO~\cite{Giannozzi2017}, Koopmans~\cite{Linscott2023}, and Yambo~\cite{Marini2009} (for electronic structure), and Phonopy~\cite{Togo2015} (for phonon dispersion) to generate EPW-ready inputs for Wannier interpolation and subsequent calculations of electron-phonon related properties.

\acknowledgments
This work was supported by the U.S. National Science Foundation through CSSI Awards OAC-2513830 and OAC-2103991. Computational resources were provided by the Texas Advanced Computing Center (TACC) at The University of Texas at Austin~\cite{TACC}, specifically the Stampede3 and Frontera \cite{Frontera} supercomputers, through Allocations TG-DMR180071 and DMR22004. This work also used the Expanse~\cite{Expanse} supercomputer at the San Diego Supercomputer Center (SDSC) through Allocation TG-DMR180071. Stampede3 is supported by NSF Award OAC-2311628, Frontera by NSF Award OAC-1818253, and Expanse by NSF Award ACI-1548562.
M.E. acknowledges Henrique Miranda for valuable discussions during the early stages of this work.

\appendix

\section{Wigner-Seitz construction}
\label{sec:WS}

The interpolation of the Hamiltonian, dynamical matrix, and electron-phonon matrix elements from real space to reciprocal space relies on the spatial decay (or localization) of these quantities in real space. Consequently, the accuracy and efficiency of the interpolation depend critically on the choice of real-space lattice vectors to be summed over in Eqs.~\eqref{eq:EPWHamBlochW}, \eqref{eq:DynmatBlochW} and \eqref{eq:ElphBloch}. In this section, we recap the procedure for optimal Wigner-Seitz (WS) construction introduced in Refs.~\cite{Pizzi2020,Ponce2021}.

For the Hamiltonian, the decay is governed by the spatial extent of the Wannier functions, and the most relevant lattice vectors are those that minimize the distance between the corresponding Wannier centers. A similar principle holds for the IFCs and electron-phonon matrix elements. In the case of IFCs, the key distances are between pairs of atomic positions, while for electron-phonon interactions, the relevant separation is between an atomic site and a Wannier center.

Given a primitive unit cell defined by $\{\ba_i\}, i = 1, 2, 3$, and a supercell expanded by $\{N_i\}$, an effective method of selecting the optimal set of lattice vectors involves minimizing the distance
\begin{equation}
\label{eq:mindist}
|\bR_{p_1 p_2 p_3} + \bT_{j_1 j_2 j_3} + \br_n - \br_m|.
\end{equation}
Here, $\br_n$ and $\br_m$ denote either atomic or Wannier center positions inside the primitive unit cell, $\bR_{p_1 p_2 p_3} = \sum_i p_i{\bf a}_i$ specifies a primitive unit cell within the BvK supercell with $p_i\in[0,N_i)$, and $\bT_{j_1 j_2 j_3} = \sum_i j_i N_i{\bf a}_i$ is a supercell lattice vector.

\begin{figure}[t!]
\centering
\includegraphics[width=1\linewidth]{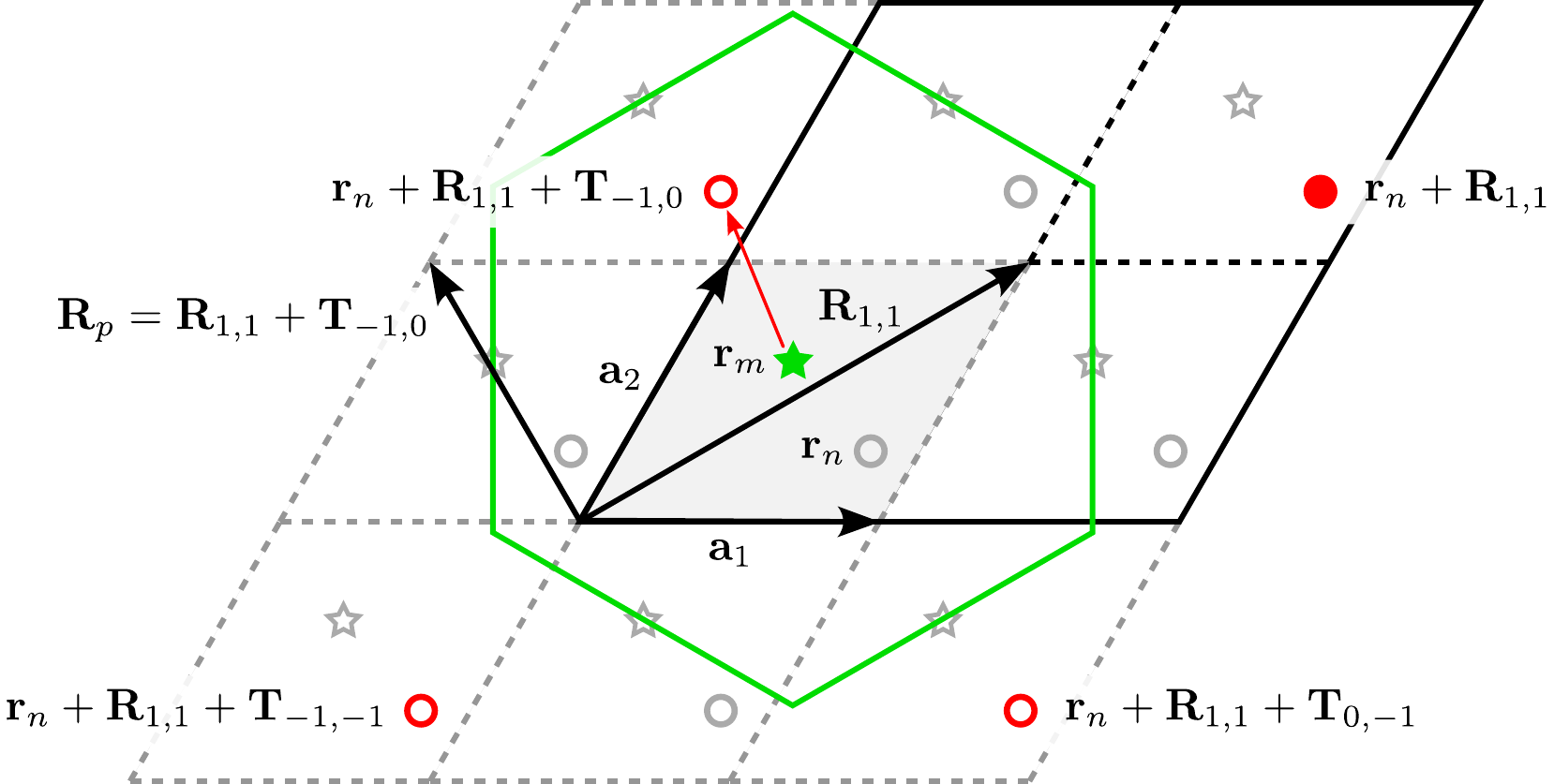}
\caption{Wigner-Seitz construction for centers $\br_m$, $\br_n$,  and unit cell $\bR_{1, 1}$ in the $2\times2$ BvK supercell. Infinite periodic grid is outlined by dashed gray lines. Points $\br_m$ (green star) and $\br_n$ (marked gray ring) are placed inside the unit cell $\bR_{0, 0}$ (light gray shade). Point $\br_n + \bR_{1,1}$ (solid red circle) is inside the top right unit cell $\bR_{1, 1}$ (black dashed lines) of the $2~\times~2$~BvK supercell (black solid lines). Among $\br_n + \bR_{1,1}$ and its replicas (red rings), only the replica falling inside the WS supercell centered around $\br_m$ (solid green lines) has minimal distance to $\br_m$ (see the red arrow). Selected $\bR_p = \bR_{1,1} + \bT_{-1, 0}$ corresponds to the vector of the unit cell this replica is located in.}
\label{fig:ws_construct}
\end{figure}

For each combination $(m,n,\{p_i\})$, the process cycles over a sufficiently large grid of $\bT_{j_1 j_2 j_3}$, e.g., $j_i\in[-3,3]$, to determine when the vector $\bR_{p_1 p_2 p_3} + \bT_{j_1 j_2 j_3} + \br_n$ falls within the WS cell centered at ${\bf r}_m$ (see Fig.~\ref{fig:ws_construct}). In case this vector is at the boundary of the WS cell, all equivalent $\mathcal{N}_{mn \{p_i\}}$ replicas are kept with the weight factor $1/\mathcal{N}_{mn \{p_i\}}$. This amounts to replacing the summation over the $\Rp$ lattice vectors in Eq.~\eqref{eq:EPWHamBlochW} with
\begin{equation}
\label{eq:EPWHamBlochWS}
\begin{aligned}
H^{\text{W}}_{mn}(\bk)  &= \sum_{\{p_i\}}^{\rm WS} \frac{1}{\mathcal{N}_{mn \{p_i\}}} \sum_{\substack{\{j_i\} \in \\ \mathcal{F}(mn \{p_i\})}} e^{i \bk \cdot \left(\bR_{\{p_i\}}+\bT_{\{j_i\}}\right)} \\
& \quad \quad \quad \times  H_{m n}(\bR_{\{p_i\}}),
\\
\mathcal{F} &: mn \{p_i\} \rightarrow \left\{\{j_i\}^{(1)}, ..., \{j_i\}^{(\mathcal{N}_{mn \{p_i\}})}\right\},
\end{aligned}
\end{equation}
where mapping $\mathcal{F}$ returns all sets of translational indices $\{j_i\}$ corresponding to given $(m, n, \{p_i\})$. A similar procedure is used in Eqs.~\eqref{eq:DynmatBlochW} and \eqref{eq:ElphBloch}.

\section{\label{sec:paw}Electron-phonon matrix elements in the PAW method}

In VASP, the computation of electron-phonon matrix elements is performed in the PAW framework.
The PAW method addresses the challenge of representing all-electron (AE) orbitals in plane-wave codes by introducing a transformation that maps smooth pseudo (PS) orbitals, suitable for expansion in plane waves, onto AE orbitals~\cite{Blochl1994,Kresse1999}.
This transformation enables efficient computation without sacrificing the ability to recover AE observables.

Unlike traditional pseudopotential approaches, which discard AE information within a core radius, the PAW method employs a dual basis representation. Each PS orbital, \(\ket{\ppsi_{n \bk}} \), is augmented inside atom-centered spheres (augmentation spheres) with AE partial waves \(\ket{\phi_{ai}} \), PS partial waves \(\ket{\tilde{\phi}_{ai}} \), and projector functions \(\ket{\tilde{p}_{ai}} \), thereby allowing reconstruction of the AE orbital, \(\ket{\psi_{n \bk}} \):
\begin{equation}
    \ket{\psi_{n \bk}} = \ket{\ppsi_{n \bk}}
    + \sum_{ai} \left( \ket{\phi_{ai}} - \ket{\tilde{\phi}_{ai}} \right)
    \braket{\tilde{p}_{ai} | \ppsi_{n \bk}}
    ,
\end{equation}
where \(i \) denotes the PAW channel, that is, a composite index that describes the local basis functions inside the augmentation sphere around the atom \(a \).
Projector functions and PS orbitals are described in the plane-wave basis while AE and PS partial waves are described on logarithmic radial grids inside the augmentation spheres.

The mapping from PS to AE orbitals is known as the PAW transformation:
\begin{equation}
    \hat{\mathcal{T}} = 1
    + \sum_{ai} \left( \ket{\phi_{ai}} - \ket{\tilde{\phi}_{ai}} \right)
    \bra{\tilde{p}_{ai}}
    .
\end{equation}
It allows us to recast the Kohn-Sham equations in terms of a generalized eigenvalue problem:
\begin{equation} \label{eq:paw-ks}
    \tilde{H} \ket{\ppsi_{n\bk}} = \varepsilon_{n\bk} \tilde{S} \ket{\ppsi_{n\bk}}
    ,
\end{equation}
with
\begin{align}
    \tilde{H} & \equiv \hat{\mathcal{T}}^\dagger \hat{H} \hat{\mathcal{T}},
    \\
    \tilde{S} & \equiv \hat{\mathcal{T}}^\dagger \hat{\mathcal{T}}.
\end{align}
The operators \(\tilde{H} \) and \(\tilde{S} \) are known as the PAW Hamiltonian and PAW overlap operator, respectively.

By simple differentiation of Eq.~\eqref{eq:paw-ks}, it is possible to define the PS electron-phonon matrix element~\cite{Engel2020}:
\begin{equation}
    g^{\text{PS}}_{mn\nu}(\bk, \bq)
    \equiv
    \braket{\ppsi_{m \bk + \bq} | \Delta_{\bq \nu} \tilde{H} | \ppsi_{n \bk}}
    ,
\end{equation}
where the phonon displacement operator, \(\Delta_{\bq \nu} \), was defined in Eq.~\eqref{eq:EPWdisplacement}.
Alternatively, one can introduce the PAW transformation after differentiation, that is, directly reformulating Eq.~\eqref{eq:EPWelph} in terms of the PAW quantities.
This leads to the AE definition of the electron-phonon matrix element~\cite{Chaput2019}, \(g^{\text{AE}}_{mn\nu}(\bk, \bq) \), which is formally equivalent to \(g_{mn\nu}(\bk, \bq) \) in the limit of a complete PAW basis.
The explicit form of \(g^{\text{AE}}_{mn\nu}(\bk, \bq) \) as computed in VASP is beyond the scope of this appendix and is given elsewhere~\cite{Engel2022}.
However, we are able to show the relationship between \(g^{\text{AE}}_{mn\nu}(\bk, \bq) \) and \(g^{\text{PS}}_{mn\nu}(\bk, \bq) \):
\begin{multline}
    g^{\text{AE}}_{mn\nu}(\bk, \bq) = g^{\text{PS}}_{mn\nu}(\bk, \bq)
    \\ +
    \left(\varepsilon_{n\bk} - \varepsilon_{m \bk + \bq}\right)
    \braket{\ppsi_{m \bk + \bq} | \hat{\mathcal{T}}^\dagger \Delta_{\bq \nu} \hat{\mathcal{T}} | \ppsi_{n \bk}}
    ,
\end{multline}
clearly highlighting the non-equivalence of the two definitions for off-diagonal elements.

Both approaches have merits. For general applications, it is best to use the AE formulation, as it is well defined in many-body perturbation theory. The PS approach is well suited for the computation of the band-gap renormalization, where a formal equivalence with the AE approach can be shown in the adiabatic approximation~\cite{Engel2022}.

\section{\label{sec:conventions}Comparison of conventions between VASP and EPW}

The conventions used by VASP for the real-space Hamiltonian, IFCs, and electron-phonon matrix elements differ from those adopted in EPW, reflecting VASP’s alignment with a finite-difference scheme~\cite{Engel2020}. As a result, the real-space arrays read from the VASP's output HDF5 file require careful reformatting to ensure compatibility with EPW’s interpolation routines.

This section addresses the case where the central cells of the electron and phonon grids do not coincide. Notation is as follows: Wannier centers are denoted by $m, n$; atoms by $\k, \kp$; and unit cell indices by $p, p', p''~\in~[0, N_p)$, with $N_p$ the number of unit cells. The absolute positions of unit cell $p$ for the electron and phonon grids, defined with respect to a random origin, are written as $\bRep(m)$ and $\bRphp(\k)$, respectively. The central (``original'', or first) unit cells correspond to $p=0$, with positions $\bReo(m)$  and $\bRpho(\k)$.

\begin{figure}[t!]
\centering
\includegraphics[width=1\columnwidth]{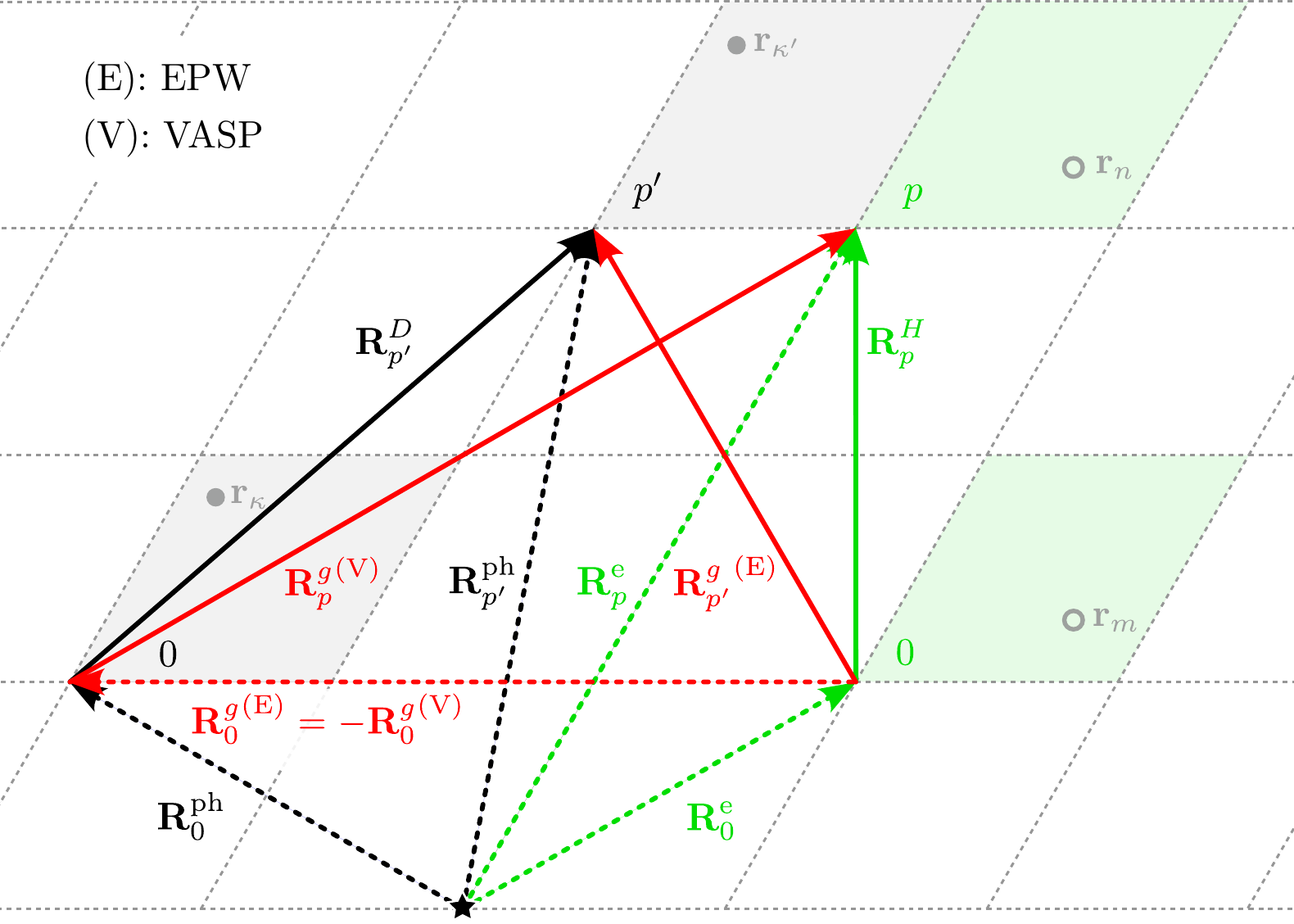}
\caption{Schematic representation of the relationships between vectors on the electron and phonon grids in EPW~(E) and VASP~(V) in the general case where the central unit cells (``0'') of the electron grid (shaded green) and phonon grid (shaded gray) do not coincide, i.e. $\bReo(m) \not\equiv \bRpho(\k)$. The two-center vectors (solid arrows) for electron-electron $\bRhp(m, n)$ (green), phonon-phonon $\bRdpp(\k, \kp)$ (black), and electron-phonon  $\bRgppE(m, \kp)$ and $\bRgpV(\k, n)$ (red) are defined according to Eqs.~\eqref{eq:H}-\eqref{eq:gV}. Random origin is indicated with a black star. Dashed green and black arrows show corresponding single-center electron and phonon vectors, respectively. A shift between the central cells of the electronic and phonon grids is represented by the dashed red arrow.
}
\label{fig:grid_ev}
\end{figure}

Grid vectors in EPW (E) and VASP (V) are then grouped into different sets depending on the centers from which they are generated.
As illustrated in Fig.~\ref{fig:grid_ev}, in both codes, the two-center electron–electron and phonon–phonon  vectors are defined in a similar way
\begin{align}
    & \bRhp(m, n) = \bRep(n) - \bReo(m \label{eq:H}), \\
    & \bRdpp(\k, \kp) = \bRphpp(\kp) - \bRpho(\k) \label{eq:D},
\end{align}
however, the two-center electron–phonon vectors differ
\begin{align}
    & \bRgpE(m, \k) = \bRphp(\k) - \bReo(m) \label{eq:gE}, \\
    & \bRgpV(\k, m) = \bRep(m) - \bRpho(\k) \label{eq:gV}.
\end{align}
This difference originates from the distinct translational symmetries each code employs to minimize the number of electron–phonon matrix elements in Wannier space.

In VASP, all phonon displacements are effectively performed in one of the cells of the phonon grid, such that
\begin{equation}
  \begin{aligned}
  g_{mn\k\a}^{\text{(V)}}&(\bRep, \bRphpp, \bReppp)
  =
  \bra{m \bReppp}
  \left(\partial_{\k\a, \bRphpp} \hat{H}\right)
  \ket{n \bRep}
  \\
  &=
  \bra{m \bReppp - \bRphpp}
  \left(\partial_{\k\a, {\bf 0}} \hat{H}\right)
  \ket{n \bRep - \bRphpp}
  ~\forall~p'.
  \\
  \end{aligned}
\end{equation}
If this reference cell is chosen as the central cell of the phonon grid, i.e. $p' = 0$, the full $N_p \times N_p \times N_p$ electron-phonon matrix reduces to an $N_p \times N_p$ form:
\begin{equation} \label{eq:gVr}
  \begin{aligned}
  g_{mn\k\a}^{\text{(V)}}&(\bRep, \bRpho, \bReppp)
  \\
  &=
  \bra{m \underbrace{\bReppp - \bRpho}_{\bRgpppV(\k,m)}}
  \left(\partial_{\k\a, {\bf 0}} \hat{H}\right)
  \ket{n \underbrace{\bRep - \bRpho}_{\bRgpV(\k,n)}}
  \\
  &=
  g_{mn\k\a}^{\text{(V)}}(\bRgpV, \bRgpppV)
  ~\text{for some fixed}~p' = 0.
  \end{aligned}
\end{equation}
With this convention, the electron-phonon grid vectors in VASP are referenced to the central cell of the phonon grid, according to Eq.~\eqref{eq:gV}.

In contrast, EPW pins one of the Wannier centers to a chosen electron-grid cell:
\begin{equation}
  \begin{aligned}
  g_{mn\k\a}^{\text{(E)}}&(\bRep, \bRphpp, \bReppp)
  =
  \bra{m \bReppp}
  \left(\partial_{\k\a, \bRphpp} \hat{H}\right)
  \ket{n \bRep}
  \\
  &=
  \bra{m {\bf 0}}
  \Big(\partial_{\k\a, \bRphpp - \bReppp} \hat{H}\Big)
  \ket{n \bRep - \bReppp}
  ~\forall~p''.
  \end{aligned}
\end{equation}
By EPW convention, this pinned cell is the central cell of the electron grid, i.e. $p'' = 0$, so the electron-phonon matrix reduces to
\begin{equation} \label{eq:gEr}
  \begin{aligned}
  g_{mn\k\a}^{\text{(E)}}&(\bRep, \bRphpp, \bReo)
  \\
  &=
  \bra{m {\bf 0}}
  \Big(\partial_{\k\a, \underbrace{\bRphpp - \bReo}_{\bRgppE(m,\k)}} \hat{H}\Big)
  \ket{n \underbrace{\bRep - \bReo}_{\bRhp(m, n)}}
  \\
  &=
  g_{mn\k\a}^{\text{(E)}}(\bRhp, \bRgppE)
  ~\text{for some fixed}~p'' = 0,
  \end{aligned}
\end{equation}
according to Eqs.~\eqref{eq:H} and \eqref{eq:gE}.

Figure \ref{fig:grid_ev} provides a schematic representation of the relationships between vectors on the electron and phonon grids in EPW and VASP:
\begin{equation}
  \begin{aligned}
  \bRgppE(m, \kp) &=
  \bRdpp(\k, \kp) + \underbrace{\bRpho(\k) - \bReo(m)}_{{\bRgo}^{\text{(E)}}(m, \k) = - {\bRgo}^{\text{(V)}}(\k, m)}
  \\
  &=
  \bRdpp(\k, \kp) - {\bRgo}^{\text{(V)}}(\k, m),\\
  \end{aligned}
  \label{eq:eph_grid_e_from_v}
\end{equation}

\begin{equation}
  \begin{aligned}
  \bRgpV(\k, n) &=
  \underbrace{- \bRpho(\k) + \bReo(m)}_{- \bRgoE(m, \k) = \bRgoV(\k, m)} + \bRhp(m, n)
  \\
  &=
  - \bRgoE(m, \k) + \bRhp(m, n).
  \end{aligned}
  \label{eq:eph_grid_v_from_e}
\end{equation}

According to Eqs.~\eqref{eq:gVr} and \eqref{eq:gEr}, the electron-phonon matrix elements in VASP and EPW are defined on differently ordered grids, namely $(\bRgpV, \bRgpppV)$ and $(\bRhp, \bRgppE)$, which involve distinct sets of two-center grid vectors. As a consequence, in the workflow illustrated in Fig.~\ref{fig:general_scheme}, two successive Fourier transformations are currently required to map the matrix elements from the VASP convention onto that used in EPW. Notably, the same index $p$, running on the electron grid, establishes a direct correspondence between specific elements of the electron-phonon matrices in Wannier space for the two codes:
\begin{equation}
  \begin{aligned}
    g_{mn\k\a}^{\text{(E)}}(\bRhp, \bRgoE)
    &=
    g_{mn\k\a}^{\text{(V)}}(\bRgpV, \bRgoV)~\\
    &\forall~p~\text{(but only for}~p' = p'' = 0 \text{)}.
  \end{aligned}
\end{equation}

Tables~\ref{tab:conventions} and \ref{tab:conventions2} highlight additional distinctions in the formulations adopted by VASP and EPW. Beyond the alternative definitions of electron–phonon grid vectors, the Hamiltonian and dynamical matrices in VASP’s Wannier representation are the Hermitian conjugates of those in EPW. Differences also arise in the stage at which divisions by $\sqrt{N_p}$ are applied within the Fourier transforms.

\begin{widetext}
\onecolumngrid
{
\begin{table}[htbp]
  \caption{\label{tab:conventions} Comparison of internal conventions in VASP (V) and EPW (E) for definitions of vectors on electron and phonon grids and their relations.
  Grid vectors: $\bRep(n)$ - electron, $\bRphp(\k)$ - phonon, $\bRhp(m, n) = \bRep(n) - \bReo(m)$ - electron-electron,
  $\bRdp(\k, \kp) = \bRphp(\kp) - \bRpho(\k)$ - phonon-phonon,
  $\bRgp(m, \k) = \bRphp(\k) - \bReo(m)$ - electron-phonon.
  See also Fig.~\ref{fig:grid_ev}.
  Note that $U^\dag_{mn\bk} \equiv \left(U^\dag_{\bk}\right)_{mn} = U^\ast_{nm\bk}$, where $U_{\bk}$ is the unitary matrix generating maximally-localized Wannier functions,
  and $\sU_{\bk} = U^\dag_{\bk}$ arises from diagonalizing the Bloch Hamiltonian.
  }
  \begin{tabular}{ccc}
  \hline
  \hline
  Quantity & VASP & EPW \\
  \hline
  \\
  \multirow{4}{*}{
  \shortstack{Wannier\\functions\\from\\Bloch\\states}
  }
  &
  \begin{math}
  \begin{aligned}
  \ket{w_{mp}} &\equiv
  \ket{m \bRep}
  = \frac{1}{\sqrt{N_p}}
  \sum_{n, \bk} e^{- i \bk \cdot \bRep}
  \ket{\psi_{n \bk}}
  U_{nm \bk}
  \end{aligned}
  \end{math}
  &
  \begin{math}
  \begin{aligned}
  \ket{w_{mp}} &\equiv
  \ket{m \bRep} =
  \frac{1}{\sqrt{N_p}}
  \sum_{n, \bk} e^{- i \bk \cdot \bRep}
  \ket{\psi_{n \bk}}
  U_{nm \bk}
  \end{aligned}
  \end{math}
  \\
  \\
  &
  \multicolumn{2}{c}
  {
  \begin{tcolorbox}[colback=gray!10!white,colframe=white,width=0.85\columnwidth]
  \begin{math}
  \begin{aligned}
    &\bRepE(m) = \bRepV(m) = \bRep(m),
    ~~~\ket{m \bRep}^{\text{(E)}} = \ket{m \bRep}^{\text{(V)}},
    ~~~U^{\text{(E)}}(\bk) = U^{\text{(V)}}(\bk)
  \end{aligned}
  \end{math}
  \end{tcolorbox}
  }
  \\
  \hline
  \\
  \multirow{2}{*}{\shortstack{Wannier\\Hamiltonian}}
  &
  \begin{math}
  \begin{aligned}
  &H_{np,m0} = H_{nm}(\bRhp) = \braket{w_{np}|\hat{H}|w_{m0}}
  \\
  &
  = \sum_{n^\prime m^\prime \bk} e^{+i\bk\cdot\bRhp} U^\dag_{nn^\prime \bk} H_{n^\prime m^\prime} (\bk) U_{m^\prime m \bk}
  \end{aligned}
  \end{math}
  &
  \begin{math}
  \begin{aligned}
  &H_{m0, np} = H_{mn}(\bRhp) = \braket{w_{m0}|\hat{H}|w_{np}}
  \\
  &=
  \frac{1}{N_p} \sum_{m^\prime n^\prime \bk} e^{-i\bk\cdot\bRhp} U^\dag_{mm^\prime \bk} H_{m^\prime n^\prime} (\bk) U_{n^\prime n \bk}
  \end{aligned}
  \end{math}
  \\
  \\
  &
  \multicolumn{2}{c}
  {
  \begin{tcolorbox}[colback=gray!10!white,colframe=white,width=0.85\columnwidth]
  \begin{math}
  \begin{aligned}
  &\bRhpE(m, n) = \bRhpV(m, n) = \bRhp(m, n),
  ~~~H^{\text{(E)}}(\bRhp) = \frac{1}{N_p} \left[H^{\text{(V)}}(\bRhp)\right]^\dag
  \end{aligned}
  \end{math}
  \end{tcolorbox}
  }
  \\
  \hline
  \\
  \multirow{4}{*}{\shortstack{Bloch\\Hamiltonian\\in\\Wannier\\gauge}}
  &
  \multirow{5}{*}{
  \begin{math}
  \begin{aligned}
  H^{\text{W}}_{n m}(\bk)
  &=
  \frac{1}{N_p}
  \sum_{p}
  e^{-i \bk \cdot \bRhp}
  H_{n m}(\bRhp)
  \end{aligned}
  \end{math}
  }
  &
  \multirow{5}{*}{
  \begin{math}
  \begin{aligned}
  H^{\text{W}}_{m n}(\bk)
  &=
  \sum_{p}
  e^{+i \bk \cdot \bRhp}
  H_{m n}(\bRhp)
  \end{aligned}
  \end{math}
  }
  \\
  \\
  \\
  \\
  &
  \multicolumn{2}{c}
  {
  \begin{tcolorbox}[colback=gray!10!white,colframe=white,width=0.85\columnwidth]
  \begin{math}
  \begin{aligned}
  &H^{\text{W(E)}}(\bk) = \left[H^{\text{W(V)}}(\bk)\right]^\dag
  \end{aligned}
  \end{math}
  \end{tcolorbox}
  }
  \\
  \hline
  \\
  \multirow{3}{*}{\shortstack{Diagonal\\Bloch\\Hamiltonian}}
  &
  \multirow{3}{*}{
  \begin{math}
  \begin{aligned}
  H_{nm}(\bk) &= \varepsilon_{m \bk} \delta_{mn}
  =
  \sum_{n^\prime m^\prime}
  \sU^\dag_{nn^\prime \bk}
  H^{\text{W}}_{n^{\prime} m^{\prime}}(\bk)
  \sU_{m^\prime m \bk}
  \end{aligned}
  \end{math}
  }
  &
  \multirow{3}{*}{
  \begin{math}
  \begin{aligned}
    H_{mn}(\bk) &= \varepsilon_{n \bk} \delta_{nm}
  =
  \sum_{m^\prime n^\prime}
  \sU^\dag_{mm^\prime \bk}
  H^{\text{W}}_{m^{\prime} n^{\prime}}(\bk)
  \sU_{n^\prime n \bk}
  \end{aligned}
  \end{math}
  }
  \\
  \\
  \\
  &
  \multicolumn{2}{c}
  {
  \begin{tcolorbox}[colback=gray!10!white,colframe=white,width=0.85\columnwidth]
  \begin{math}
  \begin{aligned}
  &\sU^{\text{(E)}}_{\bk} = U^{\dag \text{(E)}}_{\bk} = U^{\dag \text{(V)}}_{\bk} = \sU_{\bk}^{\text{(V)}},
  ~~~H^{\text{(E)}}(\bk) = \left[H^{\text{(V)}}(\bk)\right]^\dag
  \end{aligned}
  \end{math}
  \end{tcolorbox}
  }
  \\
  \hline
  \\
  \multirow{3}{*}{\shortstack{variation\\Bloch\\Hamiltonian}}
  &
  \begin{math}
  \begin{aligned}
  \Delta_{\bq \nu} \hat{H}
  &=
  \sum_{\k\a p}
  \sqrt{\frac{\hbar}{2 M_{\k} \o_{\bq \nu}}}
  e^{i \bq \cdot \bRphp}
  e_{\k\a, \nu}(\bq)
  \underbrace{
  \frac{\partial \hat{H}}{\partial \tau_{\k\a p}}
  }_{\partial_{\k\a, \bRphp} \hat{H}}
  \end{aligned}
  \end{math}
  &
  \begin{math}
  \begin{aligned}
  \Delta_{\bq \nu} \hat{H}
  &=
  \sum_{\k\a p}
  \sqrt{\frac{\hbar}{2 M_{\k} \o_{\bq \nu}}}
  e^{i \bq \cdot \bRphp}
  e_{\k\a, \nu}(\bq)
  \underbrace{
  \frac{\partial \hat{H}}{\partial \tau_{\k\a p}}
  }_{\partial_{\k\a, \bRphp} \hat{H}}
  \end{aligned}
  \end{math}
  \\
  &
  \multicolumn{2}{c}
  {
  \begin{tcolorbox}[colback=gray!10!white,colframe=white,width=0.85\columnwidth]
  \begin{math}
  \begin{aligned}
  &\bRphpE(\k) = \bRphpV(\k) = \bRphp(\k),
  ~~~\partial_{\k\a, \bRphp} \hat{H}^{\text{(E)}} = \partial_{\k\a, \bRphp} \hat{H}^{\text{(V)}}
  \end{aligned}
  \end{math}
  \end{tcolorbox}
  }
  \\
  \hline
  \hline
  \end{tabular}
\end{table}

\begin{table}[htbp]
  \caption{\label{tab:conventions2} Comparison of internal conventions in VASP (V) and EPW (E) for definitions of vectors on electron and phonon grids and their relations.
  }
  \begin{tabular}{ccc}
  \hline
  \hline
  Quantity & VASP & EPW \\
  \hline
  \\
  \multirow{2}{*}{\shortstack{IFC\\matrix}} &
  $C_{\kp\ap p, \k\a0} \equiv D_{\kp \ap, \k\a}(\bRdp)$ &
  $C_{\k\a 0, \kp\ap p}\equiv D_{\k\a,\kp \ap}(\bRdp)$
  \\
  \\
  &
  \multicolumn{2}{c}
  {
  \begin{tcolorbox}[colback=gray!10!white,colframe=white,width=0.85\columnwidth]
  \begin{math}
  \begin{aligned}
  &\bRdpE(\k, \kp) = \bRdpV(\k, \kp) = \bRdp(\k, \kp),
  ~~~D^{\text{(E)}}(\bRdp) = \left[D^{\text{(V)}}(\bRdp)\right]^\dag
  \end{aligned}
  \end{math}
  \end{tcolorbox}
  }
  \\
  \hline
  \\
  \multirow{2}{*}{\shortstack{Dynamical\\matrix}} &
  \multirow{2}{*}{
  \begin{math}
  \begin{aligned}
  D_{\k\a, \kp\ap}(\bq)
  &=
  \frac{1}{\sqrt{M_\k M_{\kp}}}
  \sum_{p}
  e^{-i \bq \cdot \bRdp}
  C_{\kp \ap p, \k \a 0}
  \end{aligned}
  \end{math}
  }
  &
  \multirow{2}{*}{
  \begin{math}
  \begin{aligned}
  D_{\k\a, \kp\ap}(\bq)
  &=
  \frac{1}{\sqrt{M_\k M_{\kp}}}
  \sum_{p}
  e^{+i \bq \cdot \bRdp}
  C_{\k \a 0, \kp \ap p}
  \end{aligned}
  \end{math}
  }
  \\
  \\
  \\
  &
  \multicolumn{2}{c}
  {
  \begin{tcolorbox}[colback=gray!10!white,colframe=white,width=0.85\columnwidth]
  \begin{math}
  \begin{aligned}
  & D^{\text{(E)}}(\bq) = \left[D^{\text{(V)}}(\bq)\right]^\dag,
  ~~~{e^{\text{(E)}}_{\k\a, \nu}(\bq)} = {e^{\text{(V)}}_{\k\a, \nu}(\bq)}
  \end{aligned}
  \end{math}
  \end{tcolorbox}
  }
  \\
\hline
  \\
  \multirow{3}{*}{\shortstack{Wannier\\electron-phonon\\matrix}}
  &
  \begin{math}
  \begin{aligned}
  g_{mn\k\a}(\bRgp, \bRgppp) &=
  \braket{w_{mp''}|\frac{\partial \hat{H}}{\partial \tau_{\k \a 0}}|w_{np}}
  \end{aligned}
  \end{math}
  &
  \begin{math}
  \begin{aligned}
  g_{mn\k\a}(\bRhp, \bRgpp) &= \braket{w_{m0}|\frac{\partial \hat{H}}{\partial \tau_{\k \a p'}}|w_{np}}
  \end{aligned}
  \end{math}
  \\
  \\
  &
  \multicolumn{2}{c}
  {
  \begin{tcolorbox}[colback=gray!10!white,colframe=white,width=0.85\columnwidth]
  \begin{math}
  \begin{aligned}
  \bRgppE(m, \kp) &= \bRdpp(\k, \kp) - \bRgoV(\k, m),\\
  \bRgpV(\k, n) &= - \bRgoE(m, \k) + \bRhp(m, n),\\
  g^{\text{(E)}}_{mn\k\a}(\bRhp, \bRgoE) &= g^{\text{(V)}}_{mn\k\a}(\bRgpV, \bRgoV)~\forall~p~\text{(but only for}~p' = p'' = 0 \text{)}
  \end{aligned}
  \end{math}
  \end{tcolorbox}
  }
  \\
  \hline
  \hline
  \end{tabular}
\end{table}
}
\end{widetext}

\clearpage


\section*{Declarations}


\subsection*{Data availability}
The data supporting this work are available within the paper, its Supplementary Information ﬁles, and from the corresponding authors upon reasonable request.

\subsection*{Code availability}
EPW and VASP codes used in present study
can be downloaded from their corresponding websites:
\url{https://epw-code.org} and \url{https://www.vasp.at},
respectively.

\section*{Author contributions}
E.R.M. and F.G. conceived the project.
A.T. and D.R. developed the EPW-VASP interface in the EPW code, and M.E. developed the matching part of the interface in VASP.
A.N.K., S.T., G.K., F.G., and E.R.M. contributed to the algorithm and methodology development through discussions.
D.R. performed all reported calculations.
D.R., E.R.M., M.E., and A.N.K. wrote the initial draft, and all authors contributed to the final version through feedback and suggestions.

\bibliography{references.bib}

@article{Giustino2007,
  title = {{Electron-phonon interaction using Wannier functions}},
  author = {Giustino, Feliciano and Cohen, Marvin L. and Louie, Steven G.},
  journal = {Phys. Rev. B},
  volume = {76},
  issue = {16},
  pages = {165108},
  numpages = {19},
  year = {2007},
  month = {Oct},
  publisher = {American Physical Society},
  doi = {10.1103/PhysRevB.76.165108},
  url = {https://link.aps.org/doi/10.1103/PhysRevB.76.165108}
}

@article{Noffsinger2010,
title = {{EPW: A program for calculating the electron–phonon coupling using maximally localized Wannier functions}},
journal = {Computer Physics Communications},
volume = {181},
number = {12},
pages = {2140-2148},
year = {2010},
issn = {0010-4655},
doi = {https://doi.org/10.1016/j.cpc.2010.08.027},
url = {https://www.sciencedirect.com/science/article/pii/S0010465510003218},
author = {Jesse Noffsinger and Feliciano Giustino and Brad D. Malone and Cheol-Hwan Park and Steven G. Louie and Marvin L. Cohen}
}

@article{Margine2013,
  title = {{Anisotropic Migdal-Eliashberg theory using Wannier functions}},
  author = {Margine, E. R. and Giustino, F.},
  journal = {Phys. Rev. B},
  volume = {87},
  issue = {2},
  pages = {024505},
  numpages = {12},
  year = {2013},
  month = {Jan},
  publisher = {American Physical Society},
  doi = {10.1103/PhysRevB.87.024505},
  url = {https://link.aps.org/doi/10.1103/PhysRevB.87.024505}
}

@article{Ponce2016,
  title={{EPW}: {Electron}-phonon coupling, transport and superconducting properties using maximally localized {Wannier} functions},
  author={Ponc{\'e}, S. and Margine, E. R. and Verdi, C. and Giustino, F.},
  journal={Computer Physics Communications},
  volume={209},
  pages={116},
  year={2016},
  url={https://www.sciencedirect.com/science/article/pii/S0010465516302260}
}

@article{Giustino2017,
  title = {Electron-phonon interactions from first principles},
  author = {Giustino, Feliciano},
  journal = {Rev. Mod. Phys.},
  volume = {89},
  issue = {1},
  pages = {015003},
  numpages = {63},
  year = {2017},
  month = {Feb},
  publisher = {American Physical Society},
  doi = {10.1103/RevModPhys.89.015003},
  url = {https://link.aps.org/doi/10.1103/RevModPhys.89.015003}
}

@Article{Lee2023,
author={Lee, Hyungjun and Bushick, Kyle and Hajinazar, Samad and Lafuente-Bartolome, Jon and Leveillee, Joshua and Lian, Chao and Lihm, Jae-Mo and Macheda, Francesco and Mori, Hitoshi and Paudyal, Hari and Sio, Weng Hong and Tiwari, Sabyasachi and Zacharias, Marios and Zhang, Xiao and Bonini, Nicola and Kioupakis, Emmanouil and Margine, Elena R. and Giustino, Feliciano},
title={{Electron--phonon physics from first principles using the EPW code}},
journal={npj Computational Materials},
year={2023},
month={Aug},
day={25},
volume={9},
number={1},
pages={156},
issn={2057-3960},
doi={10.1038/s41524-023-01107-3},
url={https://doi.org/10.1038/s41524-023-01107-3}
}

@article{Pizzi2020,
doi = {10.1088/1361-648X/ab51ff},
url = {https://dx.doi.org/10.1088/1361-648X/ab51ff},
year = {2020},
month = {jan},
publisher = {IOP Publishing},
volume = {32},
number = {16},
pages = {165902},
author = {Pizzi, Giovanni and Vitale, Valerio and Arita, Ryotaro and Blügel, Stefan and Freimuth, Frank and Géranton, Guillaume and Gibertini, Marco and Gresch, Dominik and Johnson, Charles and Koretsune, Takashi and Ibañez-Azpiroz, Julen and Lee, Hyungjun and Lihm, Jae-Mo and Marchand, Daniel and Marrazzo, Antimo and Mokrousov, Yuriy and Mustafa, Jamal I and Nohara, Yoshiro and Nomura, Yusuke and Paulatto, Lorenzo and Poncé, Samuel and Ponweiser, Thomas and Qiao, Junfeng and Thöle, Florian and Tsirkin, Stepan S and Wierzbowska, Małgorzata and Marzari, Nicola and Vanderbilt, David and Souza, Ivo and Mostofi, Arash A and Yates, Jonathan R},
title = {Wannier90 as a community code: new features and applications},
journal = {Journal of Physics: Condensed Matter}
}

@article{Marrazzo2024,
  title = {Wannier-function software ecosystem for materials simulations},
  author = {Marrazzo, Antimo and Beck, Sophie and Margine, Elena R. and Marzari, Nicola and Mostofi, Arash A. and Qiao, Junfeng and Souza, Ivo and Tsirkin, Stepan S. and Yates, Jonathan R. and Pizzi, Giovanni},
  journal = {Rev. Mod. Phys.},
  volume = {96},
  issue = {4},
  pages = {045008},
  numpages = {54},
  year = {2024},
  month = {Dec},
  publisher = {American Physical Society},
  doi = {10.1103/RevModPhys.96.045008},
  url = {https://link.aps.org/doi/10.1103/RevModPhys.96.045008}
}

@article{Frohlich1954,
author = {H. Fröhlich},
title = {Electrons in lattice fields},
journal = {Advances in Physics},
volume = {3},
number = {11},
pages = {325--361},
year = {1954},
publisher = {Taylor \& Francis},
doi = {10.1080/00018735400101213},
URL = {https://doi.org/10.1080/00018735400101213}
}

@article{Vogl1976,
  title = {{Microscopic theory of electron-phonon interaction in insulators or semiconductors}},
  author = {Vogl, P.},
  journal = {Phys. Rev. B},
  volume = {13},
  issue = {2},
  pages = {694--704},
  year = {1976},
  month = {Jan},
  publisher = {American Physical Society},
  doi = {10.1103/PhysRevB.13.694},
}

@article{Verdi2015,
  title = {Fr\"ohlich Electron-Phonon Vertex from First Principles},
  author = {Verdi, Carla and Giustino, Feliciano},
  journal = {Phys. Rev. Lett.},
  volume = {115},
  issue = {17},
  pages = {176401},
  numpages = {5},
  year = {2015},
  month = {Oct},
  publisher = {American Physical Society},
  doi = {10.1103/PhysRevLett.115.176401},
  url = {https://link.aps.org/doi/10.1103/PhysRevLett.115.176401}
}

@article{Sjakste2015,
  title = {Wannier interpolation of the electron-phonon matrix elements in polar semiconductors: Polar-optical coupling in GaAs},
  author = {Sjakste, J. and Vast, N. and Calandra, M. and Mauri, F.},
  journal = {Phys. Rev. B},
  volume = {92},
  issue = {5},
  pages = {054307},
  numpages = {8},
  year = {2015},
  month = {Aug},
  publisher = {American Physical Society},
  doi = {10.1103/PhysRevB.92.054307},
  url = {https://link.aps.org/doi/10.1103/PhysRevB.92.054307}
}

@article{Brunin2020a,
  title = {{Electron-Phonon beyond Fr\"ohlich: Dynamical Quadrupoles in Polar and Covalent Solids}},
  author = {Brunin, Guillaume and Miranda, Henrique Pereira Coutada and Giantomassi, Matteo and Royo, Miquel and Stengel, Massimiliano and Verstraete, Matthieu J. and Gonze, Xavier and Rignanese, Gian-Marco and Hautier, Geoffroy},
  journal = {Phys. Rev. Lett.},
  volume = {125},
  issue = {13},
  pages = {136601},
  numpages = {6},
  year = {2020},
  month = {Sep},
  publisher = {American Physical Society},
  doi = {10.1103/PhysRevLett.125.136601},
  url = {https://link.aps.org/doi/10.1103/PhysRevLett.125.136601}
}

@article{Brunin2020b,
  title = {{Phonon-limited electron mobility in Si, GaAs, and GaP with exact treatment of dynamical quadrupoles}},
  author = {Brunin, Guillaume and Miranda, Henrique Pereira Coutada and Giantomassi, Matteo and Royo, Miquel and Stengel, Massimiliano and Verstraete, Matthieu J. and Gonze, Xavier and Rignanese, Gian-Marco and Hautier, Geoffroy},
  journal = {Phys. Rev. B},
  volume = {102},
  issue = {9},
  pages = {094308},
  numpages = {16},
  year = {2020},
  month = {Sep},
  publisher = {American Physical Society},
  doi = {10.1103/PhysRevB.102.094308},
  url = {https://link.aps.org/doi/10.1103/PhysRevB.102.094308}
}

@article{Park2020,
  title = {Long-range quadrupole electron-phonon interaction from first principles},
  author = {Park, Jinsoo and Zhou, Jin-Jian and Jhalani, Vatsal A. and Dreyer, Cyrus E. and Bernardi, Marco},
  journal = {Phys. Rev. B},
  volume = {102},
  issue = {12},
  pages = {125203},
  numpages = {8},
  year = {2020},
  month = {Sep},
  publisher = {American Physical Society},
  doi = {10.1103/PhysRevB.102.125203},
  url = {https://link.aps.org/doi/10.1103/PhysRevB.102.125203}
}

@article{Jhalani2020,
  title = {{Piezoelectric Electron-Phonon Interaction from Ab Initio Dynamical Quadrupoles: Impact on Charge Transport in Wurtzite GaN}},
  author = {Jhalani, Vatsal A. and Zhou, Jin-Jian and Park, Jinsoo and Dreyer, Cyrus E. and Bernardi, Marco},
  journal = {Phys. Rev. Lett.},
  volume = {125},
  issue = {13},
  pages = {136602},
  numpages = {6},
  year = {2020},
  month = {Sep},
  publisher = {American Physical Society},
  doi = {10.1103/PhysRevLett.125.136602},
  url = {https://link.aps.org/doi/10.1103/PhysRevLett.125.136602}
}

@article{Royo2021,
  title = {{Exact Long-Range Dielectric Screening and Interatomic Force Constants in Quasi-Two-Dimensional Crystals}},
  author = {Royo, Miquel and Stengel, Massimiliano},
  journal = {Phys. Rev. X},
  volume = {11},
  issue = {4},
  pages = {041027},
  numpages = {22},
  year = {2021},
  month = {Nov},
  publisher = {American Physical Society},
  doi = {10.1103/PhysRevX.11.041027},
  url = {https://link.aps.org/doi/10.1103/PhysRevX.11.041027}
}

@article{Zhang2022,
  title = {{Phonon}-limited transport of two-dimensional semiconductors: {Q}uadrupole scattering and free carrier screening},
  author = {Zhang, Chenmu and Liu, Yuanyue},
  journal = {Phys. Rev. B},
  volume = {106},
  issue = {11},
  pages = {115423},
  numpages = {9},
  year = {2022},
  month = {Sep},
  publisher = {American Physical Society},
  doi = {10.1103/PhysRevB.106.115423},
}

@article{Ponce2023a,
  title = {{Accurate Prediction of Hall Mobilities in Two-Dimensional Materials through Gauge-Covariant Quadrupolar Contributions}},
  author = {Ponc\'e, Samuel and Royo, Miquel and Gibertini, Marco and Marzari, Nicola and Stengel, Massimiliano},
  journal = {Phys. Rev. Lett.},
  volume = {130},
  issue = {16},
  pages = {166301},
  numpages = {6},
  year = {2023},
  month = {Apr},
  publisher = {American Physical Society},
  doi = {10.1103/PhysRevLett.130.166301},
}

@article{Ponce2023b,
  title = {Long-range electrostatic contribution to electron-phonon couplings and mobilities of two-dimensional and bulk materials},
  author = {Ponc\'e, Samuel and Royo, Miquel and Stengel, Massimiliano and Marzari, Nicola and Gibertini, Marco},
  journal = {Phys. Rev. B},
  volume = {107},
  issue = {15},
  pages = {155424},
  numpages = {30},
  year = {2023},
  month = {Apr},
  publisher = {American Physical Society},
  doi = {10.1103/PhysRevB.107.155424},
  url = {https://link.aps.org/doi/10.1103/PhysRevB.107.155424}
}

@article{Kresse1993,
  title = {Ab initio molecular dynamics for liquid metals},
  author = {Kresse, G. and Hafner, J.},
  journal = {Phys. Rev. B},
  volume = {47},
  issue = {1},
  pages = {558--561},
  numpages = {0},
  year = {1993},
  month = {Jan},
  publisher = {American Physical Society},
  doi = {10.1103/PhysRevB.47.558},
  url = {https://link.aps.org/doi/10.1103/PhysRevB.47.558}
}

@article{Kresse1994,
  title = {Ab initio molecular-dynamics simulation of the liquid-metal--amorphous-semiconductor transition in germanium},
  author = {Kresse, G. and Hafner, J.},
  journal = {Phys. Rev. B},
  volume = {49},
  issue = {20},
  pages = {14251--14269},
  numpages = {0},
  year = {1994},
  month = {May},
  publisher = {American Physical Society},
  doi = {10.1103/PhysRevB.49.14251},
  url = {https://link.aps.org/doi/10.1103/PhysRevB.49.14251}
}

@article{Kresse1995,
doi = {10.1209/0295-5075/32/9/005},
url = {https://dx.doi.org/10.1209/0295-5075/32/9/005},
year = {1995},
month = {dec},
publisher = {},
volume = {32},
number = {9},
pages = {729},
author = {G. Kresse and J. Furthmüller and J. Hafner},
title = {Ab initio Force Constant Approach to Phonon Dispersion Relations of Diamond and Graphite},
journal = {Europhysics Letters}
}

@article{Kresse1996a,
  title = {Efficient iterative schemes for ab initio total-energy calculations using a plane-wave basis set},
  author = {Kresse, G. and Furthm\"uller, J.},
  journal = {Phys. Rev. B},
  volume = {54},
  issue = {16},
  pages = {11169--11186},
  numpages = {0},
  year = {1996},
  month = {Oct},
  publisher = {American Physical Society},
  doi = {10.1103/PhysRevB.54.11169},
  url = {https://link.aps.org/doi/10.1103/PhysRevB.54.11169}
}

@article{Kresse1996b,
title = {Efficiency of ab-initio total energy calculations for metals and semiconductors using a plane-wave basis set},
journal = {Computational Materials Science},
volume = {6},
number = {1},
pages = {15-50},
year = {1996},
issn = {0927-0256},
doi = {https://doi.org/10.1016/0927-0256(96)00008-0},
url = {https://www.sciencedirect.com/science/article/pii/0927025696000080},
author = {G. Kresse and J. Furthmüller}
}

@article{Kresse1999,
  title = {From ultrasoft pseudopotentials to the projector augmented-wave method},
  author = {Kresse, G. and Joubert, D.},
  journal = {Phys. Rev. B},
  volume = {59},
  issue = {3},
  pages = {1758--1775},
  numpages = {0},
  year = {1999},
  month = {Jan},
  publisher = {American Physical Society},
  doi = {10.1103/PhysRevB.59.1758},
  url = {https://link.aps.org/doi/10.1103/PhysRevB.59.1758}
}

@article{Giannozzi2017,
doi = {10.1088/1361-648X/aa8f79},
url = {https://dx.doi.org/10.1088/1361-648X/aa8f79},
year = {2017},
month = {oct},
publisher = {IOP Publishing},
volume = {29},
number = {46},
pages = {465901},
author = {P Giannozzi and O Andreussi and T Brumme and O Bunau and M Buongiorno Nardelli and M Calandra and R Car and C Cavazzoni and D Ceresoli and M Cococcioni and N Colonna and I Carnimeo and A Dal Corso and S de Gironcoli and P Delugas and R A DiStasio and A Ferretti and A Floris and G Fratesi and G Fugallo and R Gebauer and U Gerstmann and F Giustino and T Gorni and J Jia and M Kawamura and H-Y Ko and A Kokalj and E Küçükbenli and M Lazzeri and M Marsili and N Marzari and F Mauri and N L Nguyen and H-V Nguyen and A Otero-de-la-Roza and L Paulatto and S Poncé and D Rocca and R Sabatini and B Santra and M Schlipf and A P Seitsonen and A Smogunov and I Timrov and T Thonhauser and P Umari and N Vast and X Wu and S Baroni},
title = {Advanced capabilities for materials modelling with Quantum ESPRESSO},
journal = {Journal of Physics: Condensed Matter}
}

@software{HDF5,
author = {{The HDF Group}},
title = {Hierarchical Data Format, version 5},
url = {https://github.com/HDFGroup/hdf5}
}

@article{Blochl1994,
  title = {Projector augmented-wave method},
  author = {Bl\"ochl, P. E.},
  journal = {Phys. Rev. B},
  volume = {50},
  issue = {24},
  pages = {17953--17979},
  numpages = {0},
  year = {1994},
  month = {Dec},
  publisher = {American Physical Society},
  doi = {10.1103/PhysRevB.50.17953},
  url = {https://link.aps.org/doi/10.1103/PhysRevB.50.17953}
}

@Article{Nagamatsu2001,
author={Nagamatsu, Jun and Nakagawa, Norimasa and Muranaka, Takahiro and Zenitani, Yuji and Akimitsu, Jun},
title={{Superconductivity at 39{\thinspace}K in magnesium diboride}},
journal={Nature},
year={2001},
month={Mar},
day={01},
volume={410},
number={6824},
pages={63-64},
issn={1476-4687},
doi={10.1038/35065039},
url={https://doi.org/10.1038/35065039}
}

@article{Kafle2022,
  title = {{Ab initio study of Li-Mg-B superconductors}},
  author = {Kafle, Gyanu P. and Tomassetti, Charlsey R. and Mazin, Igor I. and Kolmogorov, Aleksey N. and Margine, Elena R.},
  journal = {Phys. Rev. Mater.},
  volume = {6},
  issue = {8},
  pages = {084801},
  numpages = {13},
  year = {2022},
  month = {Aug},
  publisher = {American Physical Society},
  doi = {10.1103/PhysRevMaterials.6.084801},
  url = {https://link.aps.org/doi/10.1103/PhysRevMaterials.6.084801}
}

@misc{Wang2024,
title={Accurate Electron-phonon Interactions from Advanced Density Functional Theory},
author={Yanyong Wang and Manuel Engel and Christopher Lane and Henrique Miranda and Lin Hou and Bernardo Barbiellini and Robert S. Markiewicz and Jian-Xin Zhu and Georg Kresse and Arun Bansil and Jianwei Sun and Ruiqi Zhang},
year={2024},
eprint={2411.08192},
archivePrefix={arXiv},
url={https://arxiv.org/abs/2411.08192},
}

@article{Ponce2021,
  title = {{First-principles predictions of Hall and drift mobilities in semiconductors}},
  author = {Ponc\'e, Samuel and Macheda, Francesco and Margine, Elena Roxana and Marzari, Nicola and Bonini, Nicola and Giustino, Feliciano},
  journal = {Phys. Rev. Research},
  volume = {3},
  issue = {4},
  pages = {043022},
  numpages = {36},
  year = {2021},
  month = {Oct},
  publisher = {American Physical Society},
  doi = {10.1103/PhysRevResearch.3.043022},
  url = {https://link.aps.org/doi/10.1103/PhysRevResearch.3.043022}
}

@article{Marzari2012,
  title = {{Maximally localized Wannier functions: Theory and applications}},
  author = {Marzari, Nicola and Mostofi, Arash A. and Yates, Jonathan R. and Souza, Ivo and Vanderbilt, David},
  journal = {Rev. Mod. Phys.},
  volume = {84},
  issue = {4},
  pages = {1419--1475},
  numpages = {0},
  year = {2012},
  month = {Oct},
  publisher = {American Physical Society},
  doi = {10.1103/RevModPhys.84.1419},
  url = {https://link.aps.org/doi/10.1103/RevModPhys.84.1419}
}

@article{Baroni1987,
  title = {Green's-function approach to linear response in solids},
  author = {Baroni, Stefano and Giannozzi, Paolo and Testa, Andrea},
  journal = {Phys. Rev. Lett.},
  volume = {58},
  issue = {18},
  pages = {1861--1864},
  numpages = {0},
  year = {1987},
  month = {May},
  publisher = {American Physical Society},
  doi = {10.1103/PhysRevLett.58.1861},
  url = {https://link.aps.org/doi/10.1103/PhysRevLett.58.1861}
}

@article{Baroni2001,
  title={{Phonons and related crystal properties from density-functional perturbation theory}},
  author={Baroni, S. and De Gironcoli, S. and Dal Corso, A. and Giannozzi, P.},
  journal={Reviews of Modern Physics},
  volume={73},
  pages={515},
  year={2001},
  url={https://journals.aps.org/rmp/abstract/10.1103/RevModPhys.73.515}
}

@article{Savrasov1992,
  title = {Linear response calculations of lattice dynamics using muffin-tin basis sets},
  author = {Savrasov, S. Yu.},
  journal = {Phys. Rev. Lett.},
  volume = {69},
  issue = {19},
  pages = {2819--2822},
  numpages = {0},
  year = {1992},
  month = {Nov},
  publisher = {American Physical Society},
  doi = {10.1103/PhysRevLett.69.2819},
  url = {https://link.aps.org/doi/10.1103/PhysRevLett.69.2819}
}

@article{Gonze1997,
  title = {Dynamical matrices, Born effective charges, dielectric permittivity tensors, and interatomic force constants from density-functional perturbation theory},
  author = {Gonze, Xavier and Lee, Changyol},
  journal = {Phys. Rev. B},
  volume = {55},
  issue = {16},
  pages = {10355--10368},
  numpages = {0},
  year = {1997},
  month = {Apr},
  publisher = {American Physical Society},
  doi = {10.1103/PhysRevB.55.10355},
  url = {https://link.aps.org/doi/10.1103/PhysRevB.55.10355}
}

@Book{Ziman1960,
  Title                    = {Electrons and Phonons},
  Author                   = {Ziman, J. M.},
  Editor                   = {Mott, N. F. and Bullard, E. C. and Wilkinson, D. H.},
  Publisher                = {Oxford University Press},
  Year                     = {1960}
}

@article{Noffsinger2012,
  title = {{Phonon-Assisted Optical Absorption in Silicon from First Principles}},
  author = {Noffsinger, Jesse and Kioupakis, Emmanouil and Van de Walle, Chris G. and Louie, Steven G. and Cohen, Marvin L.},
  journal = {Phys. Rev. Lett.},
  volume = {108},
  issue = {16},
  pages = {167402},
  numpages = {5},
  year = {2012},
  month = {Apr},
  publisher = {American Physical Society},
  doi = {10.1103/PhysRevLett.108.167402},
  url = {https://link.aps.org/doi/10.1103/PhysRevLett.108.167402}
}

@article{Tiwari2024,
  title = {Unified theory of optical absorption and luminescence including both direct and phonon-assisted processes},
  author = {Tiwari, Sabyasachi and Kioupakis, Emmanouil and Menendez, Jos\'e and Giustino, Feliciano},
  journal = {Phys. Rev. B},
  volume = {109},
  issue = {19},
  pages = {195127},
  numpages = {18},
  year = {2024},
  month = {May},
  publisher = {American Physical Society},
  doi = {10.1103/PhysRevB.109.195127},
  url = {https://link.aps.org/doi/10.1103/PhysRevB.109.195127}
}

@article{Sio2019a,
  title = {{Polarons from First Principles, without Supercells}},
  author = {Sio, Weng Hong and Verdi, Carla and Ponc\'e, Samuel and Giustino, Feliciano},
  journal = {Phys. Rev. Lett.},
  volume = {122},
  issue = {24},
  pages = {246403},
  numpages = {6},
  year = {2019},
  month = {Jun},
  publisher = {American Physical Society},
  doi = {10.1103/PhysRevLett.122.246403},
  url = {https://link.aps.org/doi/10.1103/PhysRevLett.122.246403}
}

@article{Sio2019b,
  title = {{Ab initio theory of polarons: Formalism and applications}},
  author = {Sio, Weng Hong and Verdi, Carla and Ponc\'e, Samuel and Giustino, Feliciano},
  journal = {Phys. Rev. B},
  volume = {99},
  issue = {23},
  pages = {235139},
  numpages = {21},
  year = {2019},
  month = {Jun},
  publisher = {American Physical Society},
  doi = {10.1103/PhysRevB.99.235139},
  url = {https://link.aps.org/doi/10.1103/PhysRevB.99.235139}
}

@article{Lafuente2022a,
  title = {{Unified Approach to Polarons and Phonon-Induced Band Structure Renormalization}},
  author = {Lafuente-Bartolome, Jon and Lian, Chao and Sio, Weng Hong and Gurtubay, Idoia G. and Eiguren, Asier and Giustino, Feliciano},
  journal = {Phys. Rev. Lett.},
  volume = {129},
  issue = {7},
  pages = {076402},
  numpages = {7},
  year = {2022},
  month = {Aug},
  publisher = {American Physical Society},
  doi = {10.1103/PhysRevLett.129.076402},
  url = {https://link.aps.org/doi/10.1103/PhysRevLett.129.076402}
}

@article{Lafuente2022b,
  title = {Ab initio self-consistent many-body theory of polarons at all couplings},
  author = {Lafuente-Bartolome, Jon and Lian, Chao and Sio, Weng Hong and Gurtubay, Idoia G. and Eiguren, Asier and Giustino, Feliciano},
  journal = {Phys. Rev. B},
  volume = {106},
  issue = {7},
  pages = {075119},
  numpages = {20},
  year = {2022},
  month = {Aug},
  publisher = {American Physical Society},
  doi = {10.1103/PhysRevB.106.075119},
  url = {https://link.aps.org/doi/10.1103/PhysRevB.106.075119}
}

@article{Engel2020,
  title = {Electron-phonon interactions using the projector augmented-wave method and Wannier functions},
  author = {Engel, Manuel and Marsman, Martijn and Franchini, Cesare and Kresse, Georg},
  journal = {Phys. Rev. B},
  volume = {101},
  issue = {18},
  pages = {184302},
  numpages = {15},
  year = {2020},
  month = {May},
  publisher = {American Physical Society},
  doi = {10.1103/PhysRevB.101.184302},
  url = {https://link.aps.org/doi/10.1103/PhysRevB.101.184302}
}

@article{Engel2022,
  title = {Zero-point renormalization of the band gap of semiconductors and insulators using the projector augmented wave method},
  author = {Engel, Manuel and Miranda, Henrique and Chaput, Laurent and Togo, Atsushi and Verdi, Carla and Marsman, Martijn and Kresse, Georg},
  journal = {Phys. Rev. B},
  volume = {106},
  issue = {9},
  pages = {094316},
  numpages = {19},
  year = {2022},
  month = {Sep},
  publisher = {American Physical Society},
  doi = {10.1103/PhysRevB.106.094316},
  url = {https://link.aps.org/doi/10.1103/PhysRevB.106.094316}
}

@article{Chaput2019,
  title = {Finite-displacement computation of the electron-phonon interaction within the projector augmented-wave method},
  author = {Chaput, L. and Togo, Atsushi and Tanaka, Isao},
  journal = {Phys. Rev. B},
  volume = {100},
  issue = {17},
  pages = {174304},
  numpages = {19},
  year = {2019},
  month = {Nov},
  publisher = {American Physical Society},
  doi = {10.1103/PhysRevB.100.174304},
  url = {https://link.aps.org/doi/10.1103/PhysRevB.100.174304}
}

@article{Heyd2003,
    author = {Heyd, Jochen and Scuseria, Gustavo E. and Ernzerhof, Matthias},
    title = {{Hybrid functionals based on a screened Coulomb potential}},
    journal = {The Journal of Chemical Physics},
    volume = {118},
    number = {18},
    pages = {8207-8215},
    year = {2003},
    month = {05},
    issn = {0021-9606},
    doi = {10.1063/1.1564060},
    url = {https://doi.org/10.1063/1.1564060}
}

@article{Sun2015,
  title = {Strongly Constrained and Appropriately Normed Semilocal Density Functional},
  author = {Sun, Jianwei and Ruzsinszky, Adrienn and Perdew, John P.},
  journal = {Phys. Rev. Lett.},
  volume = {115},
  issue = {3},
  pages = {036402},
  numpages = {6},
  year = {2015},
  month = {Jul},
  publisher = {American Physical Society},
  doi = {10.1103/PhysRevLett.115.036402},
  url = {https://link.aps.org/doi/10.1103/PhysRevLett.115.036402}
}

@Article{Furness2020,
author={Furness, James W. and Kaplan, Aaron D. and Ning, Jinliang and Perdew, John P. and Sun, Jianwei},
title={{Accurate and Numerically Efficient r2SCAN Meta-Generalized Gradient Approximation}},
journal={The Journal of Physical Chemistry Letters},
year={2020},
month={Oct},
day={01},
publisher={American Chemical Society},
volume={11},
number={19},
pages={8208-8215},
doi={10.1021/acs.jpclett.0c02405},
url={https://doi.org/10.1021/acs.jpclett.0c02405}
}

@article{Curtarolo2013,
title = {{AFLOW: An automatic framework for high-throughput materials discovery}},
journal = {Computational Materials Science},
volume = {58},
pages = {218-226},
year = {2012},
issn = {0927-0256},
doi = {https://doi.org/10.1016/j.commatsci.2012.02.005},
url = {https://www.sciencedirect.com/science/article/pii/S0927025612000717},
author = {Stefano Curtarolo and Wahyu Setyawan and Gus L.W. Hart and Michal Jahnatek and Roman V. Chepulskii and Richard H. Taylor and Shidong Wang and Junkai Xue and Kesong Yang and Ohad Levy and Michael J. Mehl and Harold T. Stokes and Denis O. Demchenko and Dane Morgan}
}

@Article{Saal2013,
author={Saal, James E. and Kirklin, Scott and Aykol, Muratahan and Meredig, Bryce and Wolverton, C.},
title={{Materials Design and Discovery with High-Throughput Density Functional Theory: The Open Quantum Materials Database (OQMD)}},
journal={JOM},
year={2013},
month={Nov},
day={01},
volume={65},
number={11},
pages={1501-1509},
issn={1543-1851},
doi={10.1007/s11837-013-0755-4},
url={https://doi.org/10.1007/s11837-013-0755-4}
}

@Article{Kirklin2015,
author={Kirklin, Scott and Saal, James E. and Meredig, Bryce and Thompson, Alex and Doak, Jeff W. and Aykol, Muratahan and R{\"u}hl, Stephan and Wolverton, Chris},
title={{The Open Quantum Materials Database (OQMD): assessing the accuracy of DFT formation energies}},
journal={npj Computational Materials},
year={2015},
month={Dec},
day={11},
volume={1},
number={1},
pages={15010},
issn={2057-3960},
doi={10.1038/npjcompumats.2015.10},
url={https://doi.org/10.1038/npjcompumats.2015.10}
}

@article{Jain2013,
author = {Jain, Anubhav and Ong, Shyue Ping and Hautier, Geoffroy and Chen, Wei and Richards, William Davidson and Dacek, Stephen and Cholia, Shreyas and Gunter, Dan and Skinner, David and Ceder, Gerbrand and Persson, Kristin A.},
doi = {10.1063/1.4812323},
issn = {2166532X},
journal = {APL Materials},
number = {1},
pages = {011002},
title = {{The Materials Project: A materials genome approach to accelerating materials innovation}},
url = {http://link.aip.org/link/AMPADS/v1/i1/p011002/s1\&Agg=doi},
volume = {1},
year = {2013}
}

@article{Choudhary2020,
  author  = {Choudhary, Kamal and Garrity, Kevin F. and Reid, Andrew C. E. and DeCost, Brian and Biacchi, Adam J. and Hight Walker, Angela R. and Trautt, Zachary and Hattrick-Simpers, Jason and Kusne, A. Gilad and Centrone, Andrea and Davydov, Albert and Jiang, Jie and Pachter, Ruth and Cheon, Gowoon and Reed, Evan and Agrawal, Ankit and Qian, Xiaofeng and Sharma, Vinit and Zhuang, Houlong and Kalinin, Sergei V. and Sumpter, Bobby G. and Pilania, Ghanshyam and Acar, Pinar and Mandal, Subhasish and Haule, Kristjan and Vanderbilt, David and Rabe, Karin and Tavazza, Francesca},
  title   = {The joint automated repository for various integrated simulations (JARVIS) for data-driven materials design},
  journal = {npj Computational Materials},
  year    = {2020},
  volume  = {6},
  number  = {1},
  pages   = {173},
  month   = {nov},
  issn    = {2057-3960},
  doi     = {10.1038/s41524-020-00440-1},
  url     = {https://doi.org/10.1038/s41524-020-00440-1}
}

@article{Schmidt2022,
  title   = {{A dataset of 175k stable and metastable materials calculated with the PBEsol and SCAN functionals}},
  author  = {Schmidt, Jonathan and Wang, Hai-Chen and Cerqueira, Tiago F. T. and Botti, Silvana and Marques, Miguel A. L.},
  journal = {Scientific Data},
  volume  = {9},
  pages   = {64},
  year    = {2022},
  doi     = {10.1038/s41597-022-01177-w},
  url     = {https://doi.org/10.1038/s41597-022-01177-w
}
}

@article{Frank1995,
  title = {Ab initio Force-Constant Method for Phonon Dispersions in Alkali Metals},
  author = {Frank, W. and Els\"asser, C. and F\"ahnle, M.},
  journal = {Phys. Rev. Lett.},
  volume = {74},
  issue = {10},
  pages = {1791--1794},
  numpages = {0},
  year = {1995},
  month = {Mar},
  publisher = {American Physical Society},
  doi = {10.1103/PhysRevLett.74.1791},
  url = {https://link.aps.org/doi/10.1103/PhysRevLett.74.1791}
}

@article{Parlinski1997,
  title = {First-Principles Determination of the Soft Mode in Cubic ${\mathrm{ZrO}}_{2}$},
  author = {Parlinski, K. and Li, Z. Q. and Kawazoe, Y.},
  journal = {Phys. Rev. Lett.},
  volume = {78},
  issue = {21},
  pages = {4063--4066},
  numpages = {0},
  year = {1997},
  month = {May},
  publisher = {American Physical Society},
  doi = {10.1103/PhysRevLett.78.4063},
  url = {https://link.aps.org/doi/10.1103/PhysRevLett.78.4063}
}

@article{Togo2015,
title = {First principles phonon calculations in materials science},
journal = {Scripta Materialia},
volume = {108},
pages = {1-5},
year = {2015},
issn = {1359-6462},
doi = {https://doi.org/10.1016/j.scriptamat.2015.07.021},
url = {https://www.sciencedirect.com/science/article/pii/S1359646215003127},
author = {Atsushi Togo and Isao Tanaka}
}

@article{Ewald1921,
author = {Ewald, P. P.},
title = {{Die Berechnung optischer und elektrostatischer Gitterpotentiale}},
journal = {Annalen der Physik},
volume = {369},
number = {3},
pages = {253-287},
doi = {https://doi.org/10.1002/andp.19213690304},
url = {https://onlinelibrary.wiley.com/doi/abs/10.1002/andp.19213690304},
year = {1921}
}

@BOOK{Born1998,
  title     = {Dynamical theory of crystal lattices},
  author    = {Born, Max and Huang, Kun},
  publisher = "Clarendon Press",
  series    = "Oxford Classic Texts in the Physical Sciences",
  month     =  jun,
  year      =  1998,
  address   = "Oxford, England",
}

@article{Hamann2013,
  title = {{Optimized norm-conserving Vanderbilt pseudopotentials}},
  author = {Hamann, D. R.},
  journal = {Phys. Rev. B},
  volume = {88},
  issue = {8},
  pages = {085117},
  numpages = {10},
  year = {2013},
  month = {Aug},
  publisher = {American Physical Society},
  doi = {10.1103/PhysRevB.88.085117},
  url = {https://link.aps.org/doi/10.1103/PhysRevB.88.085117}
}

@article{Setten2018,
  title={The {PseudoDojo}: {Training} and grading a 85 element optimized norm-conserving pseudopotential table},
  author={van Setten, M. J. and Giantomassi, M. and Bousquet, E. and Verstraete, M. J. and Hamann, D. R. and Gonze, X. and Rignanese, G.-M.},
  journal={Computer Physics Communications},
  volume={226},
  pages={39},
  year={2018},
  url={https://doi.org/10.1016/j.cpc.2018.01.012}
}

@BOOK{Madelung1991,
title     = {Semiconductors, Group IV Elements and III-V Compounds},
author    = {O. Madelung},
publisher = {Springer Berlin, Heidelberg},
year      = {1991},
}

@article{Sio2022,
  title = {{Unified ab initio description of Fr\"ohlich electron-phonon interactions in two-dimensional and three-dimensional materials}},
  author = {Sio, Weng Hong and Giustino, Feliciano},
  journal = {Phys. Rev. B},
  volume = {105},
  issue = {11},
  pages = {115414},
  numpages = {15},
  year = {2022},
  month = {Mar},
  publisher = {American Physical Society},
  doi = {10.1103/PhysRevB.105.115414},
  url = {https://link.aps.org/doi/10.1103/PhysRevB.105.115414}
}

@article{Sohier2016,
  title = {{Two}-dimensional {F}r\"{o}hlich interaction in transition-metal dichalcogenide monolayers: {T}heoretical modeling and first-principles calculations},
  author = {Sohier, Thibault and Calandra, Matteo and Mauri, Francesco},
  journal = {Phys. Rev. B},
  volume = {94},
  issue = {8},
  pages = {085415},
  numpages = {13},
  year = {2016},
  month = {Aug},
  publisher = {American Physical Society},
  doi = {10.1103/PhysRevB.94.085415},
}

@article{Sohier2018,
  title = {{Mobility of two-dimensional materials from first principles in an accurate and automated framework}},
  author = {Sohier, Thibault and Campi, Davide and Marzari, Nicola and Gibertini, Marco},
  journal = {Phys. Rev. Mater.},
  volume = {2},
  issue = {11},
  pages = {114010},
  numpages = {21},
  year = {2018},
  month = {Nov},
  publisher = {American Physical Society},
  doi = {10.1103/PhysRevMaterials.2.114010},
  url = {https://link.aps.org/doi/10.1103/PhysRevMaterials.2.114010}
}

@article{Deng2021,
  title = {{Ab} initio dipolar electron-phonon interactions in two-dimensional materials},
  author = {Deng, Tianqi and Wu, Gang and Shi, Wen and Wong, Zicong Marvin and Wang, Jian-Sheng and Yang, Shuo-Wang},
  journal = {Phys. Rev. B},
  volume = {103},
  issue = {7},
  pages = {075410},
  numpages = {9},
  year = {2021},
  month = {Feb},
  publisher = {American Physical Society},
  doi = {10.1103/PhysRevB.103.075410},
}

@article{Wang2006,
  title = {{Ab initio calculation of the anomalous Hall conductivity by Wannier interpolation}},
  author = {Wang, Xinjie and Yates, Jonathan R. and Souza, Ivo and Vanderbilt, David},
  journal = {Phys. Rev. B},
  volume = {74},
  issue = {19},
  pages = {195118},
  numpages = {15},
  year = {2006},
  month = {Nov},
  publisher = {American Physical Society},
  doi = {10.1103/PhysRevB.74.195118},
  url = {https://link.aps.org/doi/10.1103/PhysRevB.74.195118}
}

@article{Yates2007,
  title = {{Spectral and Fermi surface properties from Wannier interpolation}},
  author = {Yates, Jonathan R. and Wang, Xinjie and Vanderbilt, David and Souza, Ivo},
  journal = {Phys. Rev. B},
  volume = {75},
  issue = {19},
  pages = {195121},
  numpages = {11},
  year = {2007},
  month = {May},
  publisher = {American Physical Society},
  doi = {10.1103/PhysRevB.75.195121},
  url = {https://link.aps.org/doi/10.1103/PhysRevB.75.195121}
}

@incollection{Blount1962,
title = {{Formalisms of Band Theory}},
editor = {Frederick Seitz and David Turnbull},
series = {Solid State Physics},
publisher = {Academic Press},
volume = {13},
pages = {305-373},
year = {1962},
issn = {0081-1947},
doi = {https://doi.org/10.1016/S0081-1947(08)60459-2},
url = {https://www.sciencedirect.com/science/article/pii/S0081194708604592},
author = {E.I. Blount}
}

@article{Zhong2024,
  author       = {Yang Zhong and Shixu Liu and Binhua Zhang and Zhiguo Tao and Yuting Sun and Weibin Chu and Xin-Gao Gong and Ji-Hui Yang and Hongjun Xiang},
  title        = {{Accelerating the calculation of electron–phonon coupling strength with machine learning}},
  journal      = {Nature Computational Science},
  year         = {2024},
  volume       = {4},
  number       = {8},
  pages        = {615--625},
  doi          = {10.1038/s43588-024-00668-7},
  url          = {https://doi.org/10.1038/s43588-024-00668-7}
}

@article{Haldar2024,
  title = {{Machine learning electron-phonon interactions in two-dimensional semiconducting materials: The case of zero-point renormalization}},
  author = {Haldar, Anubhab and Clark, Quentin and Zacharias, Marios and Giustino, Feliciano and Sharifzadeh, Sahar},
  journal = {Phys. Rev. Mater.},
  volume = {8},
  issue = {10},
  pages = {L101001},
  numpages = {8},
  year = {2024},
  month = {Oct},
  publisher = {American Physical Society},
  doi = {10.1103/PhysRevMaterials.8.L101001},
  url = {https://link.aps.org/doi/10.1103/PhysRevMaterials.8.L101001}
}

@article{Ponce2018,
    title = {{Towards predictive many-body calculations of phonon-limited carrier mobilities in semiconductors}},
    author = {Ponc\'e, Samuel and Margine, Elena R. and Giustino, Feliciano},
    journal = {Phys. Rev. B},
    volume = {97},
    issue = {12},
    pages = {121201},
    numpages = {5},
    year = {2018},
    month = {Mar},
    publisher = {American Physical Society},
    doi = {10.1103/PhysRevB.97.121201},
}

@article{Ponce2020,
doi = {10.1088/1361-6633/ab6a43},
url = {https://dx.doi.org/10.1088/1361-6633/ab6a43},
year = {2020},
month = {feb},
publisher = {IOP Publishing},
volume = {83},
number = {3},
pages = {036501},
author = {Ponc{\'e}, Samuel and Li, Wenbin and Reichardt, Sven and Giustino, Feliciano},
title = {{First-principles calculations of charge carrier mobility and conductivity in bulk semiconductors and two-dimensional materials}},
journal = {Reports on Progress in Physics}
}

@misc{Poliukhin2025,
      title={{Carrier mobilities and electron-phonon interactions beyond DFT}},
      author={Aleksandr Poliukhin and Nicola Colonna and Francesco Libbi and Samuel Poncé and Nicola Marzari},
      year={2025},
      eprint={2508.14852},
      archivePrefix={arXiv},
      primaryClass={cond-mat.mtrl-sci},
      url={https://arxiv.org/abs/2508.14852},
}

@article{Stefanucci2023,
  title = {{In and Out-of-Equilibrium Ab Initio Theory of Electrons and Phonons}},
  author = {Stefanucci, Gianluca and van Leeuwen, Robert and Perfetto, Enrico},
  journal = {Phys. Rev. X},
  volume = {13},
  issue = {3},
  pages = {031026},
  numpages = {40},
  year = {2023},
  month = {Sep},
  publisher = {American Physical Society},
  doi = {10.1103/PhysRevX.13.031026},
  url = {https://link.aps.org/doi/10.1103/PhysRevX.13.031026}
}

@article{Samsonidze2018,
  title        = {{Accelerated Screening of Thermoelectric Materials by First-Principles Computations of Electron-Phonon Scattering}},
  author       = {Georgy Samsonidze and Boris Kozinsky},
  journal      = {Advanced Energy Materials},
  year         = {2018},
  volume       = {8},
  number       = {20},
  pages        = {1800246},
  doi          = {10.1002/aenm.201800246},
  url          = {https://onlinelibrary.wiley.com/doi/10.1002/aenm.201800246}
}

@article{Ha2024,
  title        = {{High-throughput screening of 2D materials identifies p-type monolayer WS$_2$ as potential ultra-high mobility semiconductor}},
  author       = {Viet-Anh Ha and Feliciano Giustino},
  journal      = {npj Computational Materials},
  year         = {2024},
  volume       = {10},
  number       = {229},
  doi          = {10.1038/s41524-024-01417-0},
  url          = {https://www.nature.com/articles/s41524-024-01417-0}
}

@article{Ha2025,
  title        = {{Ultrahigh Hole Mobility in Monolayer {WSe}$_2$ Enabled by Spin–Orbit Suppression of Intervalley Scattering}},
  author       = {Viet-Anh Ha and Sabyasachi Tiwari and Feliciano Giustino},
  journal      = {Nano Letters},
  year         = {2025},
  volume       = {25},
  number       = {39},
  pages        = {14304--14309},
  doi          = {10.1021/acs.nanolett.5c03258}
}

@article{Madika2025,
  title        = {{Artificial Intelligence for Materials Discovery, Development and Deployment}},
  author       = {B. Madika and A. Saha and C. Kang and B. Buyantogtokh and J. Agar and C. M. Wolverton and P. Voorhees and P. Littlewood and S. Kalinin and S. Hong},
  journal      = {ACS Nano},
  year         = {2025},
  volume       = {19},
  number       = {30},
  pages        = {27116--27158},
  doi          = {10.1021/acsnano.5c04200},
  url          = {https://pubs.acs.org/doi/full/10.1021/acsnano.5c04200}
}

@article{Li2024,
  title = {{Deep-Learning Density Functional Perturbation Theory}},
  author = {Li, He and Tang, Zechen and Fu, Jingheng and Dong, Wen-Han and Zou, Nianlong and Gong, Xiaoxun and Duan, Wenhui and Xu, Yong},
  journal = {Phys. Rev. Lett.},
  volume = {132},
  issue = {9},
  pages = {096401},
  numpages = {6},
  year = {2024},
  month = {Feb},
  publisher = {American Physical Society},
  doi = {10.1103/PhysRevLett.132.096401},
  url = {https://link.aps.org/doi/10.1103/PhysRevLett.132.096401}
}

@article{Luo2024,
  title = {{Data-Driven Compression of Electron-Phonon Interactions}},
  author = {Luo, Yao and Desai, Dhruv and Chang, Benjamin K. and Park, Jinsoo and Bernardi, Marco},
  journal = {Phys. Rev. X},
  volume = {14},
  issue = {2},
  pages = {021023},
  numpages = {12},
  year = {2024},
  month = {May},
  publisher = {American Physical Society},
  doi = {10.1103/PhysRevX.14.021023},
  url = {https://link.aps.org/doi/10.1103/PhysRevX.14.021023}
}

@article{Cerqueira2024,
  author       = {Tiago F. T. Cerqueira and Antonio Sanna and Miguel A. L. Marques},
  title        = {{Sampling the Materials Space for Conventional Superconductors via Machine Learning}},
  journal      = {Advanced Materials},
  year         = {2024},
  volume       = {36},
  number       = {7},
  pages        = {2307085},
  doi          = {10.1002/adma.202307085},
  url          = {https://doi.org/10.1002/adma.202307085},
  issn         = {0935-9648}
}

@article{Gibson2025,
  author       = {Jason B. Gibson and Ajinkya C. Hire and Philip M. Dee and Oscar Barrera and Benjamin Geisler and Peter J. Hirschfeld and Richard G. Hennig},
  title        = {{Accelerating superconductor discovery through tempered deep learning of the electron-phonon spectral function}},
  journal      = {npj Computational Materials},
  year         = {2025},
  volume       = {11},
  number       = {1},
  pages        = {7},
  doi          = {10.1038/s41524-024-01475-4},
  url          = {https://doi.org/10.1038/s41524-024-01475-4}
}

@article{Kolmogorov2010,
  title = {{New Superconducting and Semiconducting Fe-B Compounds Predicted with an Ab Initio Evolutionary Search}},
  author = {Kolmogorov, A. N. and Shah, S. and Margine, E. R. and Bialon, A. F. and Hammerschmidt, T. and Drautz, R.},
  journal = {Phys. Rev. Lett.},
  volume = {105},
  issue = {21},
  pages = {217003},
  numpages = {4},
  year = {2010},
  month = {Nov},
  publisher = {American Physical Society},
  doi = {10.1103/PhysRevLett.105.217003},
  url = {https://link.aps.org/doi/10.1103/PhysRevLett.105.217003}
}

@article{Duan2014,
  title={{Pressure-induced metallization of dense (H$_2$S)$_2$H$_2$ with high-T$_{\rm c}$ superconductivity}},
  author={Duan, Defang and Liu, Yunxian and Tian, Fubo and Li, Da and Huang, Xiaoli and Zhao, Zhonglong and Yu, Hongyu and Liu, Bingbing and Tian, Wenjing and Cui, Tian},
  journal={Sci. Rep.},
  volume={4},
  number={1},
  pages={6968},
  year={2014},
  url = {https://doi.org/10.1038/srep06968},
  publisher={Nature Publishing Group UK London}
}

@article{Shukla2003,
  title = {{Phonon Dispersion and Lifetimes in ${\mathrm{M}\mathrm{g}\mathrm{B}}_{2}$}},
  author = {Shukla, Abhay and Calandra, Matteo and d'Astuto, Matteo and Lazzeri, Michele and Mauri, Francesco and Bellin, Christophe and Krisch, Michael and Karpinski, J. and Kazakov, S. M. and Jun, J. and Daghero, D. and Parlinski, K.},
  journal = {Phys. Rev. Lett.},
  volume = {90},
  issue = {9},
  pages = {095506},
  numpages = {4},
  year = {2003},
  month = {Mar},
  publisher = {American Physical Society},
  doi = {10.1103/PhysRevLett.90.095506},
  url = {https://link.aps.org/doi/10.1103/PhysRevLett.90.095506}
}

@article{Royo2020,
  title = {{Using High Multipolar Orders to Reconstruct the Sound Velocity in Piezoelectrics from Lattice Dynamics}},
  author = {Royo, Miquel and Hahn, Konstanze R. and Stengel, Massimiliano},
  journal = {Phys. Rev. Lett.},
  volume = {125},
  issue = {21},
  pages = {217602},
  numpages = {7},
  year = {2020},
  month = {Nov},
  publisher = {American Physical Society},
  doi = {10.1103/PhysRevLett.125.217602},
  url = {https://link.aps.org/doi/10.1103/PhysRevLett.125.217602}
}

@article{Linscott2023,
  title   = {{Koopmans: An Open-Source Package for Accurately and Efficiently Predicting Spectral Properties with Koopmans Functionals}},
  author  = {Linscott, Edward B. and Colonna, Nicola and De Gennaro, Riccardo and Nguyen, Ngoc Linh and Borghi, Giovanni and Ferretti, Andrea and Dabo, Ismaila and Marzari, Nicola},
  journal = {Journal of Chemical Theory and Computation},
  year    = {2023},
  volume  = {19},
  number  = {20},
  pages   = {7097--7111},
  doi     = {10.1021/acs.jctc.3c00652},
  url     = {https://doi.org/10.1021/acs.jctc.3c00652}
}

@article{Marini2009,
title = {{yambo: An ab initio tool for excited state calculations}},
journal = {Computer Physics Communications},
volume = {180},
number = {8},
pages = {1392-1403},
year = {2009},
issn = {0010-4655},
doi = {https://doi.org/10.1016/j.cpc.2009.02.003},
url = {https://www.sciencedirect.com/science/article/pii/S0010465509000472},
author = {Andrea Marini and Conor Hogan and Myrta Grüning and Daniele Varsano}
}

@article{Dai2024a,
  title = {{Excitonic Polarons and Self-Trapped Excitons from First-Principles Exciton-Phonon Couplings}},
  author = {Dai, Zhenbang and Lian, Chao and Lafuente-Bartolome, Jon and Giustino, Feliciano},
  journal = {Phys. Rev. Lett.},
  volume = {132},
  issue = {3},
  pages = {036902},
  numpages = {7},
  year = {2024},
  month = {Jan},
  publisher = {American Physical Society},
  doi = {10.1103/PhysRevLett.132.036902},
  url = {https://link.aps.org/doi/10.1103/PhysRevLett.132.036902}
}

@article{Dai2024b,
  title = {{Theory of excitonic polarons: From models to first-principles calculations}},
  author = {Dai, Zhenbang and Lian, Chao and Lafuente-Bartolome, Jon and Giustino, Feliciano},
  journal = {Phys. Rev. B},
  volume = {109},
  issue = {4},
  pages = {045202},
  numpages = {19},
  year = {2024},
  month = {Jan},
  publisher = {American Physical Society},
  doi = {10.1103/PhysRevB.109.045202},
  url = {https://link.aps.org/doi/10.1103/PhysRevB.109.045202}
}

@article{Cuco2024,
  title = {{Intrinsic Limits of Charge Carrier Mobilities in Layered Halide Perovskites}},
  author = {Cucco, Bruno and Leveillee, Joshua and Ha, Viet-Anh and Even, Jacky and Kepenekian, Mika\"el and Giustino, Feliciano and Volonakis, George},
  journal = {PRX Energy},
  volume = {3},
  issue = {2},
  pages = {023012},
  numpages = {11},
  year = {2024},
  month = {Jun},
  publisher = {American Physical Society},
  doi = {10.1103/PRXEnergy.3.023012},
  url = {https://link.aps.org/doi/10.1103/PRXEnergy.3.023012}
}

@article{Lihm2025,
  title = {Nonperturbative Self-Consistent Electron-Phonon Spectral Functions and Transport},
  author = {Lihm, Jae-Mo and Ponc\'e, Samuel},
  journal = {Phys. Rev. Lett.},
  volume = {134},
  issue = {18},
  pages = {186401},
  numpages = {11},
  year = {2025},
  month = {May},
  publisher = {American Physical Society},
  doi = {10.1103/PhysRevLett.134.186401},
  url = {https://link.aps.org/doi/10.1103/PhysRevLett.134.186401}
}

@article{Yang2025,
  title = {{First-principles electron-phonon interactions with self-consistent Hubbard interaction: Application to transparent conducting oxides}},
  author = {Yang, Wooil and Tiwari, Sabyasachi and Giustino, Feliciano and Son, Young-Woo},
  journal = {Phys. Rev. B},
  volume = {112},
  issue = {7},
  pages = {075203},
  numpages = {15},
  year = {2025},
  month = {Aug},
  publisher = {American Physical Society},
  doi = {10.1103/w2y5-rl8s},
  url = {https://link.aps.org/doi/10.1103/w2y5-rl8s}
}

@article{Lucrezi2024,
author={Lucrezi, Roman and Ferreira, Pedro P. and Hajinazar, Samad and Mori, Hitoshi and Paudyal, Hari and Margine, Elena R. and Heil, Christoph},
title={{Full-bandwidth anisotropic Migdal-Eliashberg theory and its application to superhydrides}},
journal={Communications Physics},
year={2024},
month={Jan},
day={15},
volume={7},
number={1},
pages={33},
issn={2399-3650},
doi={10.1038/s42005-024-01528-6},
url={https://doi.org/10.1038/s42005-024-01528-6}
}

@article{Tomassetti2024,
author ="Tomassetti, Charlsey R. and Kafle, Gyanu P. and Marcial, Edan T. and Margine, Elena R. and Kolmogorov, Aleksey N.",
title  ="Prospect of high-temperature superconductivity in layered metal borocarbides",
journal  ="J. Mater. Chem. C",
year  ="2024",
volume  ="12",
issue  ="13",
pages  ="4870-4884",
publisher  ="The Royal Society of Chemistry",
doi  ="10.1039/D4TC00210E",
url  ="http://dx.doi.org/10.1039/D4TC00210E",
}

@article{Mori2024,
  title = {{Efficient anisotropic Migdal-Eliashberg calculations with an intermediate representation basis and Wannier interpolation}},
  author = {Mori, Hitoshi and Nomoto, Takuya and Arita, Ryotaro and Margine, Elena R.},
  journal = {Phys. Rev. B},
  volume = {110},
  issue = {6},
  pages = {064505},
  numpages = {13},
  year = {2024},
  month = {Aug},
  publisher = {American Physical Society},
  doi = {10.1103/PhysRevB.110.064505},
  url = {https://link.aps.org/doi/10.1103/PhysRevB.110.064505}
}

@article{Gochitashvili2025,
  author    = {Gochitashvili, Daviti and Tomassetti, Charlsey R. and Margine, Elena R. and Kolmogorov, Aleksey N.},
  title     = {{High-{Tc} {Ag\textsubscript{x}BC} and {Cu\textsubscript{x}BC} superconductors accessible via topochemical reactions}},
  journal   = {Journal of Materials Chemistry C},
  year      = {2025},
  volume    = {13},
  number    = {36},
  pages     = {18924--18935},
  publisher = {The Royal Society of Chemistry},
  doi       = {10.1039/D5TC02237A},
  url       = {http://dx.doi.org/10.1039/D5TC02237A},
}

@article{Mishra2024,
  title = {Stability-superconductivity map for compressed Na-intercalated graphite},
  author = {Mishra, Shashi B. and Marcial, Edan T. and Debata, Suryakanti and Kolmogorov, Aleksey N. and Margine, Elena R.},
  journal = {Phys. Rev. B},
  volume = {110},
  issue = {17},
  pages = {174508},
  numpages = {13},
  year = {2024},
  month = {Nov},
  publisher = {American Physical Society},
  doi = {10.1103/PhysRevB.110.174508},
  url = {https://link.aps.org/doi/10.1103/PhysRevB.110.174508}
}

@misc{Mishra2025,
      title={{Electron-phonon vertex correction effect in superconducting H$_3$S}},
      author={Shashi B. Mishra and Hitoshi Mori and Elena R. Margine},
      year={2025},
      eprint={2507.01897},
      archivePrefix={arXiv},
      primaryClass={cond-mat.supr-con},
      url={https://arxiv.org/abs/2507.01897},
}

@article{Liu2025,
  title = {{Phonon-limited carrier transport in the Weyl semimetal TaAs}},
  author = {Liu, Zhe and Mishra, Shashi B. and Lihm, Jae-Mo and Ponc\'e, Samuel and Margine, Elena R.},
  journal = {Phys. Rev. B},
  volume = {112},
  issue = {10},
  pages = {104311},
  numpages = {9},
  year = {2025},
  month = {Sep},
  publisher = {American Physical Society},
  doi = {10.1103/x8zl-w5x3},
  url = {https://link.aps.org/doi/10.1103/x8zl-w5x3}
}

@article{Bushick2023,
  title = {{Phonon-Assisted Auger-Meitner Recombination in Silicon from First Principles}},
  author = {Bushick, Kyle and Kioupakis, Emmanouil},
  journal = {Phys. Rev. Lett.},
  volume = {131},
  issue = {7},
  pages = {076902},
  numpages = {6},
  year = {2023},
  month = {Aug},
  publisher = {American Physical Society},
  doi = {10.1103/PhysRevLett.131.076902},
}

@article{ZhangX2022,
  title = {Ab initio theory of free-carrier absorption in semiconductors},
  author = {Zhang, Xiao and Shi, Guangsha and Leveillee, Joshua A. and Giustino, Feliciano and Kioupakis, Emmanouil},
  journal = {Phys. Rev. B},
  volume = {106},
  issue = {20},
  pages = {205203},
  numpages = {8},
  year = {2022},
  month = {Nov},
  publisher = {American Physical Society},
  doi = {10.1103/PhysRevB.106.205203},
  url = {https://link.aps.org/doi/10.1103/PhysRevB.106.205203}
}

@misc{Alvarez2025,
      title={Electron-phonon coupling in magnetic materials using the local spin density approximation},
      author={\'A. A. Carrasco \'Alvarez and M. Giantomassi and J. Lihm and G. E. Allemand and M. Mignolet and M. Verstraete and S. Ponc\'{e}},
      year={2025},
      eprint={2510.11350},
      archivePrefix={arXiv},
      primaryClass={cond-mat.mtrl-sci},
      url={https://arxiv.org/abs/2510.11350},
}

@article{Leveillee2023,
  title = {Ab initio calculation of carrier mobility in semiconductors including ionized-impurity scattering},
  author = {Leveillee, Joshua and Zhang, Xiao and Kioupakis, Emmanouil and Giustino, Feliciano},
  journal = {Phys. Rev. B},
  volume = {107},
  issue = {12},
  pages = {125207},
  numpages = {19},
  year = {2023},
  month = {Mar},
  publisher = {American Physical Society},
  doi = {10.1103/PhysRevB.107.125207},
  url = {https://link.aps.org/doi/10.1103/PhysRevB.107.125207}
}

@Article{Zhou2019,
author={Zhou, Jun
and Shen, Lei
and Costa, Miguel Dias
and Persson, Kristin A.
and Ong, Shyue Ping
and Huck, Patrick
and Lu, Yunhao
and Ma, Xiaoyang
and Chen, Yiming
and Tang, Hanmei
and Feng, Yuan Ping},
title={{2DMatPedia, an open computational database of two-dimensional materials from top-down and bottom-up approaches}},
journal={Scientific Data},
year={2019},
month={Jun},
day={12},
volume={6},
number={1},
pages={86},
issn={2052-4463},
doi={10.1038/s41597-019-0097-3},
url={https://doi.org/10.1038/s41597-019-0097-3}
}

@Article{Merchant2023,
author={Merchant, Amil
and Batzner, Simon
and Schoenholz, Samuel S.
and Aykol, Muratahan
and Cheon, Gowoon
and Cubuk, Ekin Dogus},
title={Scaling deep learning for materials discovery},
journal={Nature},
year={2023},
month={Dec},
day={01},
volume={624},
number={7990},
pages={80-85},
issn={1476-4687},
doi={10.1038/s41586-023-06735-9},
url={https://doi.org/10.1038/s41586-023-06735-9}
}

@misc{SI,
  note = {{See Supplementary Information at \url{} for additional details regarding convergence tests, Wannierization results, and comparison of electron-phonon matrix elements.}}
}

@misc{TACC,
  title={{Texas Advanced Computing Center (TACC)}},
  howpublished = {Service provided by the {University of Texas at Austin}, URL: \url{http://www.tacc.utexas.edu}},
  url={http://www.tacc.utexas.edu}
}

@inbook{Frontera,
author = {Stanzione, Dan and West, John and Evans, R. Todd and Minyard, Tommy and Ghattas, Omar and Panda, Dhabaleswar K.},
title = {{Frontera: The Evolution of Leadership Computing at the National Science Foundation}},
year = {2020},
isbn = {9781450366892},
publisher = {ACM},
address = {New York, NY, USA},
url = {https://doi.org/10.1145/3311790.3396656},
booktitle = {Practice and Experience in Advanced Research Computing},
pages = {106-111},
numpages = {6}
}

@inproceedings{Expanse,
author = {Strande, Shawn and Cai, Haisong and Tatineni, Mahidhar and Pfeiffer, Wayne and Irving, Christopher and Majumdar, Amit and Mishin, Dmitry and Sinkovits, Robert and Norman, Mike and Wolter, Nicole and Cooper, Trevor and Altintas, Ilkay and Kandes, Marty and Perez, Ismael and Shantharam, Manu and Thomas, Mary and Sivagnanam, Subhashini and Hutton, Thomas},
title = {{Expanse: Computing without Boundaries: Architecture, Deployment, and Early Operations Experiences of a Supercomputer Designed for the Rapid Evolution in Science and Engineering}},
year = {2021},
isbn = {9781450382922},
publisher = {Association for Computing Machinery},
address = {New York, NY, USA},
url = {https://doi.org/10.1145/3437359.3465588},
doi = {10.1145/3437359.3465588},
booktitle = {Practice and Experience in Advanced Research Computing 2021: Evolution Across All Dimensions},
articleno = {47},
numpages = {4},
location = {Boston, MA, USA},
series = {PEARC '21}
}

@article{Perdew1996,
  title = {Generalized Gradient Approximation Made Simple},
  author = {Perdew, John P. and Burke, Kieron and Ernzerhof, Matthias},
  journal = {Phys. Rev. Lett.},
  volume = {77},
  issue = {18},
  pages = {3865--3868},
  numpages = {0},
  year = {1996},
  month = {Oct},
  publisher = {American Physical Society},
  doi = {10.1103/PhysRevLett.77.3865},
  url = {https://link.aps.org/doi/10.1103/PhysRevLett.77.3865}
}

\end{document}


\title{Supplementary Information\\EPW-VASP interface for first-principles calculations of electron-phonon interactions}

\author{Danylo Radevych\footnotemark[2]{$^\dagger$}}
\thanks{These authors contributed equally.}
\affiliation{
Department of Physics, Applied Physics and Astronomy, Binghamton University-SUNY,
PO Box 6000, Binghamton, New York 13902, USA
}

\author{Aidan Thorn}
\thanks{These authors contributed equally.}
\affiliation{
Department of Physics, Applied Physics and Astronomy, Binghamton University-SUNY,
PO Box 6000, Binghamton, New York 13902, USA
}

\author{Manuel Engel}
\affiliation{University of Vienna, Faculty of Physics and Center for Computational Materials Physics, A-1090 Vienna, Austria}
\affiliation{VASP Software GmbH, A-1090, Vienna, Austria}

\author{Aleksey N. Kolmogorov}
\affiliation{
Department of Physics, Applied Physics and Astronomy, Binghamton University-SUNY,
PO Box 6000, Binghamton, New York 13902, USA
}

\author{Sabyasachi Tiwari}
\affiliation{
Oden Institute for Computational Engineering and Sciences, The University of Texas at Austin, Austin, TX 78712, USA
}
\affiliation{
Department of Physics, The University of Texas at Austin, Austin, TX 78712, USA
}

\author{Georg Kresse}
\affiliation{University of Vienna, Faculty of Physics and Center for Computational Materials Physics, A-1090 Vienna, Austria}
\affiliation{VASP Software GmbH, A-1090, Vienna, Austria}

\author{Feliciano Giustino}
\affiliation{
Oden Institute for Computational Engineering and Sciences, The University of Texas at Austin, Austin, TX 78712, USA
}
\affiliation{
Department of Physics, The University of Texas at Austin, Austin, TX 78712, USA
}

\author{Elena R. Margine\footnotemark[2]{$^\ddagger$}}
\affiliation{
Department of Physics, Applied Physics and Astronomy, Binghamton University-SUNY,
PO Box 6000, Binghamton, New York 13902, USA
}

\maketitle

\footnotetext[2]{Corresponding author: \href{mailto:dradevych@binghamton.edu}{dradevych@binghamton.edu}}
\footnotetext[3]{Corresponding author: \href{mailto:rmargine@binghamton.edu}{rmargine@binghamton.edu}}

\supplementarysection

\section{Convergence tests}

\subsection{MgB$_2$: $\bq$-grid sampling}

\begin{figure}[thbp]
\centering
\includegraphics[width=0.52\columnwidth]{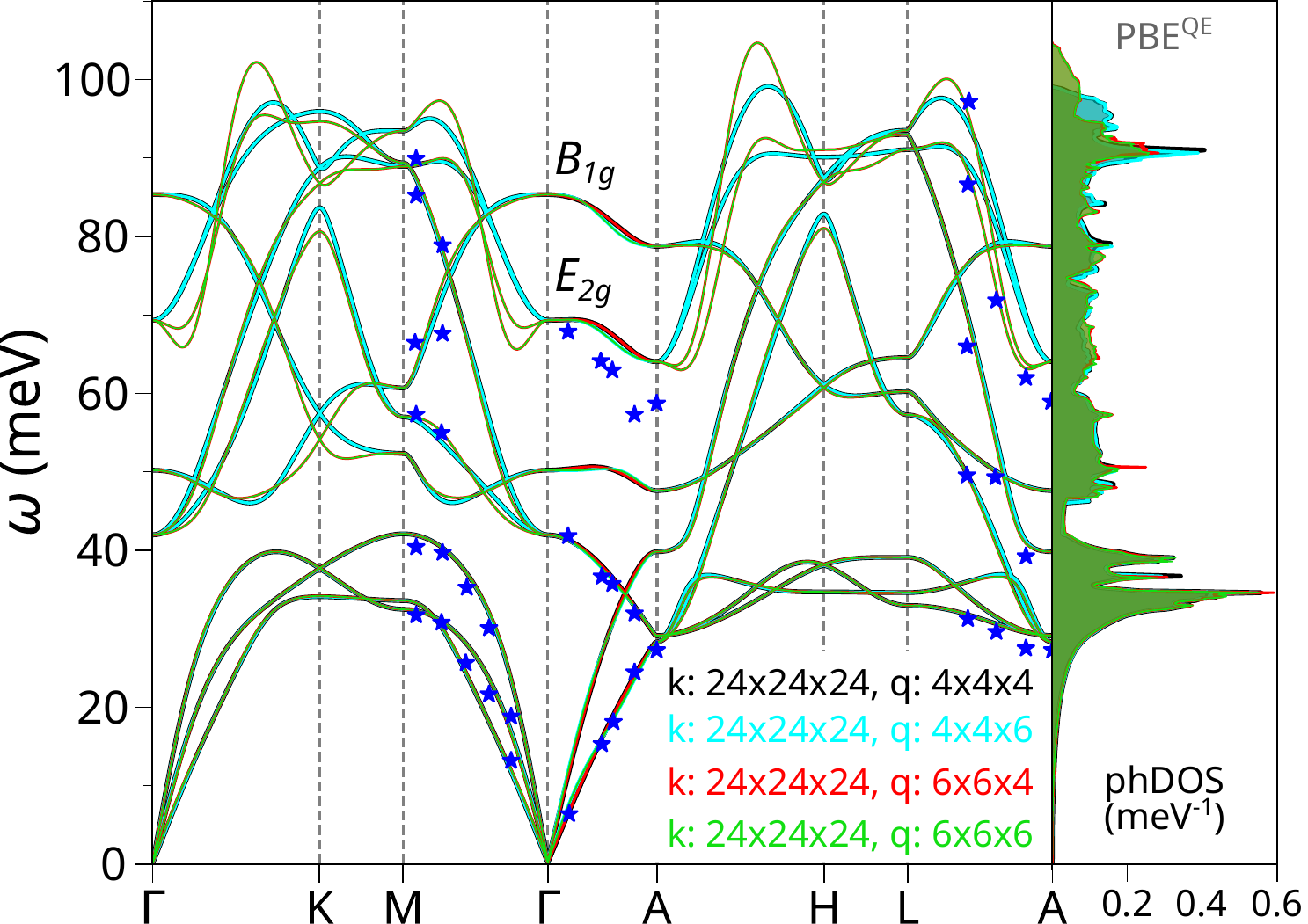}
\caption{Effect of different $\bq$-point meshes on the MgB$_2$ phonon dispersion calculated with QE-PBE. In all cases, a 24$\times$24$\times$24 $\bk$-point mesh is used. Reference experimental data (blue stars) are taken from Ref.~\cite{Shukla2003}. }
\label{fig:mgb2_qe-pbe_q-grids}
\end{figure}

The $\bq$-grid convergence tests were performed in Quantum ESPRESSO with the PBE functional (QE-PBE) within the DFPT framework. Figure~\ref{fig:mgb2_qe-pbe_q-grids} shows the MgB$_2$ phonon dispersion for different $\bq$-grids, using a fixed $\bk$-point mesh of 24$\times$24$\times$24. From Fig.~\ref{fig:mgb2_qe-pbe_q-grids}, it is evident that the choice of the in-plane $\bq$-grid sampling has a noticeable effect on the $E_{2g}$ mode splitting along the $\Gamma$-M and A-H directions. The 4$\times$4$\times$4 and 4$\times$4$\times$6 $\bq$-meshes did not reproduce the splitting accurately, whereas the denser in-plane 6$\times$6$\times$4 and 6$\times$6$\times$6 $\bq$-grids yielded similar phonon dispersions. The 6$\times$6$\times$4 $\bq$-grid was selected for final calculations, as it enabled the use of a corresponding 6$\times$6$\times$4 supercell in the finite-displacement method, while larger supercells would have been significantly more computationally demanding.

\subsection{MgB$_2$: $\bk$-grid sampling}

\begin{figure}[thbp]
\centering
\includegraphics[width=1\columnwidth]{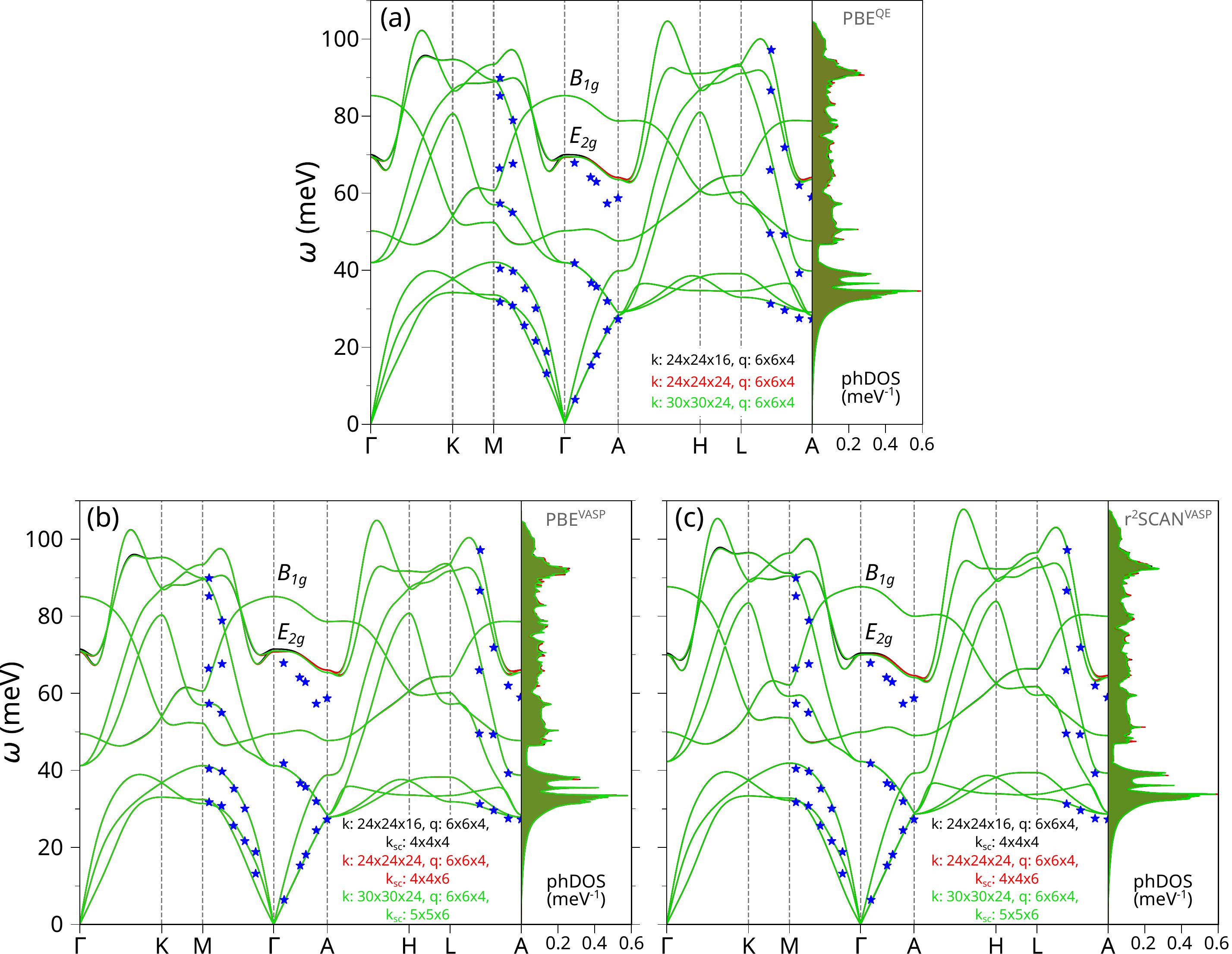}
\caption{Effect of different $\bk$-point meshes on the MgB$_2$ phonon dispersion calculated with (a) QE-PBE, (b) VASP-PBE, and (c) VASP-r$^2$SCAN. In all cases, a 6$\times$6$\times$4 $\bq$-point mesh (DFPT) or 6$\times$6$\times$4 supercell (finite-displacement method) is used.
The $\bk$-point meshes are given for the unit cells. For VASP, the corresponding supercell $\bk$-grids are indicated with the ``sc'' subscript. Reference experimental data (blue stars) are taken from Ref.~\cite{Shukla2003}. }
\label{fig:mgb2_qe-pbe_k-grids}
\end{figure}

The $\bk$-mesh convergence tests were performed for both DFPT and finite-displacement methods, using a fixed $\bq$-grid (DFPT) or supercell size (finite displacement) of 6$\times$6$\times$4. Phonon dispersions computed with QE-PBE, VASP-PBE, and VASP-r$^2$SCAN are shown in Figs.~\ref{fig:mgb2_qe-pbe_k-grids}(a)-(c). For QE, three $\bk$-grids for the unit cell were considered: 24$\times$24$\times$16, 24$\times$24$\times$24, and 30$\times$30$\times$24. For VASP, the $\bk$-grid for the supercell was defined by scaling the corresponding unit cell $\bk$-mesh in QE according to the supercell size in each direction. In other words, the $\bk$-grid density of the supercell was kept consistent with that of the unit cell. Figure~\ref{fig:mgb2_qe-pbe_k-grids} shows that the difference
between the sparsest unit cell $\bk$-mesh (24$\times$24$\times$16) and the densest (30$\times$30$\times$24) is only noticeable for the two degenerate $E_{2g}$ modes along the $\Gamma$-A direction. For all functionals, using the 30$\times$30$\times$24  unit cell $\bk$-grid shifted the $E_{2g}$ modes by approximately 1~meV downward, bringing them closer to the experimental values. Since employing the 30$\times$30$\times$24 $\bk$-mesh was not computationally demanding, it was adopted for all final calculations.

\subsection{c-BN: $\bk$- and $\bq$-grid sampling}

\begin{figure}[thbp]
\centering
\includegraphics[width=0.52\columnwidth]{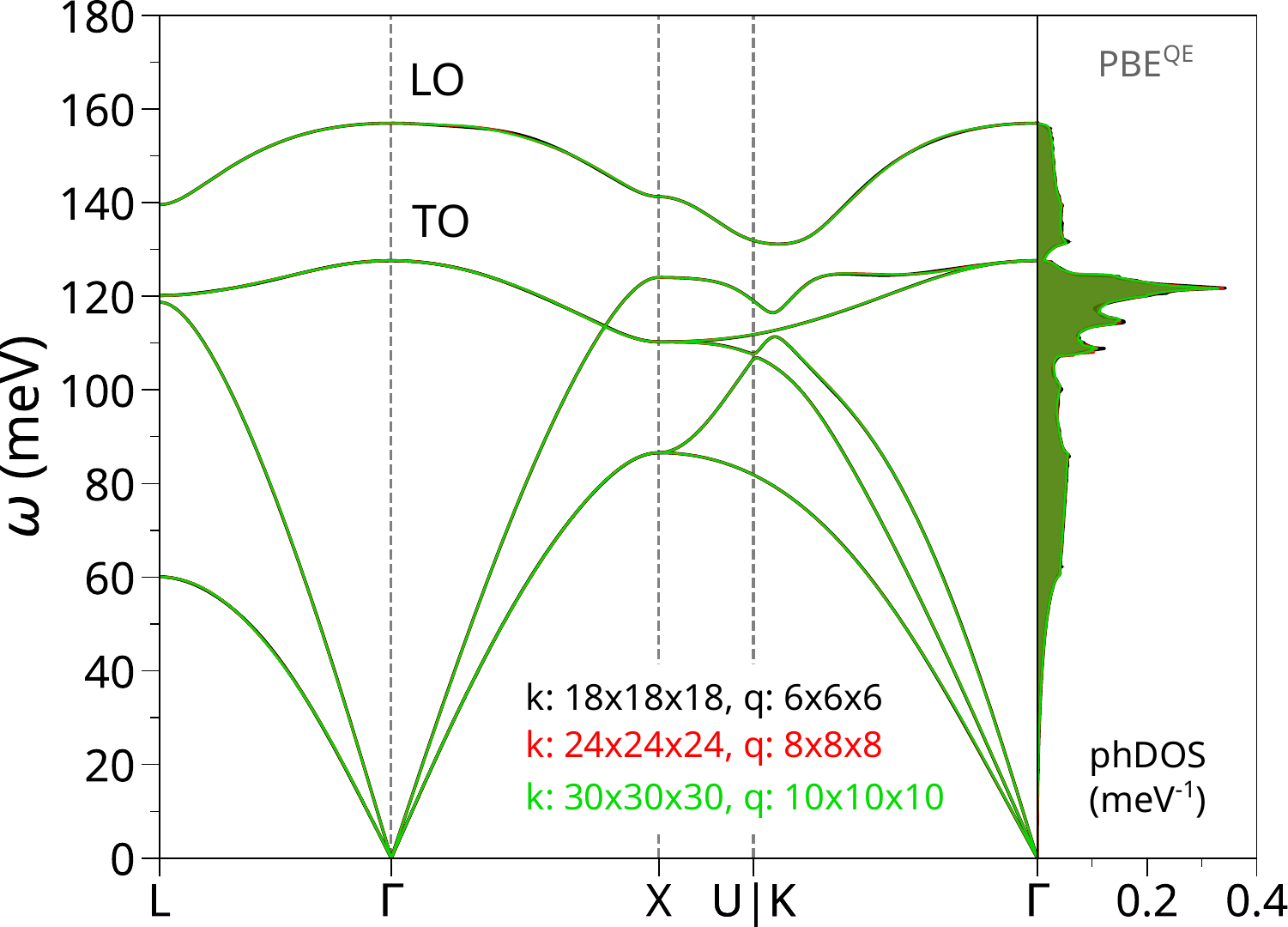}
\caption{Effect of different $\bk$, $\bq$-point meshes on the c-BN phonon dispersion calculated with QE-PBE.}
\label{fig:cbn_qe-pbe_qk-grids}
\end{figure}

Figure~\ref{fig:cbn_qe-pbe_qk-grids} shows the phonon dispersion of c-BN calculated with QE-PBE using different $\bk$ and $\bq$-grids, and demonstrates that the chosen meshes of 18$\times$18$\times$18 and 6$\times$6$\times$6 are sufficient, as increasing them to 30$\times$30$\times$30 and 10$\times$10$\times$10, respectively, does not affect the phonon modes.

\newpage
\section{Wannierization}

\begin{figure*}[htbp]
\centering
\includegraphics[width=0.6\columnwidth]{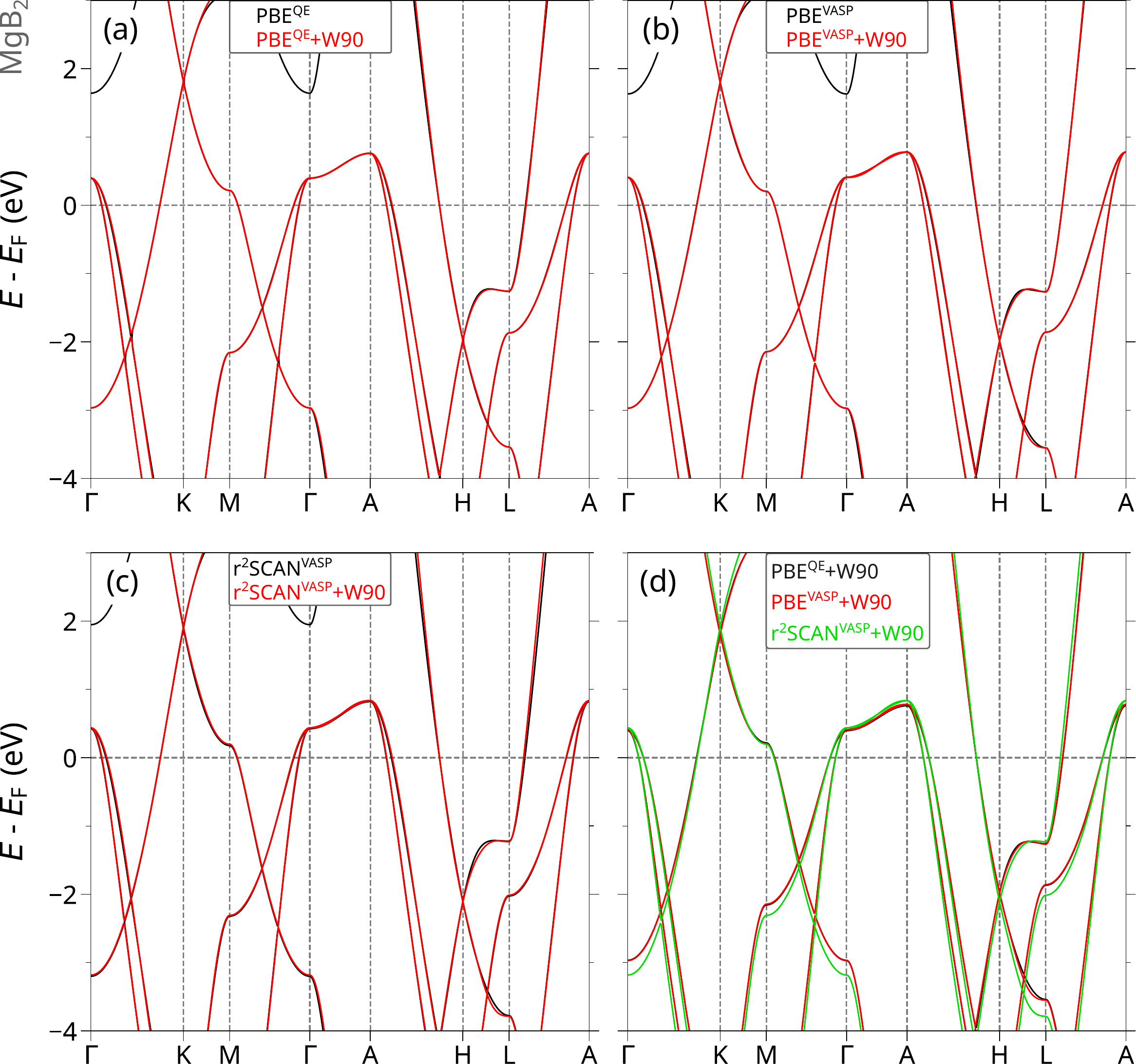}
\caption{Comparison between DFT (black) and Wannier-interpolated (red) band structures in MgB$_2$ for (a) QE-PBE, (b) VASP-PBE, and (c) VASP-r$^2$SCAN. Panel (d) shows a comparison of the Wannier-interpolated band structures from panels (a)-(c).
}
\label{fig:wannierization_mgb2}
\end{figure*}

\begin{figure*}[htbp]
\centering
\includegraphics[width=0.6\columnwidth]{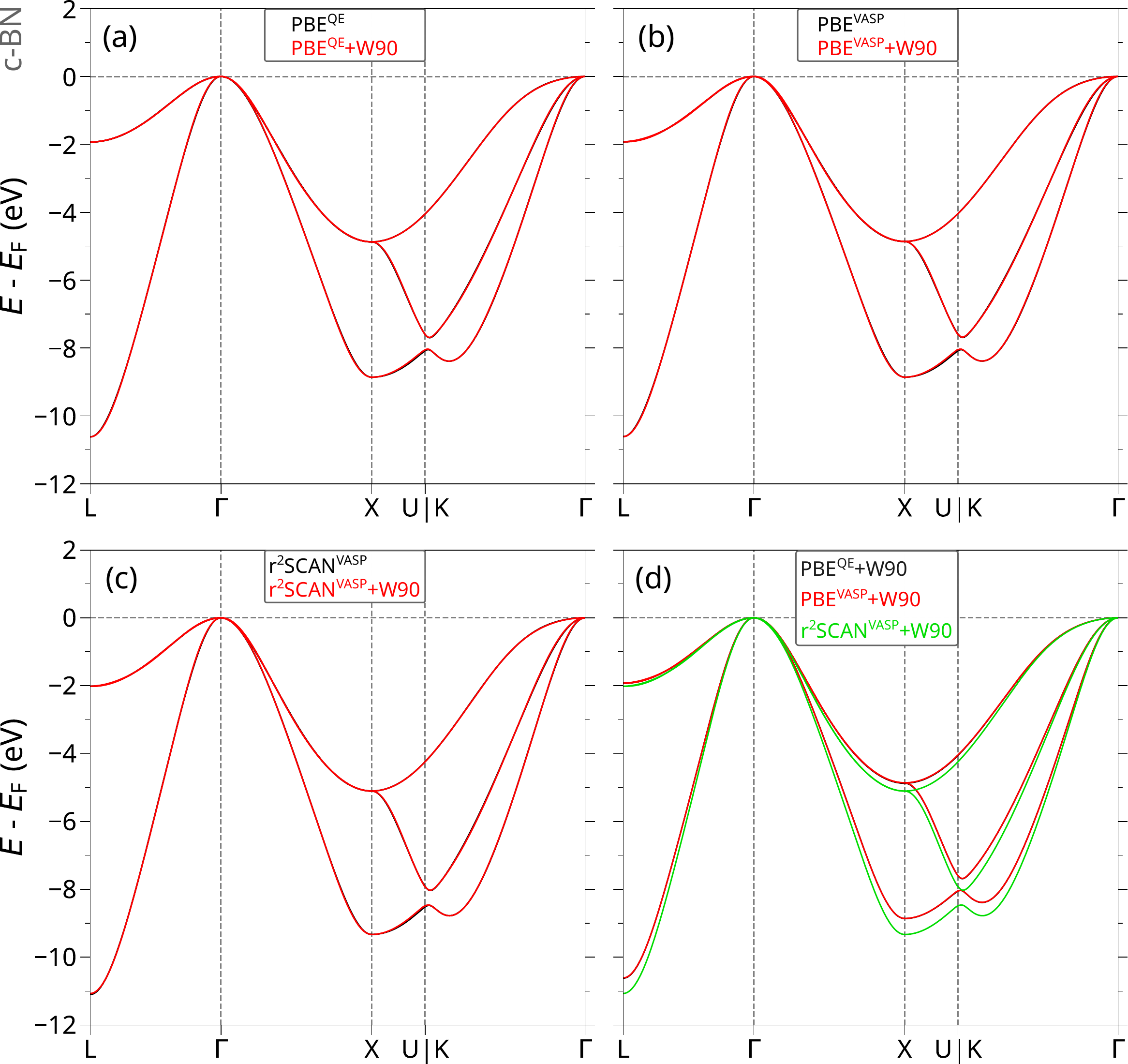}
\caption{Comparison between DFT (black) and Wannier-interpolated (red) band structures in c-BN for (a) QE-PBE, (b) VASP-PBE, and (c) VASP-r$^2$SCAN. Panel (d) shows a comparison of the Wannier-interpolated band structures from panels (a)-(c).
}
\label{fig:wannierization_cbn}
\end{figure*}

Figures~\ref{fig:wannierization_mgb2} and \ref{fig:wannierization_cbn} show comparisons between the DFT and Wannier-interpolated band structures as well as between the Wannierized bands obtained with the PBE and r$^2$SCAN functionals for MgB$_2$ and c-BN.

\section{Electron-phonon matrix elements}
\begin{figure*}[thbp]
\centering
\includegraphics[width=0.9\columnwidth]{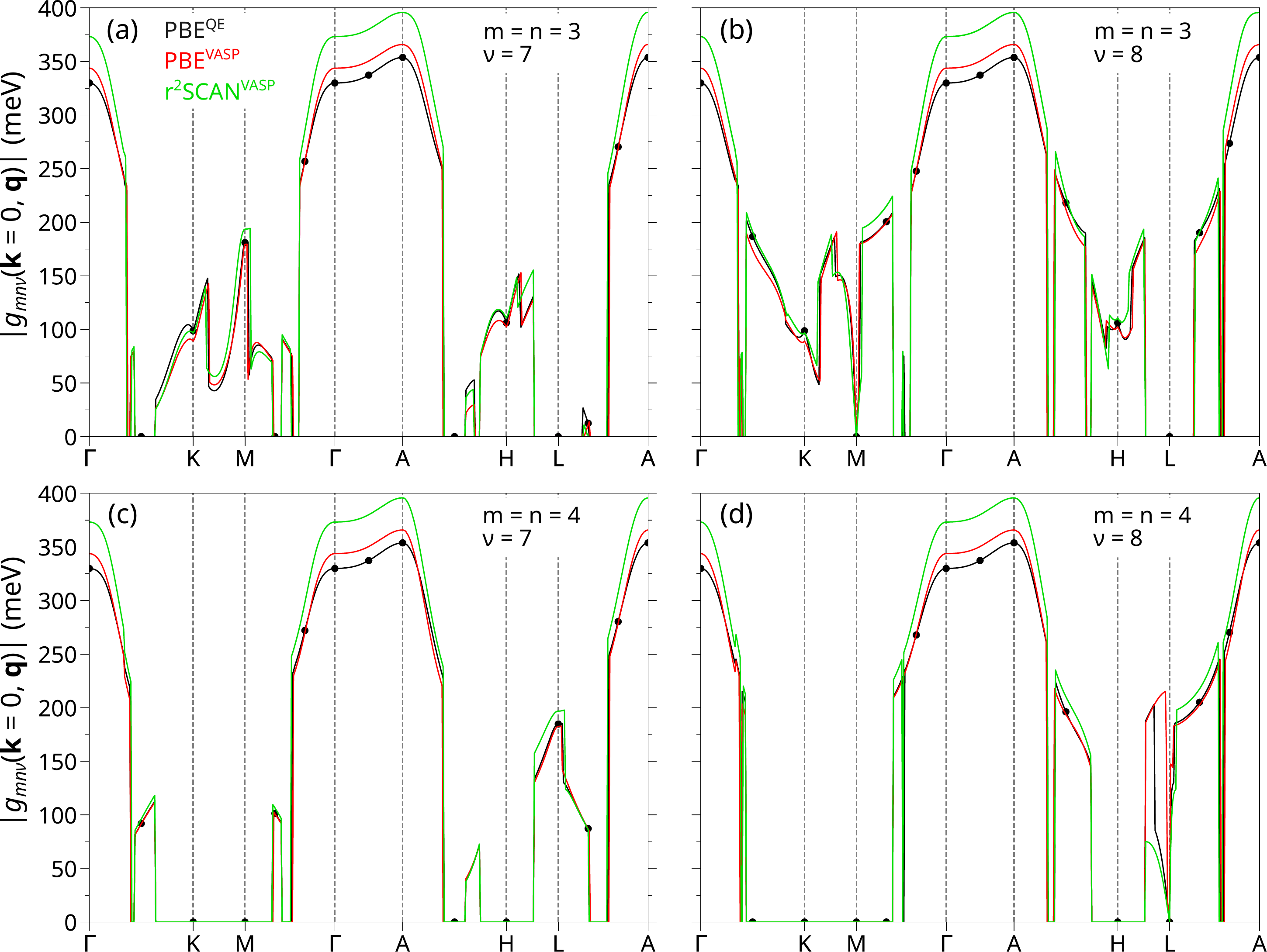}
\caption{Comparison of absolute values of the selected interpolated electron-phonon matrix elements, $|g_{mn\nu}(\bk = 0, \bq)|$, in
MgB$_2$ calculated with different functionals along the $\bq$ path for $m = n = 3,~4$ and $\nu = 7,~8$:
(a) $|g_{3,3,7}|$, (b) $|g_{3,3,8}|$, (c) $|g_{4,4,7}|$, and (d) $|g_{4,4,8}|$.
Black circles show reference QE-PBE DFPT data calculated at the points coinciding
with the coarse $\bq$-grid of $6\times6\times4$.
}
\label{fig:g_mgb2}
\end{figure*}

Figure~\ref{fig:g_mgb2} shows selected interpolated electron-phonon matrix elements, $g_{m n \nu}(\bk = 0, \bq)$, in
MgB$_2$ calculated with different functionals along the $\bq$ path for $m = n = 3,~4$ and $\nu = 7,~8$.
Note that indices $m$, $n$, and $\nu$ numerate sorted energies and frequencies and do not generally
follow bands and modes of the same representations along different high-symmetry lines.
At the $\Gamma$ point, selected indices $m = n = 3,~4$ correspond to two degenerate $\sigma$ bands
just above the Fermi energy,
and the index $\nu$ values of 7, 8 correspond to two degenerate $E_{2g}$ phonon modes.

\begin{figure*}[thbp]
\centering
\includegraphics[width=0.9\columnwidth]{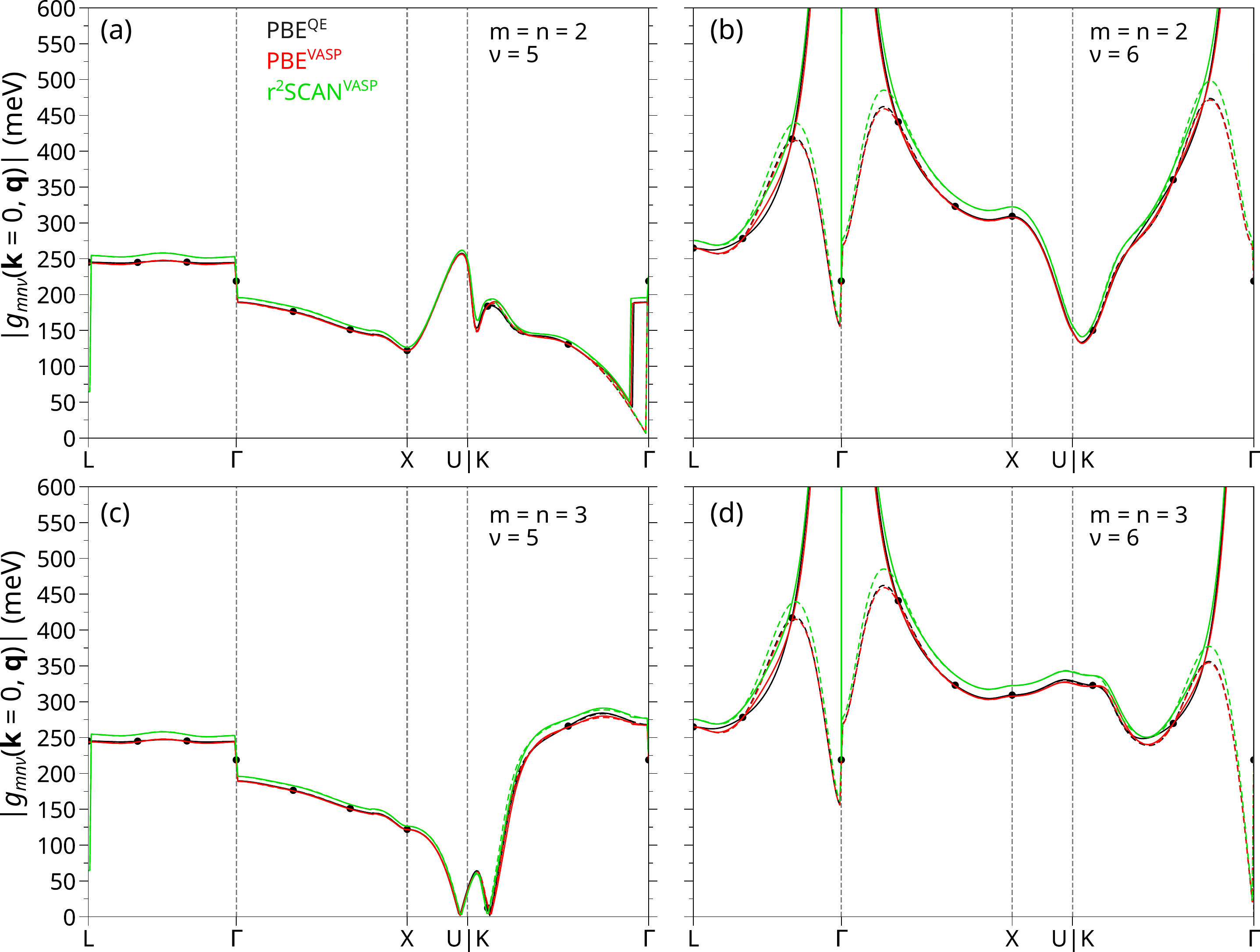}
\caption{Comparison of absolute values of the selected interpolated electron-phonon matrix elements, $|g_{mn\nu}(\bk = 0, \bq)|$, in
c-BN calculated with different functionals along the $\bq$ path for $m = n = 2,~3$ and $\nu = 5,~6$:
(a) $|g_{2,2,5}|$, (b) $|g_{2,2,6}|$, (c) $|g_{3,3,5}|$, and (d) $|g_{3,3,6}|$.
Dashed lines show values with no long-range corrections, and solid lines represent results with long-range dipole corrections included.
Black circles show reference QE-PBE DFPT data calculated at the points coinciding
with the coarse $\bq$-grid of $6\times6\times6$.
}
\label{fig:g_cbn}
\end{figure*}

Selected electron-phonon matrix elements in c-BN for $m = n = 2,~3$ and $\nu = 5,~6$  are shown in Fig.~\ref{fig:g_cbn}, where dashed lines indicate results without
long-range corrections, and solid lines are obtained with
long-range dipole corrections included. At the $\Gamma$ point, band indices $m = 2,~3$ belong to degenerate bands, and indices $\nu = 5,~6$ correspond to one of the two TO modes and one LO mode,
respectively.

\bibliography{references.bib}